\documentclass[%
 reprint,
superscriptaddress,
 amsmath,amssymb,
 aps,
]{revtex4-2}

\pdfoutput=1
\usepackage[utf8]{inputenc}
\usepackage[english]{babel}
\usepackage[T1]{fontenc}
\usepackage{amsmath}
\usepackage{hyperref}
\usepackage[braket, qm]{qcircuit}
\usepackage{braket}

\usepackage{graphicx}

\usepackage{tikz}
\usepackage{lipsum}
\usepackage[normalem]{ulem}

\begin{document}

\title{Quantum computer error structure probed by quantum error correction syndrome measurements}

\author{Spiro Gicev}\email{gicevs@unimelb.edu.au}\affiliation{School of Physics, University of Melbourne, Parkville, 3010, VIC, Australia.}
% \orcid{0000-0000-0000-0000}
\author{Lloyd C. L. Hollenberg} \email{lloydch@unimelb.edu.au} \affiliation{School of Physics, University of Melbourne, Parkville, 3010, VIC, Australia.}
% \orcid{0000-0000-0000-0000}
\author{Muhammad Usman} \email{musman@unimelb.edu.au} \affiliation{School of Physics, University of Melbourne, Parkville, 3010, VIC, Australia.} \affiliation{Data61, CSIRO, Clayton, VIC 3168, Australia.}
% \orcid{0000-0000-0000-0000}

\begin{abstract}
  With quantum devices rapidly approaching qualities and scales needed for fault tolerance, the validity of simplified error models underpinning the study of quantum error correction needs to be experimentally evaluated. In this work, we have assessed the performance of IBM superconducting quantum computer devices implementing heavy-hexagon code syndrome measurements with increasing circuit sizes up to 23 qubits, against the error assumptions underpinning code threshold calculations. Circuit operator change rate statistics in the presence of depolarizing and biased noise were modelled using analytic functions of error model parameters. Data from 16 repeated syndrome measurement cycles was found to be inconsistent with a uniform depolarizing noise model, favouring instead biased and inhomogeneous noise models. Spatial-temporal correlations investigated via $Z$ stabilizer measurements revealed significant temporal correlation in detection events. These results highlight the non-trivial structure which may be present in the noise of quantum error correction circuits, revealed by operator measurement statistics, and support the development of noise-tailored codes and decoders to adapt.\end{abstract}

\maketitle
\section{Introduction}
Quantum computing has the potential to offer significant computational advantage for many computationally intensive tasks such as quantum materials simulations \cite{Feynman1982-gq}, optimization \cite{farhi2014quantum}, machine learning \cite{huang2022quantum, West2023, Cerezo2022}, and search in large databases \cite{grover1996fast}. However, currently available Noisy Intermediate Scale Quantum (NISQ) \cite{preskill2018quantum} devices exhibit prohibitively high error rates which can significantly hinder the successful execution of sufficiently deep quantum circuits relevant for algorithms of practical interest \cite{brandhofer2021arsonisq}. Although error rates of quantum devices can be expected to gradually decrease in the next few years, the implementation of quantum error correction (QEC) will be necessary to overcome the detrimental impact of errors and noise in a scalable manner \cite{shor1995scheme}. The ultimate goal of these efforts is to execute error-corrected quantum algorithms of practical interest, a milestone known as fault-tolerant quantum computing (FTQC) \cite{shor1996fault}. This requires overcoming challenges distinct from those of focus in the NISQ era \cite{martinis2021saving, xu2022distributed}. Competitive FTQC resource estimates, usually based on surface code QEC, require millions of physical qubits and extended periods of stable operation to encode the required number of logical qubits and facilitate logical qubit operations \cite{gidney2019factor, litinski2019game, beverl2022assessing}. Accurate evaluation of device performance as progress is made towards this regime is needed to estimate how and when FTQC will be a realistic possibility.

The progression of device quality towards FTQC standards is usually indicated by comparing device error rates, such as those found with methods such as randomized benchmarking \cite{PhysRevA.77.012307, magesan2011scalable, eisert2020quantum}, to QEC code threshold error rates calculated by simulation \cite{Raussendorf_2007, wang2009threshold, wang2009graphical, fowler2012towards, stephens2014fault, chamberland2020topological}. As these simulations often make many simplifying approximations and assumptions regarding noise, they can often be treated only as order of magnitude estimates for the parameters necessary/sufficient for real physical devices performing QEC, such as $\approx1\%$ for surface codes. Additionally, the error rates and decoherence times obtained by characterization methods can also falsely suggest that noise in such devices is well described by single-qubit processes, when in reality additional effects such as cross-talk and leakage may also be present \cite{zhao2022quantum}. Simultaneous randomized benchmarking partially addresses effects arising from concurrent operations \cite{gambetta2012characterization}. However, such methods still evaluate performance using sets of circuits that are largely disjoint from those of FTQC. Metrics of device noise derived primarily from QEC-related circuits are more ideal indicators of progress made towards FTQC as they are sensitive precisely to the characteristics of noise relevant for FTQC.

Recently, Debroy et al. attempted to address the impact of context-related errors with the Context Aware Fidelity Estimation (CAFE) framework \cite{debroy2023context}. This was shown to be able to obtain knowledge of qubit error characteristics present specifically during stabilizer measurement which was not visible in individual gate characterization. There has been persistent effort towards the idea of using measurements produced by QEC circuits themselves as data for characterization protocols. The foundational results on QEC-based noise characterization initially investigated protocols theoretically and made arguments that relevant error characteristics can be extracted exclusively using (sometimes modified) stabilizer measurement circuits \cite{fowler2014scalable, combes2014insitu, fujiwara2014instantaneous}. Later, similar arguments were made when investigating the accurate tracking of time-dependent error characteristics \cite{Huo_2017, spitz2018adaptive}. Wagner et al. continued this effort by providing additional clarification as to the characteristics that can be learned using exclusively syndrome data \cite{Wagner_2022, wagner2023learning}. With the use of repetition codes, Wootton demonstrated experimentally that error rates can be found for superconducting transmon qubits based on syndrome change events of repetition codes \cite{wootton2022syndromederived}. With quantum device sizes and available control techniques continuing to progress, more of these techniques will likely see further development and testing in the near future.

% This section discusses QEC code benchmarking.
Superconducting quantum devices can now contain up to hundreds of individually addressable physical qubits \cite{borner2023classical, morvan2023phase}. However, still in the NISQ era, the performance evaluation of these devices has focused primarily on metrics of individual circuit operations \cite{knill2008randomized, magesan2012efficient}, general-purpose quantum device metrics \cite{cross2019validating}, demonstrations of quantum properties \cite{https://doi.org/10.1002/qute.202100061} and optimized instances of individual NISQ algorithms \cite{bharti2022noisy}. Multiple groups have also characterized the performance of implementations of small QEC codes, including details of operator measurement behaviour \cite{chen2021nature, krinner2022realizing, andersen2020naturephysics, marques2022logical, zhao2022realization, Sivak_2023, chen2022calibrated, sundaresan2023demonstrating, wootton2022measurements}. The conclusions drawn from these results all have the benefit of being naturally sensitive to device details that are relevant to FTQC. Recent results include work from groups that benchmark the logical error rates of codes at different distances \cite{chen2021nature, acharya2022suppressing}, perform logical qubit operations with lattice surgery \cite{erhard2021entangling, marques2022logical} and verify topological properties \cite{satzinger2021realizing, xu2023digital}. These act as effectively system-scale benchmarks but often use operator measurement performance to accurately characterize device performance in the intermediate regime between individual qubits and gates and complete quantum algorithms. Previous experimental literature featuring heavy-hexagon codes has focused on characterizing logical qubit preparation, measurement, and decoding for heavy-hexagon codes of distance-2 and distance-3 using superconducting transmon qubits \cite{chen2022calibrated, sundaresan2023demonstrating}. These investigations, based on data after the decoding process, revealed instances of leakage, cross-talk and unwanted interactions which can be present in real QEC operation, highlighting the inadequacy of the uniform depolarizing noise model at the foundation of the field. Here, we have conducted a study focused on what the statistics of quantum error correction operator measurements alone can reveal about a quantum device's noise characteristics at different circuit scales. The methodology we use allows an investigation of device characteristics relevant for quantum error correction circuits without requiring focus on the additional details of decoding, particularly relevant to the above threshold regime.

In this work, we investigated the performance of IBM transmon superconducting quantum computer devices \cite{PhysRevA.76.042319} performing syndrome measurement circuits of heavy-hexagon codes \cite{chamberland2020topological, chen2022calibrated, sundaresan2023demonstrating}. The performance across different scales of circuits relevant to heavy-hexagon codes is examined by applying techniques previously introduced for other QEC codes \cite{wootton2022syndromederived, chen2021nature, krinner2022realizing, wootton2022measurements}. We develop on these methods by showing how operator change rates under depolarizing and biased noise can be modelled analytically. Additionally, we show that this allows their relationship to be efficiently inverted to characterize noise. Applying our technique to our measurement data, we find that circuits of all scales investigated in our work required noise models beyond uniform depolarizing noise in order to be consistent with experimental results. Effects consistent with features of inhomogeneity, biased measurement errors, amplitude damping noise and a bias towards $Z$ errors are present in the noise of the system. Additionally, performance of larger circuits is not readily explicable by the results of performance of smaller circuits. When modelling these circuits with standard QEC error models, our results show that the effective error rates necessary to describe results tend to increase with respect to circuit size. We conclude by discussing the severity of these effects, possible approaches to their management, and the implications they would have on FTQC.

The remainder of this paper is organized as follows. In Sec. \ref{subsec:error_models} we briefly review error models of quantum circuits. In Sec. \ref{subsec:measurements} we give a brief overview of measurements in QEC. In Sec. \ref{subsec:gauge} we investigate the performance of individual gauge operator measurement circuits. In Sec. \ref{subsec:stabil} we investigate the performance of simultaneous measurement of multiple gauge operators to infer measurements of stabilizer generators. Repeated measurement of stabilizer operators is investigated in Sec. \ref{subsec:repeated} and Sec. \ref{subsec:spatial_temporal}, where we focus on individual measurements and correlations among measurements respectively. Throughout Sec. \ref{sec:results}, we compare experimental results with simulation to investigate whether the observed behaviour is explicable by standard error models used in QEC literature and whether the performance scales according to expectations. Finally, in Sec. \ref{sec:conclusion} we provide a summary of our work.

\section{Results}\label{sec:results}

\subsection{Error Models of Quantum 
Circuits}\label{subsec:error_models}
Noise is pervasive in quantum systems, and quickly drowns out quantum effects if left unchecked. Examples of contributions to noise include qubit frequency drift, environmental interactions, thermal relaxation and stochastic/systematic gate errors. In quantum circuits, these phenomena are often modelled by applying noise channels to qubits after every intended operation applied to them. Due to their efficient simulation, Pauli noise models, and in particular the uniform depolarizing noise model, are a popular category of error models and underlie results regarding threshold theorems in QEC \cite{fowler2012proof}. Many distinct noise models have been developed, however, which requires care to be taken when comparing different results.

\begin{figure}
    \centering
    \includegraphics[width=\columnwidth, keepaspectratio]{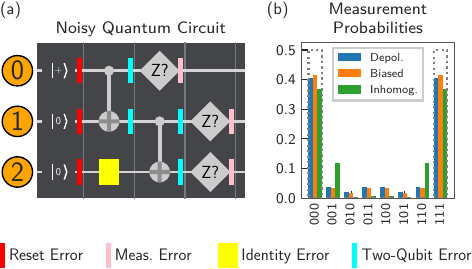}
    \caption{Simulation of quantum circuits under the influence of error models. (a) shows a quantum circuit for the preparation of a GHZ state, with errors indicated after every circuit operation. Reset, measurement, identity and two-qubit errors are represented by red, pink, yellow and cyan markers respectively. (b) shows the $Z$-basis measurement statistics obtained under no noise (dotted outline) compared with $p'=0.05$ uniform depolarizing noise, $(p,\eta)=(0.0375, 10)$ biased noise and inhomogeneous noise corresponding to $p'=0.01$ everywhere except qubit 2 which was given $p'=0.2$ for single qubit gates.} 
    \label{fig:noisy_circuit}
\end{figure}

Figure \ref{fig:noisy_circuit} (a) shows one particular example of a quantum circuit preparing and measuring a GHZ state with noise channels shown explicitly, which are similar to those used in the rest of this text. The distinct channels shown correspond to reset noise after every instance of qubit preparation, delay noise after every instance where a qubit is idle, two-qubit noise after every two-qubit gate and readout noise after every measurement operation. Single qubit depolarizing noise corresponds to the operation,
\begin{equation}
    \rho \rightarrow (1-p')\rho +\frac{p'}{4}\sum_{i=0}^{3} \sigma_i\rho \,\sigma_i,
\end{equation}
where $p'$ is the depolarization parameter (see Supplementary Sec. \ref{quantum_errors} for relation to error rate $p$) and $\sigma_i$ is an element of the set of single qubit Pauli operators \{$I, X, Y, Z\}$, respectively  \cite{nielsen_chuang_2010}. Similarly, two-qubit depolarizing noise corresponds to the operation,
\begin{equation}
    \rho \rightarrow (1-p')\rho +\frac{p'}{16}\sum_{i=0}^{3}\sum_{j=0}^{3} \sigma_i \otimes \sigma_j \rho \,\sigma_i\otimes\sigma_j.
\end{equation}
State preparation and measurement fails with probability $p'/2$, consistent with the probability of depolarizing noise events causing such processes to fail . Modifications to these channels can be made in order to better fit experimental device characteristics. For example, biased noise models increase the probability of $Z$-type errors occurring compared to other errors \cite{Ataides_NComms_2021}. When parameterized with a bias parameter $\eta$ and error rate $p$, single qubit errors perform the operation,
\begin{equation}
    \rho \rightarrow (1-p)\rho +p\sum_{i=1}^{3} r_i \sigma_i\rho\, \sigma_i,
\end{equation}
where $r_1=r_2=1/[2(\eta+1)]$ and $r_3=\eta/(\eta+1)$. Lastly, the inhomogeneous noise model modifies uniform depolarizing noise by allowing varying noise intensity across different qubits and two-qubit gates (see Supplementary Sec. \ref{quantum_errors} for further descriptions of some other channels the nuances involved with parameterization).

Multiple additional choices/assumptions are made when constructing noise models. For example, a choice must be made on whether each individual single qubit gate is modelled to take a single time step, with an associated error channel, or is treated differently. As single-qubit gates are often the fastest and lowest error rate gates in superconducting systems, they are often neglected in regimes where other gate errors dominate. Another nuance is whether to use $p$ to represent the error rate, corresponding to the probability that a nontrivial Pauli error is applied, or to use the depolarization probability $p'$, corresponding to the probability that qubits become maximally mixed. In this work, we have parameterized uniform depolarizing noise using a depolarization parameter $p'$ and biased noise using an error rate $p$. In general, a particular noise model is fully specified only when each distinct noise channel being used is provided. 

Figure \ref{fig:noisy_circuit} (b), shows how measurement probabilities differ for a pristine circuit compared to uniform depolarizing noise, $Z$-biased noise and inhomogeneous error models. We can see that some detail of the noise model is imprinted in the noise statistics of $Z$ basis measurements. Other basis measurements would also need to be performed to capture all available information.

While the measurement statistics of quantum circuits are a function of the noise present in the system, the converse is not true in general, with distinct noise models being able to give identical measurement statistics. The simplest manifestation of this phenomenon occurs for the task of distinguishing reset errors from measurement errors in a simple prepare and measure circuit for the $\ket{0}$ state. This generalizes to larger systems and other errors, even if access is given to all possible circuits \cite{Chen2023}. Nevertheless, measurement statistics can still help rule out particular classes of noise models, in favour of others, which can help develop understanding regarding the noise present in circuits and assist the design of QEC codes.

\subsection{Quantum Error Correction Measurements}\label{subsec:measurements}
\begin{figure*}[t]
  \centering
  \includegraphics{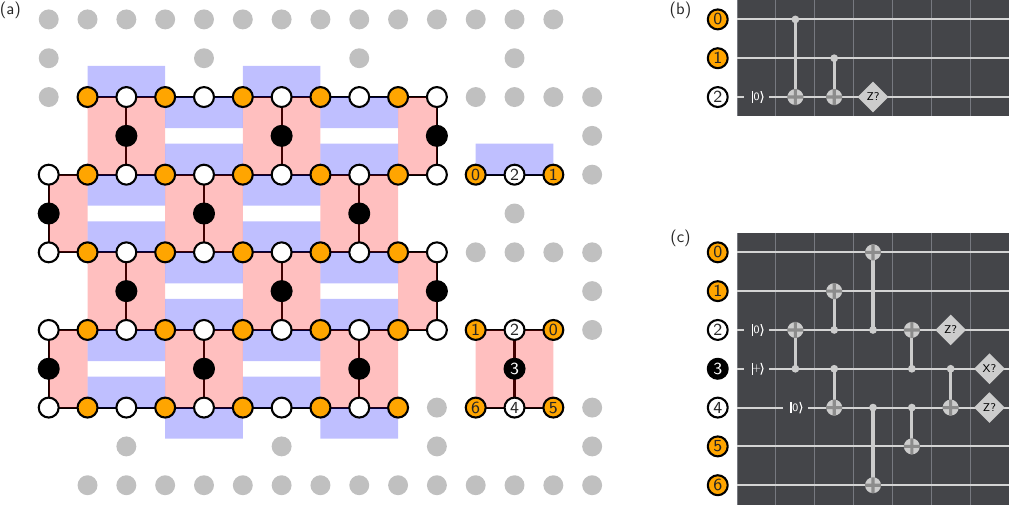}
  \caption{A heavy-hexagon code and the associated gauge operators. (a) shows qubits of a distance-5 heavy-hexagon code on a heavy-hexagon lattice, together with two-qubit $Z$ gauge operators (blue) and four-qubit $X$ flagged gauge operators (red), which also act on two qubits at boundaries. Labelled examples of a two-qubit $Z$ gauge operator and a four-qubit $X$ gauge operator are shown adjacent. (b) and (c) show circuits to measure two-qubit $Z$ gauge operators and four-qubit $X$ flagged gauge operators, respectively.}
  \label{fig:figure1}
\end{figure*}

Operator measurement circuits are expected to constitute a large portion of the instructions given to quantum devices executing fault-tolerant quantum algorithms \cite{gidney2019factor, litinski2019game, beverl2022assessing}. Figure \ref{fig:figure1} shows examples of these circuits for heavy-hexagon codes. Ideally, multiple instances of these circuits are performed in parallel across a quantum device for codes of any size, using circuits of constant depth. The measurement results of these circuits reveal information about errors which may be present at the start of the circuit or which occur throughout the circuit \cite{gottesman1997stabilizer}. Importantly, measurements made in this way do not reveal any information about the encoded quantum information itself. The accurate execution of these circuits is one of the most important contributions to the performance of a quantum error correction implementation.

In this work, we investigate methods of evaluating the performance of stabilizer measurement in quantum error correction at multiple system scales. We begin our investigation by examining the performance of the smallest elements of heavy-hexagon quantum error correction circuits. This corresponds to evaluating qubit, gate and low depth circuit performance. 

\subsection{Gauge Operator Measurement}\label{subsec:gauge}

\begin{figure*}[ht]
    \centering    \includegraphics[width=\textwidth, keepaspectratio]{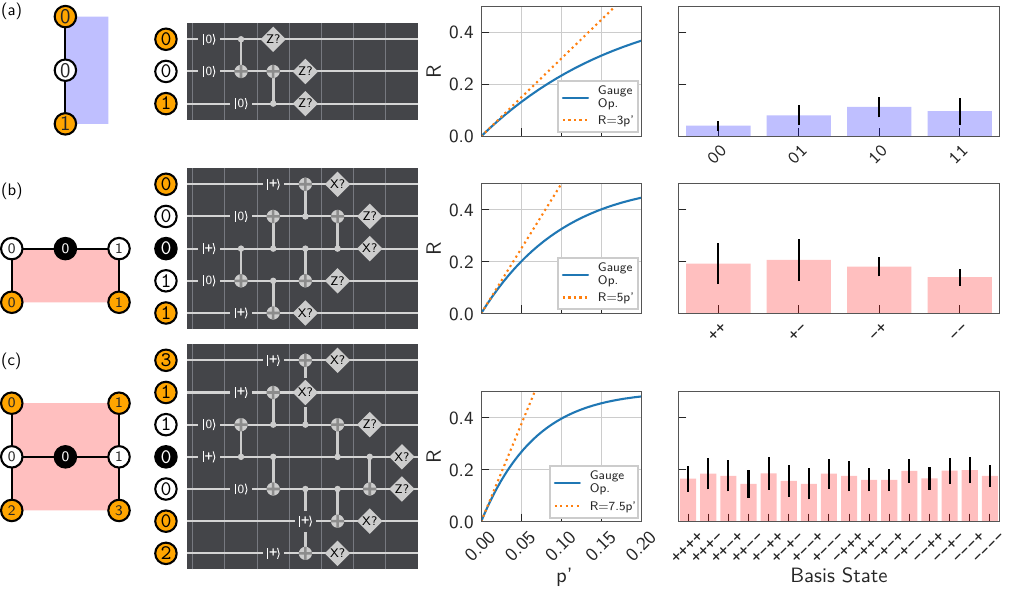}
    \caption{Gauge operator measurement circuit evaluation for the ibmq\char`_montreal device. The three rows (a), (b) and (c) correspond to the benchmarking of $ZZ$ gauge operators, $XX$ flagged gauge operators and $XXXX$ flagged gauge operators. The first column shows the diagram tile representation of each operator. The second column Shows circuits which were used for theoretical simulation. The third column shows theoretical operator change rates, $R$, as a function of depolarizing error parameter, $p'$, for each operator measurement circuit. A dotted line shows the linear component which corresponds to the contribution from single error events. The fourth column shows the operator change rates for each separable eigenstate of the operator under investigation. Values are calculated by averaging over 34 calibrations and all error bars represent one standard deviation.}
    \label{fig:fig_2}
\end{figure*}

We begin our main investigation by considering the measurement of individual operators. The individual operators able to be measured for heavy-hexagon codes are the gauge operators of the code. Their simultaneous measurement can be used to infer the result of stabilizer operator measurements, which is discussed later in Sec. \ref{subsec:stabil}. Measurement of quantum operators corresponds to projection operations applied to the wavefunction of a quantum state \cite{nielsen_chuang_2010}. The applied projection is in general stochastic, but is heralded by the eigenvalue obtained of the measured observable. As these measurements are expected to comprise a significant portion of the circuit operations applied to a fault-tolerant quantum computer, characterization and optimized execution of these circuits is of paramount importance.

A performance metric for these individual operator measurement circuits should capture the degree to which these circuits faithfully measure the required operator. This corresponds to the simultaneous process of applying a correct projection to the quantum state and returning the correct corresponding eigenvalue. Here, we investigated the accuracy of the returned eigenvalues with the use of prepare and measure circuits \cite{krinner2022realizing}. Using $27$-qubit IBM processors, we can prepare $2^n$ computational basis states to measure $Z$ operators and $2^n$ of related product states in the $X$ basis in order to measure $X$ operators. Here $n$ is the number of data qubits corresponding to the operator being measured. We assign a measurement change rate for each initial state by recording the fraction of shots for which the measured stabilizer value was inconsistent with the initially prepared state. Results are shown using $\mathrm{ibmq\char`_montreal}$ in Figure \ref{fig:fig_2}.

Figure \ref{fig:fig_2} shows the change rates obtained when measuring gauge operators which correspond to two data qubit $ZZ$ gauge operators, four data qubit flagged $XXXX$ gauge operators and two data qubit flagged $XX$ gauge operators. To resolve two-qubit gate connectivity requirements, the $X$ operators make use of additional ancilla qubits referred to as flag qubits \cite{chamberland2020topological}. Tile representations of these operators are shown in the first column of Figure \ref{fig:fig_2}. Uniform time-step circuits in the second column of Figure \ref{fig:fig_2} perform prepare and measure experiments on product state inputs. Lines between the control and target of each CNOT gate are sometimes partially covered to allow simultaneously executing gates to be drawn on the same time step.

The third column of Figure \ref{fig:fig_2} shows exact change rates which occur in simulation for each operator as a function of the depolarizing noise parameter, $p'$, as well as the lowest order dependence with a dotted line (see Supplementary Sec. \ref{quantum_errors} for relation to depolarizing error rate and Supplementary Sec. \ref{appendix:numerical_calculations} for details on calculating change rate functions). The error model used applies depolarizing noise of strength determined by parameter $p'$ after every time step, with two-qubit depolarizing noise applied after each CNOT. We note that the coefficient of the linear dependence of each operator change rate function corresponds to the number of unique time steps where a depolarizing error can change the relevant ancilla qubits measurement result divided by two. Under this error model, performance of $XXXX$ flagged operator measurements diminishes most rapidly, followed by $XX$ flagged operator measurements and lastly $ZZ$ operator measurements, which is as expected given the depth and number of qubits in each circuit. The relation between operator change rate and noise intensity supports the intuition that each quantity should be able to be readily estimated when given the other. This is especially true in the approximately linear regime corresponding to $p<1\%$, where FTQC is expected to be below threshold and hence viable.

Lastly, the fourth column of Figure \ref{fig:fig_2} shows experimental results for the change rates observed on $\mathrm{ibmq\char`_montreal}$ when different product states were prepared as input states to transpiled versions of the circuits shown in the second column of Figure \ref{fig:fig_2}. Bar heights correspond to averages observed across across 34 device calibrations and error bars correspond to the standard deviation observed across these calibrations.

We find that on average the highest accuracy is obtained when measuring $ZZ$ operators on $Z$ basis product states. The lowest average change rate occurs for $\ket{00}$, followed by $\ket{01}$, $\ket{11}$ and finally $\ket{10}$. When simulated with a uniform depolarizing error model, input states do not affect the change rate of the gauge operators. If the error rates of the $X$ gates required to prepare $\ket{1}$ states are not neglected, then, considering typical results from randomized benchmarking errors,  the experimental preparation of $\ket{1}$ states would have an error rate increased by approximately $2\times10^{-4}$. This is negligible compared to the typical preparation and readout assignment error rates on the order of $10^{-2}$. The asymmetry present in the experimental data can be understood as arising from asymmetric measurement and thermal relaxation errors. The lowest change rate occurring for $\ket{00}$ is consistent with it being the least negatively affected by the phenomena of relaxation errors and biased measurement, due to $\ket{0}$ states being lower energy than $\ket{1}$ states and the $ZZ$ eigenvalues taking a $+1$ value (see Supplementary Figure \ref{fig:sup_amp_damp}).

For the $X$ operator measurements of Figure \ref{fig:fig_2}, we find that that the average change rate of states with $+1$ eigenvalues is significantly lower than the average change rate of states with $-1$ eigenvalues. Again, such a dependence on the input state does not occur during uniform depolarizing noise simulations, but can be better explained by including noise structure such as asymmetric measurement (see Supplementary Figure \ref{fig:sup_amp_damp_XX}). The average change rates for the $ZZ$, $XX$ flagged and $XXXX$ flagged  operators observed experimentally were approximately $0.08$, $0.18$ and $0.17$ respectively. When compared with uniform depolarizing noise simulation, this corresponds to error models with depolarizing noise parameters, $p'$, of $0.030$, $0.044$, and $0.028$ respectively. Finally, for this sub-investigation, we take note of the large standard deviation observed in the results across different calibrations, consistent with effective error rate parameters changing by as much as $\pm 50\%$. Similar results are not uncommon among error rates found with randomized benchmarking experiments due to processes such as drift \cite{proctor2020detecting}. When calculating binomial confidence intervals for the mean across calibrations much smaller error bars can be achieved (see Figure \ref{fig:fig_2_comparison_2}). This suggests that differences in change rates that depend on data qubit input state on average are quite statistically significant, but can also vary significantly across different calibrations and device locations.

Our results found that two-qubit $ZZ$ gauge operators had the lowest average change rates, with the change rates of the $XX$ flagged gauge operators and $XXXX$ flagged gauge operators being slightly higher on average. Experimentally, operator change rates were found to depend on the initial state of the circuit, which does not occur for uniform depolarizing noise simulations. Additional results when gauge operator types are reversed can be found in Supplementary Figure \ref{fig:sup_gauge_ops} and give results consistent with the presence of noise beyond uniform depolarizing noise.

\subsection{Stabilizer Generator Measurement}\label{subsec:stabil}
\begin{figure}
    \centering
    \includegraphics[width=\columnwidth, keepaspectratio]{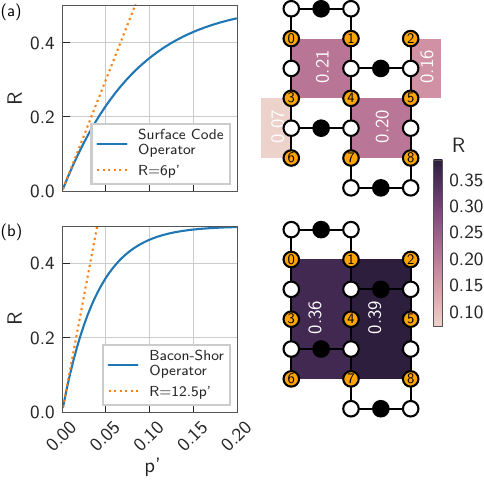} 
    \caption{Stabilizer operator evaluation for the ibmq\char`_montreal device. Stabilizer change rates, $R$, for $Z$ stabilizers and $X$ stabilizers are shown in a) and b) respectively. Theoretical change rates of each the four-qubit $Z$ stabilizers and the six-qubit $X$ stabilizers are shown as a function of $p'$, the depolarizing noise parameter. A dotted line shows the linear component of the dependence.}
    \label{fig:fig_3}
\end{figure}
We continue our investigation by investigating measurements of stabilizer operators of the heavy-hexagon code on IBM superconducting devices. For the heavy-hexagon code, stabilizer operators correspond to products of gauge operators, with the eigenvalue of the stabilizer operator corresponding to the product of the eigenvalues of the gauge operators which composed it. For example, the eigenvalue of the six-qubit Bacon-Shor type stabilizer operator of the distance-3 heavy-hexagon code $X_0X_1X_3X_4X_6X_7$, numbering data qubits top left to bottom right (see Figure \ref{fig:fig_3}), can be calculated by taking the product of results of gauge operators measurements of $X_0X_1$ and $X_3X_4X_6X_7$ \cite{chamberland2020topological}. The number of qubits the Bacon-Shor type stabilizers act upon increase with code distance, and for general code distances, $d$, is equal to $2d$ qubits. The surface-code type stabilizers, however, always act on either two or four qubits. An example of a two-qubit surface-code type stabilizer operator corresponding to a single gauge operators is $Z_2Z_5$ and a four-qubit surface-code type stabilizer operator is $Z_0Z_1Z_3Z_4$, which has its eigenvalue values calculated based on the product of the gauge operator eigenvalues of $Z_0Z_3$ and $Z_1Z_4$. The commutation of the gauge operators with the stabilizer operators and logical operators is what allows the eigenvalues of stabilizer operators to be calculated using measurements of the smaller gauge operators, without effecting subsequent operator measurements of interest \cite{aliferis2007subsystem}.

We measured individual stabilizer operators of the distance-$3$ code on the $\mathrm{ibmq\char`_montreal}$ device and calculated individual stabilizer operator change rates. This was done by performing prepare and measure circuits similar to those used for gauge operator characterization, however with the simultaneous measurement of each gauge operators needed for each stabilizer operator. Results are shown in Figure \ref{fig:fig_3}. Experimentally, we find average change rates of $0.21$ for four-qubit surface code type operators and $0.37$ for Bacon-Shor type operators. The uniform depolarizing noise parameter, $p'$, consistent with these average experimental stabilizer change rate is $0.044$ for four-qubit surface code type $Z$ operators and $0.052$ for six-qubit Bacon-Shor type $X$ operators. This is noticeably larger than the depolarizing noise parameters found from gauge operator change rates. Using $p'_{\mathrm{Z}}=0.030$, the depolarizing parameter consistent with average $Z$ gauge operator change rates, we expected surface code type $Z$ operator change rates to be approximately $0.15$. Additionally, using $p'_{X\mathrm{min}}=0.028$ and $p'_{X\mathrm{max}}=0.044$, the least and greatest depolarizing noise parameter fitted from the $X$ gauge operator measurement results, for consistent results, we required change rates  between between $0.25$ and $0.34$ for the six-qubit Bacon-Shor type operators. Experimentally, we measured larger average change rates for both stabilizer operators. This increase suggests that the larger circuits required to measure stabilizer operators have additional sources of error, which cannot be explained by a uniform depolarizing noise model. As the gauge operators which correspond to each stabilizer operator share no support, this increase in depolarizing noise parameter suggests increased effective error rates due to effects such as cross-talk. 

For the heavy-hexagon code, stabilizer change rates can also be calculated for the stabilizer operator measurements based on gauge operator measurement experimental data. As the measurement of individual $X$ and $Z$ stabilizers can be decomposed into the independent measurement of $XXXX$ flagged gauge operators, $XX$ flagged gauge operators and $ZZ$ operators, their theoretical change rates can be calculated from the change rates of of smaller operators using the formula,
\begin{equation}\label{p_axorb}
    P(A\oplus B)=P(A)(1-P(B))+P(B)(1-P(A)),
\end{equation}
where $A$ and $B$ are independently events and $P(A\oplus B)$ is the probability that exactly one has happened. This can be applied to the stabilizers composed of two gauge operators in the distance-3 heavy-hexagon code under the assumption that gauge operator changes are independent events. Under the uniform depolarizing noise model this assumption is trivially true as the sets of qubits consisting of each gauge operator measurement circuit are disjoint.

As a final comparison, we compare gauge operator measurement with stabilizer operator measurement using only experimental change rate data and Equation \ref{p_axorb}. We note that this requires the assumption that there are no significant error correlations occurring in the combined system. With this assumption, the expected change rates are 0.15 and 0.29 for the surface code type stabilizers and Bacon-Shor type stabilizers respectively, which are consistent with the theoretical predictions and lower than what is realized in experiment.

As stabilizer operators are higher weight and measured as a product of gauge operator measurements, the operator change rates are generally expected to increase compared to individual gauge operator change rates. However, experimental results show that change rates increase more than what would be expected based on models which ignored the effects of correlated errors. This is consistent with past observations of cross-talk in superconducting transmon devices in the context of gate benchmarking and heavy-hexagon code decoding in the presence of multiple simultaneous operations including mid-circuit measurement \cite{gambetta2012characterization, chen2022calibrated, sundaresan2023demonstrating}.

\subsection{Repeated Operator Measurements}\label{subsec:repeated}
\begin{figure*}
    \centering
    \includegraphics[width=\textwidth, keepaspectratio]{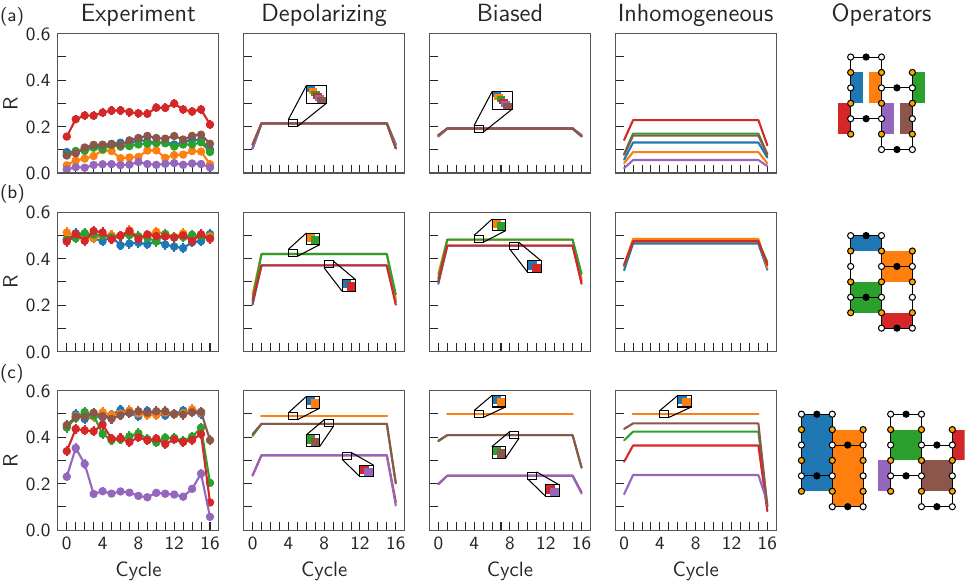}
    \caption{Operator change rates, $R$, as a function of measurement cycle for repeated operator measurement experiments on ibmq\char`_montreal and in simulation under uniform depolarizing, biased and inhomogeneous noise models. Error bars are calculated 95\% confidence intervals.  Results when only $Z$ gauge operators are measured are shown in (a). Results when only $X$ gauge operators are measured are shown in (b). Results when both $X$ and $Z$ stabilizer operators are measured are shown in (c). The uniform depolarizing noise simulation had a depolarization parameter $p'=0.039$. The biased noise simulation had error rate $p=0.045$ and bias parameter $\eta=6.5$. The inhomogeneous noise simulation had mean depolarizing noise parameter of $\Bar{p'}=0.058$. The legend on the right indicates which operator each coloured line corresponds to for each row. Overlapping simulation lines are indicated by boxes.}
    \label{fig:fig_4}
\end{figure*}
We now turn our attention to simultaneous operator measurements, as these will need to be performed repeatedly in order to be able to correct circuit errors in two-dimensional QEC codes. Results are shown in Figure \ref{fig:fig_4}, with parts (a-c) showing change rates of circuits consisting of 16 repetitions of $Z$ operator measurements, $X$ operator measurements, and full heavy-hexagon syndrome measurement respectively. When fitting experimental results to a uniform depolarizing noise model, using the mean squared error in predicted results as a cost function, the optimal simultaneous fit was found with a depolarization parameter of approximately $p'=0.039$. This is consistent with values suggested by the stabilizer generator measurement characterization of Sec. \ref{subsec:stabil}. However, for each experiment, multiple distinct phenomena arise which warrant further discussion.

In Figure \ref{fig:fig_4} (a) we show the behaviour of $Z$ gauge operator change rates across 16 cycle circuits, with data qubits prepared in the $\ket{0}^{\otimes 9}$ state (see Supplementary Figure \ref{fig:z_stabils_only}). Experimentally, we find a significant spread in change rates across different operators and different cycles. Under the uniform depolarizing noise model this does not occur, due to all $Z$ gauge operator measurement circuits being close to equivalent, up to small differences associated with order of measurement. In the experimental data we observe that operator change rates are low for the first cycle, tend to increase as more cycles are performed, and decrease for the final cycle. The reductions in change rates observed for $Z$ gauge operators in the first and final cycle are consistent with the uniform depolarizing noise model and arise due to these cycles corresponding to comparisons between operator values inferred from data qubit preparation and measurement \cite{chen2021nature, acharya2022suppressing}. However, the tendency for change rates to increase across the intermediate cycles cannot be explained by a uniform depolarizing noise model, which has constant change rates across the intermediate cycles. The phenomenon of increasing change rates has been attributed to leakage in the past but can also have contributions from measurement error asymmetry (when $-1$ eigenvalue measurements experience greater errors than $+1$ eigenvalue measurements) when all the eigenvalues are set to $+1$ \cite{chen2021nature}. 

In Figure \ref{fig:fig_4} (b) we show the behaviour of $X$ gauge operator change rates across 16 cycle circuits, measuring only the $X$ gauge operators of a distance-3 heavy-hexagon code and preparing data qubits in the $\ket{+}^{\otimes 9}$ state (see Supplementary Figure \ref{fig:x_stabils_only}). Experimentally, we find that the change rates of each operator remain close to $50\%$, which corresponds to the high error rate regime. The blue two-qubit $X$ gauge operator has a change rate slightly beneath $50\%$ for approximately half of the intermediate cycles. However, other than this, there is no consistent distinction in change rate between four- and two-qubit $X$ gauge operators. There is also no distinct reduction in experimental operator change rates at the initial and final cycle as expected from theory. In order to achieve similar effects under a uniform depolarizing noise model, a depolarizing parameter significantly larger than what is consistent with results in Figure \ref{fig:fig_4} part (a) is required. These results suggest that the increased circuit size of repeated measurements significantly increases the effective error rate compared to individual stabilizer measurement shown in Figure \ref{fig:fig_3} part (a). In the context of quantum circuits, this is consistent with the presence of cross-talk. The increased change rate of $X$ operators is consistent with mid-circuit measurement-induced phase rotations previous observed in heavy-hexagon code decoding \cite{chen2022calibrated}. Accurate modelling of cross-talk requires noise models beyond the independent depolarizing noise model, for example with the use of noise operators which apply noise channels to qubits beyond those where a gate was performed \cite{ acharya2022suppressing, fowler2014quantifying}. Further study is required to develop accurate measures of the significance of these effects on different classes of quantum circuits.

Lastly, we investigate the performance of 16 cycles of repeated full syndrome measurement circuits for both $X$ and $Z$ stabilizers of the distance-3 heavy-hexagon code. Data qubits were prepared in the $\ket{0}^{\otimes 9}$ state and $X$ stabilizers were measured first, as shown in Supplementary Figure \ref{fig:both_stabils}. Figure \ref{fig:fig_4} (c) shows the rates at which stabilizer operators changed values each cycle over a circuit of 16 cycles. Theoretically, under uniform depolarizing noise, we expect change rate curves to arrange into three groups corresponding to (in order from highest to lowest expected change rate) $X$ stabilizers, four-qubit $Z$ stabilizers and two-qubit $Z$ stabilizers. Experimentally, we found that both of the $X$ stabilizers maintained change rates close to 50\% for majority of the cycles. The $Z$ stabilizers retained more varied error rates amongst each other and across different cycle numbers. Compared to $Z$ gauge operator measurement circuits, most operator change rates are significantly higher. The exception is one operator which corresponds both to the two-qubit stabilizer coloured purple in (c) and and the gauge operator coloured red in (a). Under a uniform depolarizing noise model, this should not be the case.  Other, less readily explicable, features were also present, such as decreasing stabilizer change rates present within the first four cycles and increasing stabilizer change rates present within the last four cycles. These effects do not occur in simulations using only time-independent depolarizing noise.

Difficulty was experienced in fitting the change rate curves to the experimental data in Figure \ref{fig:fig_4} using the uniform depolarizing noise model. Increasing the depolarization parameter increases both $X$ and $Z$ operator change rates, however an improved fit required $X$ operator change rates to increase while decreasing $Z$ operator change rates. A phenomenon which can cause this effect is the presence of noise biased towards $Z$ errors. To test whether this is sufficient to explain the results observed experimentally, we modelled the system with biased uniform depolarizing noise described by error rate $p$ and bias parameter $\eta$ (see Supplementary Sec. \ref{quantum_errors}). We found an improved fit by using $p=0.045$ and $\eta=6.5$, which indicates that experimental results are better described by an error model with $Z$ errors occurring with more than six times as much probability of other possible errors (those which include only $X$ or $Y$ components). This is consistent with other results shown in the literature, where noise of superconducting qubits is also found to be biased towards $Z$ errors \cite{Aliferis_2009}. 

While the use of biased noise slightly improves the fit to experimental change rates compared to isotropic depolarizing noise, adding noise bias was unable to split the change rate curves on equivalent operators. This was expected to be caused by the common simplifying assumption that all qubits are described by identical, homogeneous, error characteristics. To investigate whether dropping this assumption can be another alternative to achieve a better fit, we fit the experimental data to an inhomogeneous noise model, where each noise channel is described by one element of a set of single-qubit and two-qubit parameters. Results are shown in the final data column of Figure \ref{fig:fig_4}. The mean squared error of predicted change rates was used as a cost function, with the addition of L2 regularization during optimization. difficulty was experience finding a reasonable and consistent simultaneous good fit to all three experiments due to an increased parameter space, the phenomenon of local minima in the cost landscape. Fitting parameters to one experiment at a time leads to poor generalization across other experiments (see Supplementary Figure \ref{fig:inhomogeneous_appendix}). This is also likely influenced by theoretical limits to the information provided by operator measurements in fitting noise models \cite{Wagner_2022}. An average depolarizing noise parameter of $\Bar{p'}=0.058$ was found after simultaneous optimization. Similar qualities of fit are likely possible with modified parameters, due to information theoretical limits as well as local minima. While using inhomogeneous noise models allows many of the orderings in experimental change rate to be made correctly modelled, important discrepancies remain which are indicative of addition features beyond bias and inhomogeneity.

We finalize our discussion of Figure \ref{fig:fig_4} by noting the remaining discrepancies between experiment and theory which cannot be resolved by inhomogeneous noise models or biased noised model as they are often described in the literature. The absence of dips in change rates at the first and last cycle of repeated $X$ gauge measurements were unable to be modelled accurately by either technique. Theoretically, these appear due to the reduced number of time steps between a stabilizer measurement and the time steps where data qubits have been prepared or measured. Experimentally, this does not appear to be the case for $X$ gauge operators, where initial and final change rates remain close to $0.5$. Further increasing the bias towards  $Z$ errors would ideally improve the fit, however we found that increasing the bias also reduced the change rate dips of the $Z$ operators, which clearly did occur in experiment. This relation between bias parameter $\eta$ and $Z$ operator initial and final change rate follows from the state preparation and measurement error rate of $p(2\eta+1)/(2\eta+2)$, which corresponds to the probability that either an $X$ or $Z$ error occurs \cite{Ataides_NComms_2021}. This effect can be understood as a conservative upper bound in the high $\eta$ regime, where state preparation and measurement error rates will approach $p$, and avoids the need of introducing another parameter. However, in our case this definition also restricts our ability to fit the experimental data. Regarding inhomogeneous error models, multiple phenomena may occur which can cause inhomogeneous depolarizing noise to be an insufficient explanation. One example is the purple $Z$ stabilizer of Figure \ref{fig:fig_4}(c) having a lower change rate than the red $Z$ stabilizer of Figure \ref{fig:fig_4}(a). As these are the same operators, one would expect the larger circuit of part (c) to cause more errors to occur and hence increase the operator change rate in (c) compared to (a). This is indeed what occurs in simulations. However, experimentally this did not occur, suggesting additional important details of the noise model of the experimental results remain, beyond bias and inhomogeneity.

\subsection{Spatial-Temporal Correlations}\label{subsec:spatial_temporal}

\begin{figure}
    \centering
    \includegraphics{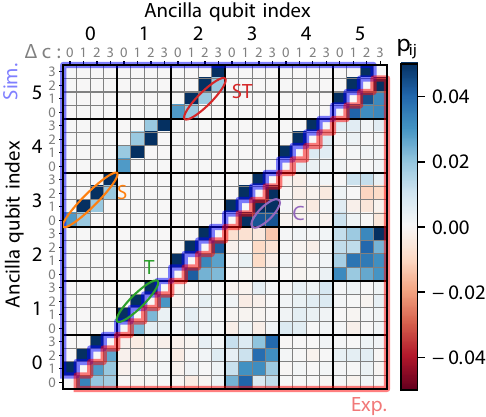}
    \caption{Correlation matrix for repeated measurements of $Z$ gauge operators of a distance-3 heavy-hexagon code on ibmq\char`_montreal (bottom right) and when simulated with $p'=0.035$ depolarizing error parameter noise (upper left). Black and grey lines separate different $Z$ operator ancilla qubits and different measurement cycles ($\Delta c$) respectively. The initial state of all data qubits was $\ket{0}$. Symbols S, T, ST and C provide examples of matrix entries indicating the presence of space-like, time-like, space-time-like and correlated errors respectively.  Matrix values with magnitude above $0.05$ have been clipped.}
    \label{fig:fig_5}
\end{figure}

Lastly, we investigated the properties of the relatively low $Z$ operator detection events by investigating their spatial and temporal correlations with the formula,
\begin{equation}
    p_{ij}=\frac{\braket{x_i x_j}-\braket{x_i}\braket{x_j}}{(1-2\braket{x_i})(1-2\braket{x_j})},
\end{equation}
which is a measure of the correlation of detection events $x_i$ and $x_j$ (with $p_{ij}$ set to $0$ for $i=j$) \cite{chen2021nature}. Figure \ref{fig:fig_5} shows this matrix, where the top left corner corresponds to a uniform depolarizing noise simulation with depolarization parameter $p'=0.035$ and the the bottom right corner corresponds to measurements on $\mathrm{ibmq\char`_montreal}$. From simulation, we expect three different classes of large correlations to be present. Space-like errors cause large correlations between simultaneous measurement changes of operators which both act on at least one shared qubit. Time-like errors cause large correlations between subsequent measurement changes of the same operator (one changes due to the measurement error and another change back if the subsequent cycle is measured correctly). Space-time-like errors cause large correlations, among operators which share a qubit, but between two subsequent measurement cycles. These are the weakest correlations in simulation due to the low number of time steps for which space-time-like errors can occur. Other correlation matrix elements were found be to vanish, as they correspond to the simultaneous, correlated, occurrence of more than one error. The $\mathrm{ibmq\char`_montreal}$ device shows signs of all three classes of errors occurring during the $Z$ operator measurement circuits. However, their relative magnitudes vary significantly across different pairs of subsequent time steps and adjacent ancilla qubits. The operator that ancilla qubit 4 measures shows much lower time-like correlations than any of the other operators (with numbering consistent with Supplementary Figure \ref{fig:falcon_layout}). Operator 4 also has much lower space-like and space-time-like correlations with operator 1 than expected in simulation. This suggests that other, more significant, error processes are present for operator 4 which act to decorrelate detection events which are expected to be strongly correlated. Finally, we also observed significant additional correlations outside of the regions which correspond to the three main error classes. Contributions to these additional correlations can include leakage, cross-talk and asymmetric readout errors. Additional results showing correlations present in different backends and larger cycle differences can be found in Supplementary Figures \ref{fig:sup_multi_correl_mat} and \ref{fig:sup_correl_mat_16_montreal}.

Finally, we investigated the relationship between the one-qubit and two-qubit inhomogeneous depolarizing noise model and the information found in correlation matrix elements. For repeated $Z$ guage operator circuits (which are essentially repetition code circuits), $X$ gauge operator circuits, and full heavy-hexagon stabilizer circuits, the system was found to be fully determined if all correlations in detection events available are considered. Supplementary Figure \ref{fig:error_rates_map_appendix} shows the performance of this inversion applied to simulated heavy-hexagon code stabilizer measurement circuits, numerically using least-squares solutions, modelling shot noise approximately by sampling from the associated binomial distribution. We find that the underlying error model can be extracted to arbitrary precision as long as error rates are below approximately 1\% and shot noise is sufficiently suppressed.

\section{Discussion and Conclusion}\label{sec:conclusion}
We investigated the performance of heavy-hexagon code stabilizer measurement circuits on IBM quantum devices and found that uniform uncorrelated depolarizing noise models fail to explain many phenomena observed. These include a dependence of operator change rates on initially prepared states and effective error rates increasing significantly as larger circuits are run. In particular, operator change rates of individual stabilizer measurement circuits are greater than expected from the behaviour of individual gauge operator measurement circuits. When investigating the temporal characteristics of circuit noise, we found increasing change rates, again indicative of noise present beyond uniform depolarizing noise. The temporal data curves also corresponded much better to simulations with noise biased towards $Z$ errors, or inhomogeneous noise. Finally, observing correlations in operator changes leads to the conclusion that significant sources of non-depolarizing noise events are present in the system.

While uniform depolarizing noise is a convenient noise model for classical simulation, as it can be efficiently simulated, can intuitively be understood as random symmetric Pauli operations, and can be used to study worst-case scenario bounds, it was not found to adequately model stabilizer statistics in IBM transmon superconducting devices. As QEC investigations predominantly focus on depolarizing noise, this may lead to uncertainty when considering the viability of FTQC as devices scale up. One way to address this is by applying noise tailoring with techniques such as randomized compiling, which would map device noise to a Pauli noise model \cite{wallman2016noise}. As non-local correlations may yet remain (distinct from those introduced by syndrome measurement circuits under circuit noise), care must be made that FTQC protocols remain robust when scaling to larger distances \cite{fowler2014quantifying}. At the same time, techniques can continue to develop which address more general noise models, such as those featuring inhomogeneous noise \cite{fowlerautotune2012}, asymmetric coherent rotations \cite{bravyi2018correcting}, biased noise \cite{Ataides_NComms_2021}, amplitude and phase damping \cite{Tomita2014Realistic}, and correlated errors \cite{Nickerson2019analysingcorrelated}.

There are multiple ways to extend our investigation. In the results shown, we did not utilize dynamical decoupling, which can be expected to combat coherent errors as well as cross-talk present in the device \cite{PhysRevApplied.18.024068}. Dynamical decoupling pulses can be implemented in any idle times between gates and the efficacy of dynamical decoupling schemes can be investigated in terms of their reduction in operator change rates or correlation matrix elements \cite{ezzell2023dynamical}. We also did not optimize scheduling beyond an ``as late as possible'' scheduler, which may have had a significant effect with the presence of mid-circuit measurement. More accurate noise models may also be pursued to fit the experimental data. These may include additional effects beyond depolarization such as leakage and cross-talk \cite{acharya2022suppressing}. It would also be very enlightening to investigate how readily randomized compiling can be used to map characteristics of noise to the well studied depolarizing noise model for various QEC codes \cite{wallman2016noise}. Alternatively, the behaviour of larger operator measurement circuits can be investigated, such as the larger Bacon-Shor stabilizers of larger distance codes or investigating evidence of logical error correlations in systems of multiple logical qubits. 

Our results show how simulations can be used to efficiently study the behaviour of independent Pauli noise in quantum error correction circuits. Comparing these simulations to measurements obtained from IBM devices supports the need to further develop techniques to tailor quantum error correction protocols to particular characteristics of noise expected to be present in future devices running circuits of fault-tolerant quantum computing.

% \bibliographystyle{quantum}
% \def\bibsection{\subsection*{\refname}}
% \bibliographystyle{plainnat}
% \bibliography{sample}

\noindent
\\ \\
\noindent
\textbf{Data Availability:}
The data that support the findings of this study are available within the article and the Supplementary Material. Further information can be provided upon reasonable request to the corresponding author. 
\noindent
\\ \\
\noindent
\textbf{Code Availability:}
The source code used to generate figures in this work can be provided upon reasonable request to the corresponding author. 
\noindent
\\ \\
\noindent
\textbf{Acknowledgements:} The research was supported by the University of Melbourne through the establishment of the IBM Quantum Network Hub at the University. Computational resources were provided by the National Computing Infrastructure (NCI) and the Pawsey Supercomputing Research Center through the National Computational Merit Allocation Scheme (NCMAS). This research was supported by The University of Melbourne’s Research Computing Services and the Peta-scale Campus Initiative.
\\ \\
\noindent
\textbf{Author Contributions:} M.U. and L.C.L.H. planned and supervised the project. S.G. carried out the simulations and ran the circuits on IBM devices. All authors contributed to the analysis of data. S.G wrote the manuscript with input from M.U. and L.C.L.H.
\\ \\
\noindent
\textbf{Competing Financial Interests:} The authors declare no competing financial or non-financial interests.

% \bibliography{sample}

\begin{thebibliography}{74}%
\makeatletter
\providecommand \@ifxundefined [1]{%
 \@ifx{#1\undefined}
}%
\providecommand \@ifnum [1]{%
 \ifnum #1\expandafter \@firstoftwo
 \else \expandafter \@secondoftwo
 \fi
}%
\providecommand \@ifx [1]{%
 \ifx #1\expandafter \@firstoftwo
 \else \expandafter \@secondoftwo
 \fi
}%
\providecommand \natexlab [1]{#1}%
\providecommand \enquote  [1]{``#1''}%
\providecommand \bibnamefont  [1]{#1}%
\providecommand \bibfnamefont [1]{#1}%
\providecommand \citenamefont [1]{#1}%
\providecommand \href@noop [0]{\@secondoftwo}%
\providecommand \href [0]{\begingroup \@sanitize@url \@href}%
\providecommand \@href[1]{\@@startlink{#1}\@@href}%
\providecommand \@@href[1]{\endgroup#1\@@endlink}%
\providecommand \@sanitize@url [0]{\catcode `\\12\catcode `\$12\catcode
  `\&12\catcode `\#12\catcode `\^12\catcode `\_12\catcode `\%12\relax}%
\providecommand \@@startlink[1]{}%
\providecommand \@@endlink[0]{}%
\providecommand \url  [0]{\begingroup\@sanitize@url \@url }%
\providecommand \@url [1]{\endgroup\@href {#1}{\urlprefix }}%
\providecommand \urlprefix  [0]{URL }%
\providecommand \Eprint [0]{\href }%
\providecommand \doibase [0]{https://doi.org/}%
\providecommand \selectlanguage [0]{\@gobble}%
\providecommand \bibinfo  [0]{\@secondoftwo}%
\providecommand \bibfield  [0]{\@secondoftwo}%
\providecommand \translation [1]{[#1]}%
\providecommand \BibitemOpen [0]{}%
\providecommand \bibitemStop [0]{}%
\providecommand \bibitemNoStop [0]{.\EOS\space}%
\providecommand \EOS [0]{\spacefactor3000\relax}%
\providecommand \BibitemShut  [1]{\csname bibitem#1\endcsname}%
\let\auto@bib@innerbib\@empty
%</preamble>
\bibitem [{\citenamefont {Feynman}(1982)}]{Feynman1982-gq}%
  \BibitemOpen
  \bibfield  {author} {\bibinfo {author} {\bibfnamefont {R.~P.}\ \bibnamefont
  {Feynman}},\ }\bibfield  {title} {\bibinfo {title} {Simulating physics with
  computers},\ }\href {https://doi.org/https://doi.org/10.1007/BF02650179}
  {\bibfield  {journal} {\bibinfo  {journal} {International Journal of
  Theoretical Physics}\ }\textbf {\bibinfo {volume} {21}},\ \bibinfo {pages}
  {467} (\bibinfo {year} {1982})}\BibitemShut {NoStop}%
\bibitem [{\citenamefont {Farhi}\ \emph {et~al.}(2014)\citenamefont {Farhi},
  \citenamefont {Goldstone},\ and\ \citenamefont {Gutmann}}]{farhi2014quantum}%
  \BibitemOpen
  \bibfield  {author} {\bibinfo {author} {\bibfnamefont {E.}~\bibnamefont
  {Farhi}}, \bibinfo {author} {\bibfnamefont {J.}~\bibnamefont {Goldstone}},\
  and\ \bibinfo {author} {\bibfnamefont {S.}~\bibnamefont {Gutmann}},\
  }\href@noop {} {\bibinfo {title} {A quantum approximate optimization
  algorithm}} (\bibinfo {year} {2014}),\ \Eprint
  {https://arxiv.org/abs/1411.4028} {arXiv:1411.4028 [quant-ph]} \BibitemShut
  {NoStop}%
\bibitem [{\citenamefont {Huang}\ \emph {et~al.}(2022)\citenamefont {Huang},
  \citenamefont {Broughton}, \citenamefont {Cotler}, \citenamefont {Chen},
  \citenamefont {Li}, \citenamefont {Mohseni}, \citenamefont {Neven},
  \citenamefont {Babbush}, \citenamefont {Kueng}, \citenamefont {Preskill},\
  and\ \citenamefont {McClean}}]{huang2022quantum}%
  \BibitemOpen
  \bibfield  {author} {\bibinfo {author} {\bibfnamefont {H.-Y.}\ \bibnamefont
  {Huang}}, \bibinfo {author} {\bibfnamefont {M.}~\bibnamefont {Broughton}},
  \bibinfo {author} {\bibfnamefont {J.}~\bibnamefont {Cotler}}, \bibinfo
  {author} {\bibfnamefont {S.}~\bibnamefont {Chen}}, \bibinfo {author}
  {\bibfnamefont {J.}~\bibnamefont {Li}}, \bibinfo {author} {\bibfnamefont
  {M.}~\bibnamefont {Mohseni}}, \bibinfo {author} {\bibfnamefont
  {H.}~\bibnamefont {Neven}}, \bibinfo {author} {\bibfnamefont
  {R.}~\bibnamefont {Babbush}}, \bibinfo {author} {\bibfnamefont
  {R.}~\bibnamefont {Kueng}}, \bibinfo {author} {\bibfnamefont
  {J.}~\bibnamefont {Preskill}},\ and\ \bibinfo {author} {\bibfnamefont
  {J.~R.}\ \bibnamefont {McClean}},\ }\bibfield  {title} {\bibinfo {title}
  {Quantum advantage in learning from experiments},\ }\href
  {https://doi.org/https://doi.org/10.1126/science.abn7293} {\bibfield
  {journal} {\bibinfo  {journal} {Science}\ }\textbf {\bibinfo {volume}
  {376}},\ \bibinfo {pages} {1182} (\bibinfo {year} {2022})}\BibitemShut
  {NoStop}%
\bibitem [{\citenamefont {West}\ \emph {et~al.}(2023)\citenamefont {West},
  \citenamefont {Tsang}, \citenamefont {Low}, \citenamefont {Hill},
  \citenamefont {Leckie}, \citenamefont {Hollenberg}, \citenamefont {Erfani},\
  and\ \citenamefont {Usman}}]{West2023}%
  \BibitemOpen
  \bibfield  {author} {\bibinfo {author} {\bibfnamefont {M.~T.}\ \bibnamefont
  {West}}, \bibinfo {author} {\bibfnamefont {S.-L.}\ \bibnamefont {Tsang}},
  \bibinfo {author} {\bibfnamefont {J.~S.}\ \bibnamefont {Low}}, \bibinfo
  {author} {\bibfnamefont {C.~D.}\ \bibnamefont {Hill}}, \bibinfo {author}
  {\bibfnamefont {C.}~\bibnamefont {Leckie}}, \bibinfo {author} {\bibfnamefont
  {L.~C.~L.}\ \bibnamefont {Hollenberg}}, \bibinfo {author} {\bibfnamefont
  {S.~M.}\ \bibnamefont {Erfani}},\ and\ \bibinfo {author} {\bibfnamefont
  {M.}~\bibnamefont {Usman}},\ }\bibfield  {title} {\bibinfo {title} {Towards
  quantum enhanced adversarial robustness in machine learning},\ }\href
  {https://doi.org/10.1038/s42256-023-00661-1} {\bibfield  {journal} {\bibinfo
  {journal} {Nature Machine Intelligence}\ }\textbf {\bibinfo {volume} {5}},\
  \bibinfo {pages} {581} (\bibinfo {year} {2023})}\BibitemShut {NoStop}%
\bibitem [{\citenamefont {Cerezo}\ \emph {et~al.}(2022)\citenamefont {Cerezo},
  \citenamefont {Verdon}, \citenamefont {Huang}, \citenamefont {Cincio},\ and\
  \citenamefont {Coles}}]{Cerezo2022}%
  \BibitemOpen
  \bibfield  {author} {\bibinfo {author} {\bibfnamefont {M.}~\bibnamefont
  {Cerezo}}, \bibinfo {author} {\bibfnamefont {G.}~\bibnamefont {Verdon}},
  \bibinfo {author} {\bibfnamefont {H.-Y.}\ \bibnamefont {Huang}}, \bibinfo
  {author} {\bibfnamefont {L.}~\bibnamefont {Cincio}},\ and\ \bibinfo {author}
  {\bibfnamefont {P.~J.}\ \bibnamefont {Coles}},\ }\bibfield  {title} {\bibinfo
  {title} {Challenges and opportunities in quantum machine learning},\ }\href
  {https://doi.org/10.1038/s43588-022-00311-3} {\bibfield  {journal} {\bibinfo
  {journal} {Nature Computational Science}\ }\textbf {\bibinfo {volume} {2}},\
  \bibinfo {pages} {567} (\bibinfo {year} {2022})}\BibitemShut {NoStop}%
\bibitem [{\citenamefont {Grover}(1996)}]{grover1996fast}%
  \BibitemOpen
  \bibfield  {author} {\bibinfo {author} {\bibfnamefont {L.~K.}\ \bibnamefont
  {Grover}},\ }\bibfield  {title} {\bibinfo {title} {A fast quantum mechanical
  algorithm for database search},\ }in\ \href
  {https://doi.org/https://doi.org/10.1145/237814.237866} {\emph {\bibinfo
  {booktitle} {Proceedings of the Twenty-Eighth Annual ACM Symposium on Theory
  of Computing}}},\ \bibinfo {series and number} {STOC '96}\ (\bibinfo
  {publisher} {Association for Computing Machinery},\ \bibinfo {address} {New
  York, NY, USA},\ \bibinfo {year} {1996})\ p.\ \bibinfo {pages}
  {212–219}\BibitemShut {NoStop}%
\bibitem [{\citenamefont {Preskill}(2018)}]{preskill2018quantum}%
  \BibitemOpen
  \bibfield  {author} {\bibinfo {author} {\bibfnamefont {J.}~\bibnamefont
  {Preskill}},\ }\bibfield  {title} {\bibinfo {title} {Quantum computing in the
  {NISQ} era and beyond},\ }\href {https://doi.org/10.22331/q-2018-08-06-79}
  {\bibfield  {journal} {\bibinfo  {journal} {Quantum}\ }\textbf {\bibinfo
  {volume} {2}},\ \bibinfo {pages} {79} (\bibinfo {year} {2018})}\BibitemShut
  {NoStop}%
\bibitem [{\citenamefont {Brandhofer}\ \emph {et~al.}(2021)\citenamefont
  {Brandhofer}, \citenamefont {Devitt},\ and\ \citenamefont
  {Polian}}]{brandhofer2021arsonisq}%
  \BibitemOpen
  \bibfield  {author} {\bibinfo {author} {\bibfnamefont {S.}~\bibnamefont
  {Brandhofer}}, \bibinfo {author} {\bibfnamefont {S.}~\bibnamefont {Devitt}},\
  and\ \bibinfo {author} {\bibfnamefont {I.}~\bibnamefont {Polian}},\
  }\bibfield  {title} {\bibinfo {title} {Arsonisq: Analyzing quantum algorithms
  on near-term architectures},\ }in\ \href
  {https://doi.org/10.1109/ETS50041.2021.9465414} {\emph {\bibinfo {booktitle}
  {2021 IEEE European Test Symposium (ETS)}}}\ (\bibinfo {year} {2021})\ pp.\
  \bibinfo {pages} {1--6}\BibitemShut {NoStop}%
\bibitem [{\citenamefont {Shor}(1995)}]{shor1995scheme}%
  \BibitemOpen
  \bibfield  {author} {\bibinfo {author} {\bibfnamefont {P.~W.}\ \bibnamefont
  {Shor}},\ }\bibfield  {title} {\bibinfo {title} {Scheme for reducing
  decoherence in quantum computer memory},\ }\href
  {https://doi.org/10.1103/PhysRevA.52.R2493} {\bibfield  {journal} {\bibinfo
  {journal} {Phys. Rev. A}\ }\textbf {\bibinfo {volume} {52}},\ \bibinfo
  {pages} {R2493} (\bibinfo {year} {1995})}\BibitemShut {NoStop}%
\bibitem [{\citenamefont {Shor}(1996)}]{shor1996fault}%
  \BibitemOpen
  \bibfield  {author} {\bibinfo {author} {\bibfnamefont {P.~W.}\ \bibnamefont
  {Shor}},\ }\bibfield  {title} {\bibinfo {title} {Fault-tolerant quantum
  computation},\ }in\ \href
  {https://doi.org/https://doi.org/10.1109/SFCS.1996.548464} {\emph {\bibinfo
  {booktitle} {Proceedings of 37th Conference on Foundations of Computer
  Science}}}\ (\bibinfo {year} {1996})\ pp.\ \bibinfo {pages}
  {56--65}\BibitemShut {NoStop}%
\bibitem [{\citenamefont {Martinis}(2021)}]{martinis2021saving}%
  \BibitemOpen
  \bibfield  {author} {\bibinfo {author} {\bibfnamefont {J.~M.}\ \bibnamefont
  {Martinis}},\ }\bibfield  {title} {\bibinfo {title} {Saving superconducting
  quantum processors from decay and correlated errors generated by gamma and
  cosmic rays},\ }\href {https://doi.org/10.1038/s41534-021-00431-0} {\bibfield
   {journal} {\bibinfo  {journal} {npj Quantum Information}\ }\textbf {\bibinfo
  {volume} {7}},\ \bibinfo {pages} {90} (\bibinfo {year} {2021})}\BibitemShut
  {NoStop}%
\bibitem [{\citenamefont {Xu}\ \emph {et~al.}(2022)\citenamefont {Xu},
  \citenamefont {Seif}, \citenamefont {Yan}, \citenamefont {Mannucci},
  \citenamefont {Sane}, \citenamefont {Van~Meter}, \citenamefont {Cleland},\
  and\ \citenamefont {Jiang}}]{xu2022distributed}%
  \BibitemOpen
  \bibfield  {author} {\bibinfo {author} {\bibfnamefont {Q.}~\bibnamefont
  {Xu}}, \bibinfo {author} {\bibfnamefont {A.}~\bibnamefont {Seif}}, \bibinfo
  {author} {\bibfnamefont {H.}~\bibnamefont {Yan}}, \bibinfo {author}
  {\bibfnamefont {N.}~\bibnamefont {Mannucci}}, \bibinfo {author}
  {\bibfnamefont {B.~O.}\ \bibnamefont {Sane}}, \bibinfo {author}
  {\bibfnamefont {R.}~\bibnamefont {Van~Meter}}, \bibinfo {author}
  {\bibfnamefont {A.~N.}\ \bibnamefont {Cleland}},\ and\ \bibinfo {author}
  {\bibfnamefont {L.}~\bibnamefont {Jiang}},\ }\bibfield  {title} {\bibinfo
  {title} {Distributed quantum error correction for chip-level catastrophic
  errors},\ }\href {https://doi.org/10.1103/PhysRevLett.129.240502} {\bibfield
  {journal} {\bibinfo  {journal} {Phys. Rev. Lett.}\ }\textbf {\bibinfo
  {volume} {129}},\ \bibinfo {pages} {240502} (\bibinfo {year}
  {2022})}\BibitemShut {NoStop}%
\bibitem [{\citenamefont {Gidney}\ and\ \citenamefont
  {Eker{\aa}}(2021)}]{gidney2019factor}%
  \BibitemOpen
  \bibfield  {author} {\bibinfo {author} {\bibfnamefont {C.}~\bibnamefont
  {Gidney}}\ and\ \bibinfo {author} {\bibfnamefont {M.}~\bibnamefont
  {Eker{\aa}}},\ }\bibfield  {title} {\bibinfo {title} {How to factor 2048 bit
  {RSA} integers in 8 hours using 20 million noisy qubits},\ }\href
  {https://doi.org/10.22331/q-2021-04-15-433} {\bibfield  {journal} {\bibinfo
  {journal} {Quantum}\ }\textbf {\bibinfo {volume} {5}},\ \bibinfo {pages}
  {433} (\bibinfo {year} {2021})}\BibitemShut {NoStop}%
\bibitem [{\citenamefont {Litinski}(2019)}]{litinski2019game}%
  \BibitemOpen
  \bibfield  {author} {\bibinfo {author} {\bibfnamefont {D.}~\bibnamefont
  {Litinski}},\ }\bibfield  {title} {\bibinfo {title} {A game of surface codes:
  Large-scale quantum computing with lattice surgery},\ }\href
  {https://doi.org/10.22331/q-2019-03-05-128} {\bibfield  {journal} {\bibinfo
  {journal} {Quantum}\ }\textbf {\bibinfo {volume} {3}},\ \bibinfo {pages}
  {128} (\bibinfo {year} {2019})}\BibitemShut {NoStop}%
\bibitem [{\citenamefont {Beverland}\ \emph {et~al.}(2022)\citenamefont
  {Beverland}, \citenamefont {Murali}, \citenamefont {Troyer}, \citenamefont
  {Svore}, \citenamefont {Hoefler}, \citenamefont {Kliuchnikov}, \citenamefont
  {Low}, \citenamefont {Soeken}, \citenamefont {Sundaram},\ and\ \citenamefont
  {Vaschillo}}]{beverl2022assessing}%
  \BibitemOpen
  \bibfield  {author} {\bibinfo {author} {\bibfnamefont {M.~E.}\ \bibnamefont
  {Beverland}}, \bibinfo {author} {\bibfnamefont {P.}~\bibnamefont {Murali}},
  \bibinfo {author} {\bibfnamefont {M.}~\bibnamefont {Troyer}}, \bibinfo
  {author} {\bibfnamefont {K.~M.}\ \bibnamefont {Svore}}, \bibinfo {author}
  {\bibfnamefont {T.}~\bibnamefont {Hoefler}}, \bibinfo {author} {\bibfnamefont
  {V.}~\bibnamefont {Kliuchnikov}}, \bibinfo {author} {\bibfnamefont {G.~H.}\
  \bibnamefont {Low}}, \bibinfo {author} {\bibfnamefont {M.}~\bibnamefont
  {Soeken}}, \bibinfo {author} {\bibfnamefont {A.}~\bibnamefont {Sundaram}},\
  and\ \bibinfo {author} {\bibfnamefont {A.}~\bibnamefont {Vaschillo}},\
  }\href@noop {} {\bibinfo {title} {Assessing requirements to scale to
  practical quantum advantage}} (\bibinfo {year} {2022}),\ \Eprint
  {https://arxiv.org/abs/2211.07629} {arXiv:2211.07629 [quant-ph]} \BibitemShut
  {NoStop}%
\bibitem [{\citenamefont {Knill}\ \emph
  {et~al.}(2008{\natexlab{a}})\citenamefont {Knill}, \citenamefont {Leibfried},
  \citenamefont {Reichle}, \citenamefont {Britton}, \citenamefont {Blakestad},
  \citenamefont {Jost}, \citenamefont {Langer}, \citenamefont {Ozeri},
  \citenamefont {Seidelin},\ and\ \citenamefont
  {Wineland}}]{PhysRevA.77.012307}%
  \BibitemOpen
  \bibfield  {author} {\bibinfo {author} {\bibfnamefont {E.}~\bibnamefont
  {Knill}}, \bibinfo {author} {\bibfnamefont {D.}~\bibnamefont {Leibfried}},
  \bibinfo {author} {\bibfnamefont {R.}~\bibnamefont {Reichle}}, \bibinfo
  {author} {\bibfnamefont {J.}~\bibnamefont {Britton}}, \bibinfo {author}
  {\bibfnamefont {R.~B.}\ \bibnamefont {Blakestad}}, \bibinfo {author}
  {\bibfnamefont {J.~D.}\ \bibnamefont {Jost}}, \bibinfo {author}
  {\bibfnamefont {C.}~\bibnamefont {Langer}}, \bibinfo {author} {\bibfnamefont
  {R.}~\bibnamefont {Ozeri}}, \bibinfo {author} {\bibfnamefont
  {S.}~\bibnamefont {Seidelin}},\ and\ \bibinfo {author} {\bibfnamefont
  {D.~J.}\ \bibnamefont {Wineland}},\ }\bibfield  {title} {\bibinfo {title}
  {Randomized benchmarking of quantum gates},\ }\href
  {https://doi.org/10.1103/PhysRevA.77.012307} {\bibfield  {journal} {\bibinfo
  {journal} {Phys. Rev. A}\ }\textbf {\bibinfo {volume} {77}},\ \bibinfo
  {pages} {012307} (\bibinfo {year} {2008}{\natexlab{a}})}\BibitemShut
  {NoStop}%
\bibitem [{\citenamefont {Magesan}\ \emph {et~al.}(2011)\citenamefont
  {Magesan}, \citenamefont {Gambetta},\ and\ \citenamefont
  {Emerson}}]{magesan2011scalable}%
  \BibitemOpen
  \bibfield  {author} {\bibinfo {author} {\bibfnamefont {E.}~\bibnamefont
  {Magesan}}, \bibinfo {author} {\bibfnamefont {J.~M.}\ \bibnamefont
  {Gambetta}},\ and\ \bibinfo {author} {\bibfnamefont {J.}~\bibnamefont
  {Emerson}},\ }\bibfield  {title} {\bibinfo {title} {Scalable and robust
  randomized benchmarking of quantum processes},\ }\href
  {https://doi.org/10.1103/PhysRevLett.106.180504} {\bibfield  {journal}
  {\bibinfo  {journal} {Phys. Rev. Lett.}\ }\textbf {\bibinfo {volume} {106}},\
  \bibinfo {pages} {180504} (\bibinfo {year} {2011})}\BibitemShut {NoStop}%
\bibitem [{\citenamefont {Eisert}\ \emph {et~al.}(2020)\citenamefont {Eisert},
  \citenamefont {Hangleiter}, \citenamefont {Walk}, \citenamefont {Roth},
  \citenamefont {Markham}, \citenamefont {Parekh}, \citenamefont {Chabaud},\
  and\ \citenamefont {Kashefi}}]{eisert2020quantum}%
  \BibitemOpen
  \bibfield  {author} {\bibinfo {author} {\bibfnamefont {J.}~\bibnamefont
  {Eisert}}, \bibinfo {author} {\bibfnamefont {D.}~\bibnamefont {Hangleiter}},
  \bibinfo {author} {\bibfnamefont {N.}~\bibnamefont {Walk}}, \bibinfo {author}
  {\bibfnamefont {I.}~\bibnamefont {Roth}}, \bibinfo {author} {\bibfnamefont
  {D.}~\bibnamefont {Markham}}, \bibinfo {author} {\bibfnamefont
  {R.}~\bibnamefont {Parekh}}, \bibinfo {author} {\bibfnamefont
  {U.}~\bibnamefont {Chabaud}},\ and\ \bibinfo {author} {\bibfnamefont
  {E.}~\bibnamefont {Kashefi}},\ }\bibfield  {title} {\bibinfo {title} {Quantum
  certification and benchmarking},\ }\href
  {https://doi.org/10.1038/s42254-020-0186-4} {\bibfield  {journal} {\bibinfo
  {journal} {Nature Reviews Physics}\ }\textbf {\bibinfo {volume} {2}},\
  \bibinfo {pages} {382} (\bibinfo {year} {2020})}\BibitemShut {NoStop}%
\bibitem [{\citenamefont {Raussendorf}\ \emph {et~al.}(2007)\citenamefont
  {Raussendorf}, \citenamefont {Harrington},\ and\ \citenamefont
  {Goyal}}]{Raussendorf_2007}%
  \BibitemOpen
  \bibfield  {author} {\bibinfo {author} {\bibfnamefont {R.}~\bibnamefont
  {Raussendorf}}, \bibinfo {author} {\bibfnamefont {J.}~\bibnamefont
  {Harrington}},\ and\ \bibinfo {author} {\bibfnamefont {K.}~\bibnamefont
  {Goyal}},\ }\bibfield  {title} {\bibinfo {title} {Topological fault-tolerance
  in cluster state quantum computation},\ }\href
  {https://doi.org/10.1088/1367-2630/9/6/199} {\bibfield  {journal} {\bibinfo
  {journal} {New Journal of Physics}\ }\textbf {\bibinfo {volume} {9}},\
  \bibinfo {pages} {199} (\bibinfo {year} {2007})}\BibitemShut {NoStop}%
\bibitem [{\citenamefont {Wang}\ \emph
  {et~al.}(2010{\natexlab{a}})\citenamefont {Wang}, \citenamefont {Fowler},
  \citenamefont {Stephens},\ and\ \citenamefont
  {Hollenberg}}]{wang2009threshold}%
  \BibitemOpen
  \bibfield  {author} {\bibinfo {author} {\bibfnamefont {D.~S.}\ \bibnamefont
  {Wang}}, \bibinfo {author} {\bibfnamefont {A.~G.}\ \bibnamefont {Fowler}},
  \bibinfo {author} {\bibfnamefont {A.~M.}\ \bibnamefont {Stephens}},\ and\
  \bibinfo {author} {\bibfnamefont {L.~C.~L.}\ \bibnamefont {Hollenberg}},\
  }\bibfield  {title} {\bibinfo {title} {Threshold error rates for the toric
  and planar codes},\ }\href
  {https://doi.org/https://doi.org/10.48550/arXiv.0905.0531} {\bibfield
  {journal} {\bibinfo  {journal} {Quantum Info. Comput.}\ }\textbf {\bibinfo
  {volume} {10}},\ \bibinfo {pages} {456–469} (\bibinfo {year}
  {2010}{\natexlab{a}})}\BibitemShut {NoStop}%
\bibitem [{\citenamefont {Wang}\ \emph
  {et~al.}(2010{\natexlab{b}})\citenamefont {Wang}, \citenamefont {Fowler},
  \citenamefont {Hill},\ and\ \citenamefont {Hollenberg}}]{wang2009graphical}%
  \BibitemOpen
  \bibfield  {author} {\bibinfo {author} {\bibfnamefont {D.~S.}\ \bibnamefont
  {Wang}}, \bibinfo {author} {\bibfnamefont {A.~G.}\ \bibnamefont {Fowler}},
  \bibinfo {author} {\bibfnamefont {C.~D.}\ \bibnamefont {Hill}},\ and\
  \bibinfo {author} {\bibfnamefont {L.~C.~L.}\ \bibnamefont {Hollenberg}},\
  }\bibfield  {title} {\bibinfo {title} {Graphical algorithms and threshold
  error rates for the 2d color code},\ }\href
  {https://doi.org/https://doi.org/10.48550/arXiv.0907.1708} {\bibfield
  {journal} {\bibinfo  {journal} {Quantum Info. Comput.}\ }\textbf {\bibinfo
  {volume} {10}},\ \bibinfo {pages} {780–802} (\bibinfo {year}
  {2010}{\natexlab{b}})}\BibitemShut {NoStop}%
\bibitem [{\citenamefont {Fowler}\ \emph
  {et~al.}(2012{\natexlab{a}})\citenamefont {Fowler}, \citenamefont
  {Mariantoni}, \citenamefont {Martinis},\ and\ \citenamefont
  {Cleland}}]{fowler2012towards}%
  \BibitemOpen
  \bibfield  {author} {\bibinfo {author} {\bibfnamefont {A.~G.}\ \bibnamefont
  {Fowler}}, \bibinfo {author} {\bibfnamefont {M.}~\bibnamefont {Mariantoni}},
  \bibinfo {author} {\bibfnamefont {J.~M.}\ \bibnamefont {Martinis}},\ and\
  \bibinfo {author} {\bibfnamefont {A.~N.}\ \bibnamefont {Cleland}},\
  }\bibfield  {title} {\bibinfo {title} {Surface codes: Towards practical
  large-scale quantum computation},\ }\href
  {https://doi.org/10.1103/PhysRevA.86.032324} {\bibfield  {journal} {\bibinfo
  {journal} {Phys. Rev. A}\ }\textbf {\bibinfo {volume} {86}},\ \bibinfo
  {pages} {032324} (\bibinfo {year} {2012}{\natexlab{a}})}\BibitemShut
  {NoStop}%
\bibitem [{\citenamefont {Stephens}(2014)}]{stephens2014fault}%
  \BibitemOpen
  \bibfield  {author} {\bibinfo {author} {\bibfnamefont {A.~M.}\ \bibnamefont
  {Stephens}},\ }\bibfield  {title} {\bibinfo {title} {Fault-tolerant
  thresholds for quantum error correction with the surface code},\ }\href
  {https://doi.org/10.1103/PhysRevA.89.022321} {\bibfield  {journal} {\bibinfo
  {journal} {Phys. Rev. A}\ }\textbf {\bibinfo {volume} {89}},\ \bibinfo
  {pages} {022321} (\bibinfo {year} {2014})}\BibitemShut {NoStop}%
\bibitem [{\citenamefont {Chamberland}\ \emph {et~al.}(2020)\citenamefont
  {Chamberland}, \citenamefont {Zhu}, \citenamefont {Yoder}, \citenamefont
  {Hertzberg},\ and\ \citenamefont {Cross}}]{chamberland2020topological}%
  \BibitemOpen
  \bibfield  {author} {\bibinfo {author} {\bibfnamefont {C.}~\bibnamefont
  {Chamberland}}, \bibinfo {author} {\bibfnamefont {G.}~\bibnamefont {Zhu}},
  \bibinfo {author} {\bibfnamefont {T.~J.}\ \bibnamefont {Yoder}}, \bibinfo
  {author} {\bibfnamefont {J.~B.}\ \bibnamefont {Hertzberg}},\ and\ \bibinfo
  {author} {\bibfnamefont {A.~W.}\ \bibnamefont {Cross}},\ }\bibfield  {title}
  {\bibinfo {title} {Topological and subsystem codes on low-degree graphs with
  flag qubits},\ }\href {https://doi.org/10.1103/PhysRevX.10.011022} {\bibfield
   {journal} {\bibinfo  {journal} {Phys. Rev. X}\ }\textbf {\bibinfo {volume}
  {10}},\ \bibinfo {pages} {011022} (\bibinfo {year} {2020})}\BibitemShut
  {NoStop}%
\bibitem [{\citenamefont {Zhao}\ \emph
  {et~al.}(2022{\natexlab{a}})\citenamefont {Zhao}, \citenamefont {Linghu},
  \citenamefont {Li}, \citenamefont {Xu}, \citenamefont {Wang}, \citenamefont
  {Xue}, \citenamefont {Jin},\ and\ \citenamefont {Yu}}]{zhao2022quantum}%
  \BibitemOpen
  \bibfield  {author} {\bibinfo {author} {\bibfnamefont {P.}~\bibnamefont
  {Zhao}}, \bibinfo {author} {\bibfnamefont {K.}~\bibnamefont {Linghu}},
  \bibinfo {author} {\bibfnamefont {Z.}~\bibnamefont {Li}}, \bibinfo {author}
  {\bibfnamefont {P.}~\bibnamefont {Xu}}, \bibinfo {author} {\bibfnamefont
  {R.}~\bibnamefont {Wang}}, \bibinfo {author} {\bibfnamefont {G.}~\bibnamefont
  {Xue}}, \bibinfo {author} {\bibfnamefont {Y.}~\bibnamefont {Jin}},\ and\
  \bibinfo {author} {\bibfnamefont {H.}~\bibnamefont {Yu}},\ }\bibfield
  {title} {\bibinfo {title} {Quantum crosstalk analysis for simultaneous gate
  operations on superconducting qubits},\ }\href
  {https://doi.org/10.1103/PRXQuantum.3.020301} {\bibfield  {journal} {\bibinfo
   {journal} {PRX Quantum}\ }\textbf {\bibinfo {volume} {3}},\ \bibinfo {pages}
  {020301} (\bibinfo {year} {2022}{\natexlab{a}})}\BibitemShut {NoStop}%
\bibitem [{\citenamefont {Gambetta}\ \emph {et~al.}(2012)\citenamefont
  {Gambetta}, \citenamefont {C\'orcoles}, \citenamefont {Merkel}, \citenamefont
  {Johnson}, \citenamefont {Smolin}, \citenamefont {Chow}, \citenamefont
  {Ryan}, \citenamefont {Rigetti}, \citenamefont {Poletto}, \citenamefont
  {Ohki}, \citenamefont {Ketchen},\ and\ \citenamefont
  {Steffen}}]{gambetta2012characterization}%
  \BibitemOpen
  \bibfield  {author} {\bibinfo {author} {\bibfnamefont {J.~M.}\ \bibnamefont
  {Gambetta}}, \bibinfo {author} {\bibfnamefont {A.~D.}\ \bibnamefont
  {C\'orcoles}}, \bibinfo {author} {\bibfnamefont {S.~T.}\ \bibnamefont
  {Merkel}}, \bibinfo {author} {\bibfnamefont {B.~R.}\ \bibnamefont {Johnson}},
  \bibinfo {author} {\bibfnamefont {J.~A.}\ \bibnamefont {Smolin}}, \bibinfo
  {author} {\bibfnamefont {J.~M.}\ \bibnamefont {Chow}}, \bibinfo {author}
  {\bibfnamefont {C.~A.}\ \bibnamefont {Ryan}}, \bibinfo {author}
  {\bibfnamefont {C.}~\bibnamefont {Rigetti}}, \bibinfo {author} {\bibfnamefont
  {S.}~\bibnamefont {Poletto}}, \bibinfo {author} {\bibfnamefont {T.~A.}\
  \bibnamefont {Ohki}}, \bibinfo {author} {\bibfnamefont {M.~B.}\ \bibnamefont
  {Ketchen}},\ and\ \bibinfo {author} {\bibfnamefont {M.}~\bibnamefont
  {Steffen}},\ }\bibfield  {title} {\bibinfo {title} {Characterization of
  addressability by simultaneous randomized benchmarking},\ }\href
  {https://doi.org/10.1103/PhysRevLett.109.240504} {\bibfield  {journal}
  {\bibinfo  {journal} {Phys. Rev. Lett.}\ }\textbf {\bibinfo {volume} {109}},\
  \bibinfo {pages} {240504} (\bibinfo {year} {2012})}\BibitemShut {NoStop}%
\bibitem [{\citenamefont {Debroy}\ \emph {et~al.}(2023)\citenamefont {Debroy},
  \citenamefont {Genois}, \citenamefont {Gross}, \citenamefont {Mruczkiewicz},
  \citenamefont {Lee}, \citenamefont {Hong}, \citenamefont {Chen},
  \citenamefont {Smelyanskiy},\ and\ \citenamefont
  {Jiang}}]{debroy2023context}%
  \BibitemOpen
  \bibfield  {author} {\bibinfo {author} {\bibfnamefont {D.~M.}\ \bibnamefont
  {Debroy}}, \bibinfo {author} {\bibfnamefont {E.}~\bibnamefont {Genois}},
  \bibinfo {author} {\bibfnamefont {J.~A.}\ \bibnamefont {Gross}}, \bibinfo
  {author} {\bibfnamefont {W.}~\bibnamefont {Mruczkiewicz}}, \bibinfo {author}
  {\bibfnamefont {K.}~\bibnamefont {Lee}}, \bibinfo {author} {\bibfnamefont
  {S.}~\bibnamefont {Hong}}, \bibinfo {author} {\bibfnamefont {Z.}~\bibnamefont
  {Chen}}, \bibinfo {author} {\bibfnamefont {V.}~\bibnamefont {Smelyanskiy}},\
  and\ \bibinfo {author} {\bibfnamefont {Z.}~\bibnamefont {Jiang}},\
  }\href@noop {} {\bibinfo {title} {Context aware fidelity estimation}}
  (\bibinfo {year} {2023}),\ \Eprint {https://arxiv.org/abs/2303.17565}
  {arXiv:2303.17565 [quant-ph]} \BibitemShut {NoStop}%
\bibitem [{\citenamefont {Fowler}\ \emph {et~al.}(2014)\citenamefont {Fowler},
  \citenamefont {Sank}, \citenamefont {Kelly}, \citenamefont {Barends},\ and\
  \citenamefont {Martinis}}]{fowler2014scalable}%
  \BibitemOpen
  \bibfield  {author} {\bibinfo {author} {\bibfnamefont {A.~G.}\ \bibnamefont
  {Fowler}}, \bibinfo {author} {\bibfnamefont {D.}~\bibnamefont {Sank}},
  \bibinfo {author} {\bibfnamefont {J.}~\bibnamefont {Kelly}}, \bibinfo
  {author} {\bibfnamefont {R.}~\bibnamefont {Barends}},\ and\ \bibinfo {author}
  {\bibfnamefont {J.~M.}\ \bibnamefont {Martinis}},\ }\href@noop {} {\bibinfo
  {title} {Scalable extraction of error models from the output of error
  detection circuits}} (\bibinfo {year} {2014}),\ \Eprint
  {https://arxiv.org/abs/1405.1454} {arXiv:1405.1454 [quant-ph]} \BibitemShut
  {NoStop}%
\bibitem [{\citenamefont {Combes}\ \emph {et~al.}(2014)\citenamefont {Combes},
  \citenamefont {Ferrie}, \citenamefont {Cesare}, \citenamefont {Tiersch},
  \citenamefont {Milburn}, \citenamefont {Briegel},\ and\ \citenamefont
  {Caves}}]{combes2014insitu}%
  \BibitemOpen
  \bibfield  {author} {\bibinfo {author} {\bibfnamefont {J.}~\bibnamefont
  {Combes}}, \bibinfo {author} {\bibfnamefont {C.}~\bibnamefont {Ferrie}},
  \bibinfo {author} {\bibfnamefont {C.}~\bibnamefont {Cesare}}, \bibinfo
  {author} {\bibfnamefont {M.}~\bibnamefont {Tiersch}}, \bibinfo {author}
  {\bibfnamefont {G.~J.}\ \bibnamefont {Milburn}}, \bibinfo {author}
  {\bibfnamefont {H.~J.}\ \bibnamefont {Briegel}},\ and\ \bibinfo {author}
  {\bibfnamefont {C.~M.}\ \bibnamefont {Caves}},\ }\href@noop {} {\bibinfo
  {title} {In-situ characterization of quantum devices with error correction}}
  (\bibinfo {year} {2014}),\ \Eprint {https://arxiv.org/abs/1405.5656}
  {arXiv:1405.5656 [quant-ph]} \BibitemShut {NoStop}%
\bibitem [{\citenamefont {Fujiwara}(2014)}]{fujiwara2014instantaneous}%
  \BibitemOpen
  \bibfield  {author} {\bibinfo {author} {\bibfnamefont {Y.}~\bibnamefont
  {Fujiwara}},\ }\href@noop {} {\bibinfo {title} {Instantaneous quantum channel
  estimation during quantum information processing}} (\bibinfo {year} {2014}),\
  \Eprint {https://arxiv.org/abs/1405.6267} {arXiv:1405.6267 [quant-ph]}
  \BibitemShut {NoStop}%
\bibitem [{\citenamefont {Huo}\ and\ \citenamefont {Li}(2017)}]{Huo_2017}%
  \BibitemOpen
  \bibfield  {author} {\bibinfo {author} {\bibfnamefont {M.-X.}\ \bibnamefont
  {Huo}}\ and\ \bibinfo {author} {\bibfnamefont {Y.}~\bibnamefont {Li}},\
  }\bibfield  {title} {\bibinfo {title} {Learning time-dependent noise to
  reduce logical errors: real time error rate estimation in quantum error
  correction},\ }\href {https://doi.org/10.1088/1367-2630/aa916e} {\bibfield
  {journal} {\bibinfo  {journal} {New Journal of Physics}\ }\textbf {\bibinfo
  {volume} {19}},\ \bibinfo {pages} {123032} (\bibinfo {year}
  {2017})}\BibitemShut {NoStop}%
\bibitem [{\citenamefont {Spitz}\ \emph {et~al.}(2018)\citenamefont {Spitz},
  \citenamefont {Tarasinski}, \citenamefont {Beenakker},\ and\ \citenamefont
  {O'Brien}}]{spitz2018adaptive}%
  \BibitemOpen
  \bibfield  {author} {\bibinfo {author} {\bibfnamefont {S.~T.}\ \bibnamefont
  {Spitz}}, \bibinfo {author} {\bibfnamefont {B.}~\bibnamefont {Tarasinski}},
  \bibinfo {author} {\bibfnamefont {C.~W.~J.}\ \bibnamefont {Beenakker}},\ and\
  \bibinfo {author} {\bibfnamefont {T.~E.}\ \bibnamefont {O'Brien}},\
  }\bibfield  {title} {\bibinfo {title} {Adaptive weight estimator for quantum
  error correction in a time-dependent environment},\ }\href
  {https://doi.org/https://doi.org/10.1002/qute.201800012} {\bibfield
  {journal} {\bibinfo  {journal} {Advanced Quantum Technologies}\ }\textbf
  {\bibinfo {volume} {1}},\ \bibinfo {pages} {1800012} (\bibinfo {year}
  {2018})}\BibitemShut {NoStop}%
\bibitem [{\citenamefont {Wagner}\ \emph {et~al.}(2022)\citenamefont {Wagner},
  \citenamefont {Kampermann}, \citenamefont {Bru{\ss}},\ and\ \citenamefont
  {Kliesch}}]{Wagner_2022}%
  \BibitemOpen
  \bibfield  {author} {\bibinfo {author} {\bibfnamefont {T.}~\bibnamefont
  {Wagner}}, \bibinfo {author} {\bibfnamefont {H.}~\bibnamefont {Kampermann}},
  \bibinfo {author} {\bibfnamefont {D.}~\bibnamefont {Bru{\ss}}},\ and\
  \bibinfo {author} {\bibfnamefont {M.}~\bibnamefont {Kliesch}},\ }\bibfield
  {title} {\bibinfo {title} {Pauli channels can be estimated from syndrome
  measurements in quantum error correction},\ }\href
  {https://doi.org/10.22331/q-2022-09-19-809} {\bibfield  {journal} {\bibinfo
  {journal} {Quantum}\ }\textbf {\bibinfo {volume} {6}},\ \bibinfo {pages}
  {809} (\bibinfo {year} {2022})}\BibitemShut {NoStop}%
\bibitem [{\citenamefont {Wagner}\ \emph {et~al.}(2023)\citenamefont {Wagner},
  \citenamefont {Kampermann}, \citenamefont {Bru\ss{}},\ and\ \citenamefont
  {Kliesch}}]{wagner2023learning}%
  \BibitemOpen
  \bibfield  {author} {\bibinfo {author} {\bibfnamefont {T.}~\bibnamefont
  {Wagner}}, \bibinfo {author} {\bibfnamefont {H.}~\bibnamefont {Kampermann}},
  \bibinfo {author} {\bibfnamefont {D.}~\bibnamefont {Bru\ss{}}},\ and\
  \bibinfo {author} {\bibfnamefont {M.}~\bibnamefont {Kliesch}},\ }\bibfield
  {title} {\bibinfo {title} {Learning logical pauli noise in quantum error
  correction},\ }\href {https://doi.org/10.1103/PhysRevLett.130.200601}
  {\bibfield  {journal} {\bibinfo  {journal} {Phys. Rev. Lett.}\ }\textbf
  {\bibinfo {volume} {130}},\ \bibinfo {pages} {200601} (\bibinfo {year}
  {2023})}\BibitemShut {NoStop}%
\bibitem [{\citenamefont
  {Wootton}(2022{\natexlab{a}})}]{wootton2022syndromederived}%
  \BibitemOpen
  \bibfield  {author} {\bibinfo {author} {\bibfnamefont {J.~R.}\ \bibnamefont
  {Wootton}},\ }\href@noop {} {\bibinfo {title} {Syndrome-derived error rates
  as a benchmark of quantum hardware}} (\bibinfo {year} {2022}{\natexlab{a}}),\
  \Eprint {https://arxiv.org/abs/2207.00553} {arXiv:2207.00553 [quant-ph]}
  \BibitemShut {NoStop}%
\bibitem [{\citenamefont {Börner}\ \emph {et~al.}(2023)\citenamefont
  {Börner}, \citenamefont {Berke}, \citenamefont {DiVincenzo}, \citenamefont
  {Trebst},\ and\ \citenamefont {Altland}}]{borner2023classical}%
  \BibitemOpen
  \bibfield  {author} {\bibinfo {author} {\bibfnamefont {S.-D.}\ \bibnamefont
  {Börner}}, \bibinfo {author} {\bibfnamefont {C.}~\bibnamefont {Berke}},
  \bibinfo {author} {\bibfnamefont {D.~P.}\ \bibnamefont {DiVincenzo}},
  \bibinfo {author} {\bibfnamefont {S.}~\bibnamefont {Trebst}},\ and\ \bibinfo
  {author} {\bibfnamefont {A.}~\bibnamefont {Altland}},\ }\href@noop {}
  {\bibinfo {title} {Classical chaos in quantum computers}} (\bibinfo {year}
  {2023}),\ \Eprint {https://arxiv.org/abs/2304.14435} {arXiv:2304.14435
  [quant-ph]} \BibitemShut {NoStop}%
\bibitem [{\citenamefont {Morvan}\ \emph {et~al.}(2023)\citenamefont {Morvan}
  \emph {et~al.}}]{morvan2023phase}%
  \BibitemOpen
  \bibfield  {author} {\bibinfo {author} {\bibfnamefont {A.}~\bibnamefont
  {Morvan}} \emph {et~al.},\ }\href@noop {} {\bibinfo {title} {Phase transition
  in random circuit sampling}} (\bibinfo {year} {2023}),\ \Eprint
  {https://arxiv.org/abs/2304.11119} {arXiv:2304.11119 [quant-ph]} \BibitemShut
  {NoStop}%
\bibitem [{\citenamefont {Knill}\ \emph
  {et~al.}(2008{\natexlab{b}})\citenamefont {Knill}, \citenamefont {Leibfried},
  \citenamefont {Reichle}, \citenamefont {Britton}, \citenamefont {Blakestad},
  \citenamefont {Jost}, \citenamefont {Langer}, \citenamefont {Ozeri},
  \citenamefont {Seidelin},\ and\ \citenamefont
  {Wineland}}]{knill2008randomized}%
  \BibitemOpen
  \bibfield  {author} {\bibinfo {author} {\bibfnamefont {E.}~\bibnamefont
  {Knill}}, \bibinfo {author} {\bibfnamefont {D.}~\bibnamefont {Leibfried}},
  \bibinfo {author} {\bibfnamefont {R.}~\bibnamefont {Reichle}}, \bibinfo
  {author} {\bibfnamefont {J.}~\bibnamefont {Britton}}, \bibinfo {author}
  {\bibfnamefont {R.~B.}\ \bibnamefont {Blakestad}}, \bibinfo {author}
  {\bibfnamefont {J.~D.}\ \bibnamefont {Jost}}, \bibinfo {author}
  {\bibfnamefont {C.}~\bibnamefont {Langer}}, \bibinfo {author} {\bibfnamefont
  {R.}~\bibnamefont {Ozeri}}, \bibinfo {author} {\bibfnamefont
  {S.}~\bibnamefont {Seidelin}},\ and\ \bibinfo {author} {\bibfnamefont
  {D.~J.}\ \bibnamefont {Wineland}},\ }\bibfield  {title} {\bibinfo {title}
  {Randomized benchmarking of quantum gates},\ }\href
  {https://doi.org/10.1103/PhysRevA.77.012307} {\bibfield  {journal} {\bibinfo
  {journal} {Phys. Rev. A}\ }\textbf {\bibinfo {volume} {77}},\ \bibinfo
  {pages} {012307} (\bibinfo {year} {2008}{\natexlab{b}})}\BibitemShut
  {NoStop}%
\bibitem [{\citenamefont {Magesan}\ \emph {et~al.}(2012)\citenamefont
  {Magesan}, \citenamefont {Gambetta}, \citenamefont {Johnson}, \citenamefont
  {Ryan}, \citenamefont {Chow}, \citenamefont {Merkel}, \citenamefont
  {da~Silva}, \citenamefont {Keefe}, \citenamefont {Rothwell}, \citenamefont
  {Ohki}, \citenamefont {Ketchen},\ and\ \citenamefont
  {Steffen}}]{magesan2012efficient}%
  \BibitemOpen
  \bibfield  {author} {\bibinfo {author} {\bibfnamefont {E.}~\bibnamefont
  {Magesan}}, \bibinfo {author} {\bibfnamefont {J.~M.}\ \bibnamefont
  {Gambetta}}, \bibinfo {author} {\bibfnamefont {B.~R.}\ \bibnamefont
  {Johnson}}, \bibinfo {author} {\bibfnamefont {C.~A.}\ \bibnamefont {Ryan}},
  \bibinfo {author} {\bibfnamefont {J.~M.}\ \bibnamefont {Chow}}, \bibinfo
  {author} {\bibfnamefont {S.~T.}\ \bibnamefont {Merkel}}, \bibinfo {author}
  {\bibfnamefont {M.~P.}\ \bibnamefont {da~Silva}}, \bibinfo {author}
  {\bibfnamefont {G.~A.}\ \bibnamefont {Keefe}}, \bibinfo {author}
  {\bibfnamefont {M.~B.}\ \bibnamefont {Rothwell}}, \bibinfo {author}
  {\bibfnamefont {T.~A.}\ \bibnamefont {Ohki}}, \bibinfo {author}
  {\bibfnamefont {M.~B.}\ \bibnamefont {Ketchen}},\ and\ \bibinfo {author}
  {\bibfnamefont {M.}~\bibnamefont {Steffen}},\ }\bibfield  {title} {\bibinfo
  {title} {Efficient measurement of quantum gate error by interleaved
  randomized benchmarking},\ }\href
  {https://doi.org/10.1103/PhysRevLett.109.080505} {\bibfield  {journal}
  {\bibinfo  {journal} {Phys. Rev. Lett.}\ }\textbf {\bibinfo {volume} {109}},\
  \bibinfo {pages} {080505} (\bibinfo {year} {2012})}\BibitemShut {NoStop}%
\bibitem [{\citenamefont {Cross}\ \emph {et~al.}(2019)\citenamefont {Cross},
  \citenamefont {Bishop}, \citenamefont {Sheldon}, \citenamefont {Nation},\
  and\ \citenamefont {Gambetta}}]{cross2019validating}%
  \BibitemOpen
  \bibfield  {author} {\bibinfo {author} {\bibfnamefont {A.~W.}\ \bibnamefont
  {Cross}}, \bibinfo {author} {\bibfnamefont {L.~S.}\ \bibnamefont {Bishop}},
  \bibinfo {author} {\bibfnamefont {S.}~\bibnamefont {Sheldon}}, \bibinfo
  {author} {\bibfnamefont {P.~D.}\ \bibnamefont {Nation}},\ and\ \bibinfo
  {author} {\bibfnamefont {J.~M.}\ \bibnamefont {Gambetta}},\ }\bibfield
  {title} {\bibinfo {title} {Validating quantum computers using randomized
  model circuits},\ }\href {https://doi.org/10.1103/PhysRevA.100.032328}
  {\bibfield  {journal} {\bibinfo  {journal} {Phys. Rev. A}\ }\textbf {\bibinfo
  {volume} {100}},\ \bibinfo {pages} {032328} (\bibinfo {year}
  {2019})}\BibitemShut {NoStop}%
\bibitem [{\citenamefont {Mooney}\ \emph {et~al.}(2021)\citenamefont {Mooney},
  \citenamefont {White}, \citenamefont {Hill},\ and\ \citenamefont
  {Hollenberg}}]{https://doi.org/10.1002/qute.202100061}%
  \BibitemOpen
  \bibfield  {author} {\bibinfo {author} {\bibfnamefont {G.~J.}\ \bibnamefont
  {Mooney}}, \bibinfo {author} {\bibfnamefont {G.~A.~L.}\ \bibnamefont
  {White}}, \bibinfo {author} {\bibfnamefont {C.~D.}\ \bibnamefont {Hill}},\
  and\ \bibinfo {author} {\bibfnamefont {L.~C.~L.}\ \bibnamefont
  {Hollenberg}},\ }\bibfield  {title} {\bibinfo {title} {Whole-device
  entanglement in a 65-qubit superconducting quantum computer},\ }\href
  {https://doi.org/https://doi.org/10.1002/qute.202100061} {\bibfield
  {journal} {\bibinfo  {journal} {Advanced Quantum Technologies}\ }\textbf
  {\bibinfo {volume} {4}},\ \bibinfo {pages} {2100061} (\bibinfo {year}
  {2021})}\BibitemShut {NoStop}%
\bibitem [{\citenamefont {Bharti}\ \emph {et~al.}(2022)\citenamefont {Bharti},
  \citenamefont {Cervera-Lierta}, \citenamefont {Kyaw}, \citenamefont {Haug},
  \citenamefont {Alperin-Lea}, \citenamefont {Anand}, \citenamefont {Degroote},
  \citenamefont {Heimonen}, \citenamefont {Kottmann}, \citenamefont {Menke},
  \citenamefont {Mok}, \citenamefont {Sim}, \citenamefont {Kwek},\ and\
  \citenamefont {Aspuru-Guzik}}]{bharti2022noisy}%
  \BibitemOpen
  \bibfield  {author} {\bibinfo {author} {\bibfnamefont {K.}~\bibnamefont
  {Bharti}}, \bibinfo {author} {\bibfnamefont {A.}~\bibnamefont
  {Cervera-Lierta}}, \bibinfo {author} {\bibfnamefont {T.~H.}\ \bibnamefont
  {Kyaw}}, \bibinfo {author} {\bibfnamefont {T.}~\bibnamefont {Haug}}, \bibinfo
  {author} {\bibfnamefont {S.}~\bibnamefont {Alperin-Lea}}, \bibinfo {author}
  {\bibfnamefont {A.}~\bibnamefont {Anand}}, \bibinfo {author} {\bibfnamefont
  {M.}~\bibnamefont {Degroote}}, \bibinfo {author} {\bibfnamefont
  {H.}~\bibnamefont {Heimonen}}, \bibinfo {author} {\bibfnamefont {J.~S.}\
  \bibnamefont {Kottmann}}, \bibinfo {author} {\bibfnamefont {T.}~\bibnamefont
  {Menke}}, \bibinfo {author} {\bibfnamefont {W.-K.}\ \bibnamefont {Mok}},
  \bibinfo {author} {\bibfnamefont {S.}~\bibnamefont {Sim}}, \bibinfo {author}
  {\bibfnamefont {L.-C.}\ \bibnamefont {Kwek}},\ and\ \bibinfo {author}
  {\bibfnamefont {A.}~\bibnamefont {Aspuru-Guzik}},\ }\bibfield  {title}
  {\bibinfo {title} {Noisy intermediate-scale quantum algorithms},\ }\href
  {https://doi.org/10.1103/RevModPhys.94.015004} {\bibfield  {journal}
  {\bibinfo  {journal} {Rev. Mod. Phys.}\ }\textbf {\bibinfo {volume} {94}},\
  \bibinfo {pages} {015004} (\bibinfo {year} {2022})}\BibitemShut {NoStop}%
\bibitem [{\citenamefont {Chen}\ \emph {et~al.}(2021)\citenamefont {Chen} \emph
  {et~al.}}]{chen2021nature}%
  \BibitemOpen
  \bibfield  {author} {\bibinfo {author} {\bibfnamefont {Z.}~\bibnamefont
  {Chen}} \emph {et~al.},\ }\bibfield  {title} {\bibinfo {title} {Exponential
  suppression of bit or phase errors with cyclic error correction},\ }\href
  {https://doi.org/10.1038/s41586-021-03588-y} {\bibfield  {journal} {\bibinfo
  {journal} {Nature}\ }\textbf {\bibinfo {volume} {595}},\ \bibinfo {pages}
  {383} (\bibinfo {year} {2021})}\BibitemShut {NoStop}%
\bibitem [{\citenamefont {Krinner}\ \emph {et~al.}(2022)\citenamefont
  {Krinner}, \citenamefont {Lacroix}, \citenamefont {Remm}, \citenamefont
  {Di~Paolo}, \citenamefont {Genois}, \citenamefont {Leroux}, \citenamefont
  {Hellings}, \citenamefont {Lazar}, \citenamefont {Swiadek}, \citenamefont
  {Herrmann}, \citenamefont {Norris}, \citenamefont {Andersen}, \citenamefont
  {M{\"u}ller}, \citenamefont {Blais}, \citenamefont {Eichler},\ and\
  \citenamefont {Wallraff}}]{krinner2022realizing}%
  \BibitemOpen
  \bibfield  {author} {\bibinfo {author} {\bibfnamefont {S.}~\bibnamefont
  {Krinner}}, \bibinfo {author} {\bibfnamefont {N.}~\bibnamefont {Lacroix}},
  \bibinfo {author} {\bibfnamefont {A.}~\bibnamefont {Remm}}, \bibinfo {author}
  {\bibfnamefont {A.}~\bibnamefont {Di~Paolo}}, \bibinfo {author}
  {\bibfnamefont {E.}~\bibnamefont {Genois}}, \bibinfo {author} {\bibfnamefont
  {C.}~\bibnamefont {Leroux}}, \bibinfo {author} {\bibfnamefont
  {C.}~\bibnamefont {Hellings}}, \bibinfo {author} {\bibfnamefont
  {S.}~\bibnamefont {Lazar}}, \bibinfo {author} {\bibfnamefont
  {F.}~\bibnamefont {Swiadek}}, \bibinfo {author} {\bibfnamefont
  {J.}~\bibnamefont {Herrmann}}, \bibinfo {author} {\bibfnamefont {G.~J.}\
  \bibnamefont {Norris}}, \bibinfo {author} {\bibfnamefont {C.~K.}\
  \bibnamefont {Andersen}}, \bibinfo {author} {\bibfnamefont {M.}~\bibnamefont
  {M{\"u}ller}}, \bibinfo {author} {\bibfnamefont {A.}~\bibnamefont {Blais}},
  \bibinfo {author} {\bibfnamefont {C.}~\bibnamefont {Eichler}},\ and\ \bibinfo
  {author} {\bibfnamefont {A.}~\bibnamefont {Wallraff}},\ }\bibfield  {title}
  {\bibinfo {title} {Realizing repeated quantum error correction in a
  distance-three surface code},\ }\href
  {https://doi.org/10.1038/s41586-022-04566-8} {\bibfield  {journal} {\bibinfo
  {journal} {Nature}\ }\textbf {\bibinfo {volume} {605}},\ \bibinfo {pages}
  {669} (\bibinfo {year} {2022})}\BibitemShut {NoStop}%
\bibitem [{\citenamefont {Andersen}\ \emph {et~al.}(2020)\citenamefont
  {Andersen}, \citenamefont {Remm}, \citenamefont {Lazar}, \citenamefont
  {Krinner}, \citenamefont {Lacroix}, \citenamefont {Norris}, \citenamefont
  {Gabureac}, \citenamefont {Eichler},\ and\ \citenamefont
  {Wallraff}}]{andersen2020naturephysics}%
  \BibitemOpen
  \bibfield  {author} {\bibinfo {author} {\bibfnamefont {C.~K.}\ \bibnamefont
  {Andersen}}, \bibinfo {author} {\bibfnamefont {A.}~\bibnamefont {Remm}},
  \bibinfo {author} {\bibfnamefont {S.}~\bibnamefont {Lazar}}, \bibinfo
  {author} {\bibfnamefont {S.}~\bibnamefont {Krinner}}, \bibinfo {author}
  {\bibfnamefont {N.}~\bibnamefont {Lacroix}}, \bibinfo {author} {\bibfnamefont
  {G.~J.}\ \bibnamefont {Norris}}, \bibinfo {author} {\bibfnamefont
  {M.}~\bibnamefont {Gabureac}}, \bibinfo {author} {\bibfnamefont
  {C.}~\bibnamefont {Eichler}},\ and\ \bibinfo {author} {\bibfnamefont
  {A.}~\bibnamefont {Wallraff}},\ }\bibfield  {title} {\bibinfo {title}
  {Repeated quantum error detection in a surface code},\ }\href
  {https://doi.org/10.1038/s41567-020-0920-y} {\bibfield  {journal} {\bibinfo
  {journal} {Nature Physics}\ }\textbf {\bibinfo {volume} {16}},\ \bibinfo
  {pages} {875} (\bibinfo {year} {2020})}\BibitemShut {NoStop}%
\bibitem [{\citenamefont {Marques}\ \emph {et~al.}(2022)\citenamefont
  {Marques}, \citenamefont {Varbanov}, \citenamefont {Moreira}, \citenamefont
  {Ali}, \citenamefont {Muthusubramanian}, \citenamefont {Zachariadis},
  \citenamefont {Battistel}, \citenamefont {Beekman}, \citenamefont {Haider},
  \citenamefont {Vlothuizen}, \citenamefont {Bruno}, \citenamefont {Terhal},\
  and\ \citenamefont {DiCarlo}}]{marques2022logical}%
  \BibitemOpen
  \bibfield  {author} {\bibinfo {author} {\bibfnamefont {J.~F.}\ \bibnamefont
  {Marques}}, \bibinfo {author} {\bibfnamefont {B.~M.}\ \bibnamefont
  {Varbanov}}, \bibinfo {author} {\bibfnamefont {M.~S.}\ \bibnamefont
  {Moreira}}, \bibinfo {author} {\bibfnamefont {H.}~\bibnamefont {Ali}},
  \bibinfo {author} {\bibfnamefont {N.}~\bibnamefont {Muthusubramanian}},
  \bibinfo {author} {\bibfnamefont {C.}~\bibnamefont {Zachariadis}}, \bibinfo
  {author} {\bibfnamefont {F.}~\bibnamefont {Battistel}}, \bibinfo {author}
  {\bibfnamefont {M.}~\bibnamefont {Beekman}}, \bibinfo {author} {\bibfnamefont
  {N.}~\bibnamefont {Haider}}, \bibinfo {author} {\bibfnamefont
  {W.}~\bibnamefont {Vlothuizen}}, \bibinfo {author} {\bibfnamefont
  {A.}~\bibnamefont {Bruno}}, \bibinfo {author} {\bibfnamefont {B.~M.}\
  \bibnamefont {Terhal}},\ and\ \bibinfo {author} {\bibfnamefont
  {L.}~\bibnamefont {DiCarlo}},\ }\bibfield  {title} {\bibinfo {title}
  {Logical-qubit operations in an error-detecting surface code},\ }\href
  {https://doi.org/10.1038/s41567-021-01423-9} {\bibfield  {journal} {\bibinfo
  {journal} {Nature Physics}\ }\textbf {\bibinfo {volume} {18}},\ \bibinfo
  {pages} {80} (\bibinfo {year} {2022})}\BibitemShut {NoStop}%
\bibitem [{\citenamefont {Zhao}\ \emph
  {et~al.}(2022{\natexlab{b}})\citenamefont {Zhao} \emph
  {et~al.}}]{zhao2022realization}%
  \BibitemOpen
  \bibfield  {author} {\bibinfo {author} {\bibfnamefont {Y.}~\bibnamefont
  {Zhao}} \emph {et~al.},\ }\bibfield  {title} {\bibinfo {title} {Realization
  of an error-correcting surface code with superconducting qubits},\ }\href
  {https://doi.org/10.1103/PhysRevLett.129.030501} {\bibfield  {journal}
  {\bibinfo  {journal} {Phys. Rev. Lett.}\ }\textbf {\bibinfo {volume} {129}},\
  \bibinfo {pages} {030501} (\bibinfo {year} {2022}{\natexlab{b}})}\BibitemShut
  {NoStop}%
\bibitem [{\citenamefont {Sivak}\ \emph {et~al.}(2023)\citenamefont {Sivak},
  \citenamefont {Eickbusch}, \citenamefont {Royer}, \citenamefont {Singh},
  \citenamefont {Tsioutsios}, \citenamefont {Ganjam}, \citenamefont {Miano},
  \citenamefont {Brock}, \citenamefont {Ding}, \citenamefont {Frunzio},
  \citenamefont {Girvin}, \citenamefont {Schoelkopf},\ and\ \citenamefont
  {Devoret}}]{Sivak_2023}%
  \BibitemOpen
  \bibfield  {author} {\bibinfo {author} {\bibfnamefont {V.~V.}\ \bibnamefont
  {Sivak}}, \bibinfo {author} {\bibfnamefont {A.}~\bibnamefont {Eickbusch}},
  \bibinfo {author} {\bibfnamefont {B.}~\bibnamefont {Royer}}, \bibinfo
  {author} {\bibfnamefont {S.}~\bibnamefont {Singh}}, \bibinfo {author}
  {\bibfnamefont {I.}~\bibnamefont {Tsioutsios}}, \bibinfo {author}
  {\bibfnamefont {S.}~\bibnamefont {Ganjam}}, \bibinfo {author} {\bibfnamefont
  {A.}~\bibnamefont {Miano}}, \bibinfo {author} {\bibfnamefont {B.~L.}\
  \bibnamefont {Brock}}, \bibinfo {author} {\bibfnamefont {A.~Z.}\ \bibnamefont
  {Ding}}, \bibinfo {author} {\bibfnamefont {L.}~\bibnamefont {Frunzio}},
  \bibinfo {author} {\bibfnamefont {S.~M.}\ \bibnamefont {Girvin}}, \bibinfo
  {author} {\bibfnamefont {R.~J.}\ \bibnamefont {Schoelkopf}},\ and\ \bibinfo
  {author} {\bibfnamefont {M.~H.}\ \bibnamefont {Devoret}},\ }\bibfield
  {title} {\bibinfo {title} {Real-time quantum error correction beyond
  break-even},\ }\href {https://doi.org/10.1038/s41586-023-05782-6} {\bibfield
  {journal} {\bibinfo  {journal} {Nature}\ }\textbf {\bibinfo {volume} {616}},\
  \bibinfo {pages} {50} (\bibinfo {year} {2023})}\BibitemShut {NoStop}%
\bibitem [{\citenamefont {Chen}\ \emph {et~al.}(2022)\citenamefont {Chen},
  \citenamefont {Yoder}, \citenamefont {Kim}, \citenamefont {Sundaresan},
  \citenamefont {Srinivasan}, \citenamefont {Li}, \citenamefont {C\'orcoles},
  \citenamefont {Cross},\ and\ \citenamefont {Takita}}]{chen2022calibrated}%
  \BibitemOpen
  \bibfield  {author} {\bibinfo {author} {\bibfnamefont {E.~H.}\ \bibnamefont
  {Chen}}, \bibinfo {author} {\bibfnamefont {T.~J.}\ \bibnamefont {Yoder}},
  \bibinfo {author} {\bibfnamefont {Y.}~\bibnamefont {Kim}}, \bibinfo {author}
  {\bibfnamefont {N.}~\bibnamefont {Sundaresan}}, \bibinfo {author}
  {\bibfnamefont {S.}~\bibnamefont {Srinivasan}}, \bibinfo {author}
  {\bibfnamefont {M.}~\bibnamefont {Li}}, \bibinfo {author} {\bibfnamefont
  {A.~D.}\ \bibnamefont {C\'orcoles}}, \bibinfo {author} {\bibfnamefont
  {A.~W.}\ \bibnamefont {Cross}},\ and\ \bibinfo {author} {\bibfnamefont
  {M.}~\bibnamefont {Takita}},\ }\bibfield  {title} {\bibinfo {title}
  {Calibrated decoders for experimental quantum error correction},\ }\href
  {https://doi.org/10.1103/PhysRevLett.128.110504} {\bibfield  {journal}
  {\bibinfo  {journal} {Phys. Rev. Lett.}\ }\textbf {\bibinfo {volume} {128}},\
  \bibinfo {pages} {110504} (\bibinfo {year} {2022})}\BibitemShut {NoStop}%
\bibitem [{\citenamefont {Sundaresan}\ \emph {et~al.}(2023)\citenamefont
  {Sundaresan}, \citenamefont {Yoder}, \citenamefont {Kim}, \citenamefont {Li},
  \citenamefont {Chen}, \citenamefont {Harper}, \citenamefont {Thorbeck},
  \citenamefont {Cross}, \citenamefont {C{\'o}rcoles},\ and\ \citenamefont
  {Takita}}]{sundaresan2023demonstrating}%
  \BibitemOpen
  \bibfield  {author} {\bibinfo {author} {\bibfnamefont {N.}~\bibnamefont
  {Sundaresan}}, \bibinfo {author} {\bibfnamefont {T.~J.}\ \bibnamefont
  {Yoder}}, \bibinfo {author} {\bibfnamefont {Y.}~\bibnamefont {Kim}}, \bibinfo
  {author} {\bibfnamefont {M.}~\bibnamefont {Li}}, \bibinfo {author}
  {\bibfnamefont {E.~H.}\ \bibnamefont {Chen}}, \bibinfo {author}
  {\bibfnamefont {G.}~\bibnamefont {Harper}}, \bibinfo {author} {\bibfnamefont
  {T.}~\bibnamefont {Thorbeck}}, \bibinfo {author} {\bibfnamefont {A.~W.}\
  \bibnamefont {Cross}}, \bibinfo {author} {\bibfnamefont {A.~D.}\ \bibnamefont
  {C{\'o}rcoles}},\ and\ \bibinfo {author} {\bibfnamefont {M.}~\bibnamefont
  {Takita}},\ }\bibfield  {title} {\bibinfo {title} {Demonstrating multi-round
  subsystem quantum error correction using matching and maximum likelihood
  decoders},\ }\href {https://doi.org/10.1038/s41467-023-38247-5} {\bibfield
  {journal} {\bibinfo  {journal} {Nature Communications}\ }\textbf {\bibinfo
  {volume} {14}},\ \bibinfo {pages} {2852} (\bibinfo {year}
  {2023})}\BibitemShut {NoStop}%
\bibitem [{\citenamefont
  {Wootton}(2022{\natexlab{b}})}]{wootton2022measurements}%
  \BibitemOpen
  \bibfield  {author} {\bibinfo {author} {\bibfnamefont {J.~R.}\ \bibnamefont
  {Wootton}},\ }\href@noop {} {\bibinfo {title} {Measurements of floquet code
  plaquette stabilizers}} (\bibinfo {year} {2022}{\natexlab{b}}),\ \Eprint
  {https://arxiv.org/abs/2210.13154} {arXiv:2210.13154 [quant-ph]} \BibitemShut
  {NoStop}%
\bibitem [{\citenamefont {Acharya}\ \emph {et~al.}(2023)\citenamefont {Acharya}
  \emph {et~al.}}]{acharya2022suppressing}%
  \BibitemOpen
  \bibfield  {author} {\bibinfo {author} {\bibfnamefont {R.}~\bibnamefont
  {Acharya}} \emph {et~al.},\ }\bibfield  {title} {\bibinfo {title}
  {Suppressing quantum errors by scaling a surface code logical qubit},\ }\href
  {https://doi.org/10.1038/s41586-022-05434-1} {\bibfield  {journal} {\bibinfo
  {journal} {Nature}\ }\textbf {\bibinfo {volume} {614}},\ \bibinfo {pages}
  {676} (\bibinfo {year} {2023})}\BibitemShut {NoStop}%
\bibitem [{\citenamefont {Erhard}\ \emph {et~al.}(2021)\citenamefont {Erhard},
  \citenamefont {Poulsen~Nautrup}, \citenamefont {Meth}, \citenamefont
  {Postler}, \citenamefont {Stricker}, \citenamefont {Stadler}, \citenamefont
  {Negnevitsky}, \citenamefont {Ringbauer}, \citenamefont {Schindler},
  \citenamefont {Briegel}, \citenamefont {Blatt}, \citenamefont {Friis},\ and\
  \citenamefont {Monz}}]{erhard2021entangling}%
  \BibitemOpen
  \bibfield  {author} {\bibinfo {author} {\bibfnamefont {A.}~\bibnamefont
  {Erhard}}, \bibinfo {author} {\bibfnamefont {H.}~\bibnamefont
  {Poulsen~Nautrup}}, \bibinfo {author} {\bibfnamefont {M.}~\bibnamefont
  {Meth}}, \bibinfo {author} {\bibfnamefont {L.}~\bibnamefont {Postler}},
  \bibinfo {author} {\bibfnamefont {R.}~\bibnamefont {Stricker}}, \bibinfo
  {author} {\bibfnamefont {M.}~\bibnamefont {Stadler}}, \bibinfo {author}
  {\bibfnamefont {V.}~\bibnamefont {Negnevitsky}}, \bibinfo {author}
  {\bibfnamefont {M.}~\bibnamefont {Ringbauer}}, \bibinfo {author}
  {\bibfnamefont {P.}~\bibnamefont {Schindler}}, \bibinfo {author}
  {\bibfnamefont {H.~J.}\ \bibnamefont {Briegel}}, \bibinfo {author}
  {\bibfnamefont {R.}~\bibnamefont {Blatt}}, \bibinfo {author} {\bibfnamefont
  {N.}~\bibnamefont {Friis}},\ and\ \bibinfo {author} {\bibfnamefont
  {T.}~\bibnamefont {Monz}},\ }\bibfield  {title} {\bibinfo {title} {Entangling
  logical qubits with lattice surgery},\ }\href
  {https://doi.org/10.1038/s41586-020-03079-6} {\bibfield  {journal} {\bibinfo
  {journal} {Nature}\ }\textbf {\bibinfo {volume} {589}},\ \bibinfo {pages}
  {220} (\bibinfo {year} {2021})}\BibitemShut {NoStop}%
\bibitem [{\citenamefont {Satzinger}\ \emph {et~al.}(2021)\citenamefont
  {Satzinger} \emph {et~al.}}]{satzinger2021realizing}%
  \BibitemOpen
  \bibfield  {author} {\bibinfo {author} {\bibfnamefont {K.~J.}\ \bibnamefont
  {Satzinger}} \emph {et~al.},\ }\bibfield  {title} {\bibinfo {title}
  {Realizing topologically ordered states on a quantum processor},\ }\href
  {https://doi.org/10.1126/science.abi8378} {\bibfield  {journal} {\bibinfo
  {journal} {Science}\ }\textbf {\bibinfo {volume} {374}},\ \bibinfo {pages}
  {1237} (\bibinfo {year} {2021})}\BibitemShut {NoStop}%
\bibitem [{\citenamefont {Xu}\ \emph {et~al.}(2023)\citenamefont {Xu} \emph
  {et~al.}}]{xu2023digital}%
  \BibitemOpen
  \bibfield  {author} {\bibinfo {author} {\bibfnamefont {S.}~\bibnamefont {Xu}}
  \emph {et~al.},\ }\bibfield  {title} {\bibinfo {title} {Digital simulation of
  projective non-abelian anyons with 68 superconducting qubits},\ }\href
  {https://doi.org/10.1088/0256-307X/40/6/060301} {\bibfield  {journal}
  {\bibinfo  {journal} {Chinese Physics Letters}\ }\textbf {\bibinfo {volume}
  {40}},\ \bibinfo {pages} {060301} (\bibinfo {year} {2023})}\BibitemShut
  {NoStop}%
\bibitem [{\citenamefont {Koch}\ \emph {et~al.}(2007)\citenamefont {Koch},
  \citenamefont {Yu}, \citenamefont {Gambetta}, \citenamefont {Houck},
  \citenamefont {Schuster}, \citenamefont {Majer}, \citenamefont {Blais},
  \citenamefont {Devoret}, \citenamefont {Girvin},\ and\ \citenamefont
  {Schoelkopf}}]{PhysRevA.76.042319}%
  \BibitemOpen
  \bibfield  {author} {\bibinfo {author} {\bibfnamefont {J.}~\bibnamefont
  {Koch}}, \bibinfo {author} {\bibfnamefont {T.~M.}\ \bibnamefont {Yu}},
  \bibinfo {author} {\bibfnamefont {J.}~\bibnamefont {Gambetta}}, \bibinfo
  {author} {\bibfnamefont {A.~A.}\ \bibnamefont {Houck}}, \bibinfo {author}
  {\bibfnamefont {D.~I.}\ \bibnamefont {Schuster}}, \bibinfo {author}
  {\bibfnamefont {J.}~\bibnamefont {Majer}}, \bibinfo {author} {\bibfnamefont
  {A.}~\bibnamefont {Blais}}, \bibinfo {author} {\bibfnamefont {M.~H.}\
  \bibnamefont {Devoret}}, \bibinfo {author} {\bibfnamefont {S.~M.}\
  \bibnamefont {Girvin}},\ and\ \bibinfo {author} {\bibfnamefont {R.~J.}\
  \bibnamefont {Schoelkopf}},\ }\bibfield  {title} {\bibinfo {title}
  {Charge-insensitive qubit design derived from the cooper pair box},\ }\href
  {https://doi.org/10.1103/PhysRevA.76.042319} {\bibfield  {journal} {\bibinfo
  {journal} {Phys. Rev. A}\ }\textbf {\bibinfo {volume} {76}},\ \bibinfo
  {pages} {042319} (\bibinfo {year} {2007})}\BibitemShut {NoStop}%
\bibitem [{\citenamefont {Fowler}(2012)}]{fowler2012proof}%
  \BibitemOpen
  \bibfield  {author} {\bibinfo {author} {\bibfnamefont {A.~G.}\ \bibnamefont
  {Fowler}},\ }\bibfield  {title} {\bibinfo {title} {Proof of finite surface
  code threshold for matching},\ }\href
  {https://doi.org/10.1103/PhysRevLett.109.180502} {\bibfield  {journal}
  {\bibinfo  {journal} {Phys. Rev. Lett.}\ }\textbf {\bibinfo {volume} {109}},\
  \bibinfo {pages} {180502} (\bibinfo {year} {2012})}\BibitemShut {NoStop}%
\bibitem [{\citenamefont {Nielsen}\ and\ \citenamefont
  {Chuang}(2010)}]{nielsen_chuang_2010}%
  \BibitemOpen
  \bibfield  {author} {\bibinfo {author} {\bibfnamefont {M.~A.}\ \bibnamefont
  {Nielsen}}\ and\ \bibinfo {author} {\bibfnamefont {I.~L.}\ \bibnamefont
  {Chuang}},\ }\href {https://doi.org/10.1017/CBO9780511976667} {\emph
  {\bibinfo {title} {Quantum Computation and Quantum Information: 10th
  Anniversary Edition}}}\ (\bibinfo  {publisher} {Cambridge University Press},\
  \bibinfo {year} {2010})\BibitemShut {NoStop}%
\bibitem [{\citenamefont {Bonilla~Ataides}\ \emph {et~al.}(2021)\citenamefont
  {Bonilla~Ataides}, \citenamefont {Tuckett}, \citenamefont {Bartlett},
  \citenamefont {Flammia},\ and\ \citenamefont {Brown}}]{Ataides_NComms_2021}%
  \BibitemOpen
  \bibfield  {author} {\bibinfo {author} {\bibfnamefont {J.~P.}\ \bibnamefont
  {Bonilla~Ataides}}, \bibinfo {author} {\bibfnamefont {D.~K.}\ \bibnamefont
  {Tuckett}}, \bibinfo {author} {\bibfnamefont {S.~D.}\ \bibnamefont
  {Bartlett}}, \bibinfo {author} {\bibfnamefont {S.~T.}\ \bibnamefont
  {Flammia}},\ and\ \bibinfo {author} {\bibfnamefont {B.~J.}\ \bibnamefont
  {Brown}},\ }\bibfield  {title} {\bibinfo {title} {The xzzx surface code},\
  }\href {https://doi.org/10.1038/s41467-021-22274-1} {\bibfield  {journal}
  {\bibinfo  {journal} {Nature Communications}\ }\textbf {\bibinfo {volume}
  {12}},\ \bibinfo {pages} {2172} (\bibinfo {year} {2021})}\BibitemShut
  {NoStop}%
\bibitem [{\citenamefont {Chen}\ \emph {et~al.}(2023)\citenamefont {Chen},
  \citenamefont {Liu}, \citenamefont {Otten}, \citenamefont {Seif},
  \citenamefont {Fefferman},\ and\ \citenamefont {Jiang}}]{Chen2023}%
  \BibitemOpen
  \bibfield  {author} {\bibinfo {author} {\bibfnamefont {S.}~\bibnamefont
  {Chen}}, \bibinfo {author} {\bibfnamefont {Y.}~\bibnamefont {Liu}}, \bibinfo
  {author} {\bibfnamefont {M.}~\bibnamefont {Otten}}, \bibinfo {author}
  {\bibfnamefont {A.}~\bibnamefont {Seif}}, \bibinfo {author} {\bibfnamefont
  {B.}~\bibnamefont {Fefferman}},\ and\ \bibinfo {author} {\bibfnamefont
  {L.}~\bibnamefont {Jiang}},\ }\bibfield  {title} {\bibinfo {title} {The
  learnability of pauli noise},\ }\href
  {https://doi.org/10.1038/s41467-022-35759-4} {\bibfield  {journal} {\bibinfo
  {journal} {Nature Communications}\ }\textbf {\bibinfo {volume} {14}},\
  \bibinfo {pages} {52} (\bibinfo {year} {2023})}\BibitemShut {NoStop}%
\bibitem [{\citenamefont {Gottesman}(1997)}]{gottesman1997stabilizer}%
  \BibitemOpen
  \bibfield  {author} {\bibinfo {author} {\bibfnamefont {D.}~\bibnamefont
  {Gottesman}},\ }\href@noop {} {\bibinfo {title} {Stabilizer codes and quantum
  error correction}} (\bibinfo {year} {1997}),\ \Eprint
  {https://arxiv.org/abs/quant-ph/9705052} {arXiv:quant-ph/9705052 [quant-ph]}
  \BibitemShut {NoStop}%
\bibitem [{\citenamefont {Proctor}\ \emph {et~al.}(2020)\citenamefont
  {Proctor}, \citenamefont {Revelle}, \citenamefont {Nielsen}, \citenamefont
  {Rudinger}, \citenamefont {Lobser}, \citenamefont {Maunz}, \citenamefont
  {Blume-Kohout},\ and\ \citenamefont {Young}}]{proctor2020detecting}%
  \BibitemOpen
  \bibfield  {author} {\bibinfo {author} {\bibfnamefont {T.}~\bibnamefont
  {Proctor}}, \bibinfo {author} {\bibfnamefont {M.}~\bibnamefont {Revelle}},
  \bibinfo {author} {\bibfnamefont {E.}~\bibnamefont {Nielsen}}, \bibinfo
  {author} {\bibfnamefont {K.}~\bibnamefont {Rudinger}}, \bibinfo {author}
  {\bibfnamefont {D.}~\bibnamefont {Lobser}}, \bibinfo {author} {\bibfnamefont
  {P.}~\bibnamefont {Maunz}}, \bibinfo {author} {\bibfnamefont
  {R.}~\bibnamefont {Blume-Kohout}},\ and\ \bibinfo {author} {\bibfnamefont
  {K.}~\bibnamefont {Young}},\ }\bibfield  {title} {\bibinfo {title} {Detecting
  and tracking drift in quantum information processors},\ }\href
  {https://doi.org/10.1038/s41467-020-19074-4} {\bibfield  {journal} {\bibinfo
  {journal} {Nature Communications}\ }\textbf {\bibinfo {volume} {11}},\
  \bibinfo {pages} {5396} (\bibinfo {year} {2020})}\BibitemShut {NoStop}%
\bibitem [{\citenamefont {Aliferis}\ and\ \citenamefont
  {Cross}(2007)}]{aliferis2007subsystem}%
  \BibitemOpen
  \bibfield  {author} {\bibinfo {author} {\bibfnamefont {P.}~\bibnamefont
  {Aliferis}}\ and\ \bibinfo {author} {\bibfnamefont {A.~W.}\ \bibnamefont
  {Cross}},\ }\bibfield  {title} {\bibinfo {title} {Subsystem fault tolerance
  with the bacon-shor code},\ }\href
  {https://doi.org/10.1103/PhysRevLett.98.220502} {\bibfield  {journal}
  {\bibinfo  {journal} {Phys. Rev. Lett.}\ }\textbf {\bibinfo {volume} {98}},\
  \bibinfo {pages} {220502} (\bibinfo {year} {2007})}\BibitemShut {NoStop}%
\bibitem [{\citenamefont {Fowler}\ and\ \citenamefont
  {Martinis}(2014)}]{fowler2014quantifying}%
  \BibitemOpen
  \bibfield  {author} {\bibinfo {author} {\bibfnamefont {A.~G.}\ \bibnamefont
  {Fowler}}\ and\ \bibinfo {author} {\bibfnamefont {J.~M.}\ \bibnamefont
  {Martinis}},\ }\bibfield  {title} {\bibinfo {title} {Quantifying the effects
  of local many-qubit errors and nonlocal two-qubit errors on the surface
  code},\ }\href {https://doi.org/10.1103/PhysRevA.89.032316} {\bibfield
  {journal} {\bibinfo  {journal} {Phys. Rev. A}\ }\textbf {\bibinfo {volume}
  {89}},\ \bibinfo {pages} {032316} (\bibinfo {year} {2014})}\BibitemShut
  {NoStop}%
\bibitem [{\citenamefont {Aliferis}\ \emph {et~al.}(2009)\citenamefont
  {Aliferis}, \citenamefont {Brito}, \citenamefont {DiVincenzo}, \citenamefont
  {Preskill}, \citenamefont {Steffen},\ and\ \citenamefont
  {Terhal}}]{Aliferis_2009}%
  \BibitemOpen
  \bibfield  {author} {\bibinfo {author} {\bibfnamefont {P.}~\bibnamefont
  {Aliferis}}, \bibinfo {author} {\bibfnamefont {F.}~\bibnamefont {Brito}},
  \bibinfo {author} {\bibfnamefont {D.~P.}\ \bibnamefont {DiVincenzo}},
  \bibinfo {author} {\bibfnamefont {J.}~\bibnamefont {Preskill}}, \bibinfo
  {author} {\bibfnamefont {M.}~\bibnamefont {Steffen}},\ and\ \bibinfo {author}
  {\bibfnamefont {B.~M.}\ \bibnamefont {Terhal}},\ }\bibfield  {title}
  {\bibinfo {title} {Fault-tolerant computing with biased-noise superconducting
  qubits: a case study},\ }\href
  {https://doi.org/10.1088/1367-2630/11/1/013061} {\bibfield  {journal}
  {\bibinfo  {journal} {New Journal of Physics}\ }\textbf {\bibinfo {volume}
  {11}},\ \bibinfo {pages} {013061} (\bibinfo {year} {2009})}\BibitemShut
  {NoStop}%
\bibitem [{\citenamefont {Wallman}\ and\ \citenamefont
  {Emerson}(2016)}]{wallman2016noise}%
  \BibitemOpen
  \bibfield  {author} {\bibinfo {author} {\bibfnamefont {J.~J.}\ \bibnamefont
  {Wallman}}\ and\ \bibinfo {author} {\bibfnamefont {J.}~\bibnamefont
  {Emerson}},\ }\bibfield  {title} {\bibinfo {title} {Noise tailoring for
  scalable quantum computation via randomized compiling},\ }\href
  {https://doi.org/10.1103/PhysRevA.94.052325} {\bibfield  {journal} {\bibinfo
  {journal} {Phys. Rev. A}\ }\textbf {\bibinfo {volume} {94}},\ \bibinfo
  {pages} {052325} (\bibinfo {year} {2016})}\BibitemShut {NoStop}%
\bibitem [{\citenamefont {Fowler}\ \emph
  {et~al.}(2012{\natexlab{b}})\citenamefont {Fowler}, \citenamefont
  {Whiteside}, \citenamefont {McInnes},\ and\ \citenamefont
  {Rabbani}}]{fowlerautotune2012}%
  \BibitemOpen
  \bibfield  {author} {\bibinfo {author} {\bibfnamefont {A.~G.}\ \bibnamefont
  {Fowler}}, \bibinfo {author} {\bibfnamefont {A.~C.}\ \bibnamefont
  {Whiteside}}, \bibinfo {author} {\bibfnamefont {A.~L.}\ \bibnamefont
  {McInnes}},\ and\ \bibinfo {author} {\bibfnamefont {A.}~\bibnamefont
  {Rabbani}},\ }\bibfield  {title} {\bibinfo {title} {Topological code
  autotune},\ }\href {https://doi.org/10.1103/PhysRevX.2.041003} {\bibfield
  {journal} {\bibinfo  {journal} {Phys. Rev. X}\ }\textbf {\bibinfo {volume}
  {2}},\ \bibinfo {pages} {041003} (\bibinfo {year}
  {2012}{\natexlab{b}})}\BibitemShut {NoStop}%
\bibitem [{\citenamefont {Bravyi}\ \emph {et~al.}(2018)\citenamefont {Bravyi},
  \citenamefont {Englbrecht}, \citenamefont {K{\"o}nig},\ and\ \citenamefont
  {Peard}}]{bravyi2018correcting}%
  \BibitemOpen
  \bibfield  {author} {\bibinfo {author} {\bibfnamefont {S.}~\bibnamefont
  {Bravyi}}, \bibinfo {author} {\bibfnamefont {M.}~\bibnamefont {Englbrecht}},
  \bibinfo {author} {\bibfnamefont {R.}~\bibnamefont {K{\"o}nig}},\ and\
  \bibinfo {author} {\bibfnamefont {N.}~\bibnamefont {Peard}},\ }\bibfield
  {title} {\bibinfo {title} {Correcting coherent errors with surface codes},\
  }\href {https://doi.org/10.1038/s41534-018-0106-y} {\bibfield  {journal}
  {\bibinfo  {journal} {npj Quantum Information}\ }\textbf {\bibinfo {volume}
  {4}},\ \bibinfo {pages} {55} (\bibinfo {year} {2018})}\BibitemShut {NoStop}%
\bibitem [{\citenamefont {Tomita}\ and\ \citenamefont
  {Svore}(2014)}]{Tomita2014Realistic}%
  \BibitemOpen
  \bibfield  {author} {\bibinfo {author} {\bibfnamefont {Y.}~\bibnamefont
  {Tomita}}\ and\ \bibinfo {author} {\bibfnamefont {K.~M.}\ \bibnamefont
  {Svore}},\ }\bibfield  {title} {\bibinfo {title} {Low-distance surface codes
  under realistic quantum noise},\ }\href
  {https://doi.org/10.1103/PhysRevA.90.062320} {\bibfield  {journal} {\bibinfo
  {journal} {Phys. Rev. A}\ }\textbf {\bibinfo {volume} {90}},\ \bibinfo
  {pages} {062320} (\bibinfo {year} {2014})}\BibitemShut {NoStop}%
\bibitem [{\citenamefont {Nickerson}\ and\ \citenamefont
  {Brown}(2019)}]{Nickerson2019analysingcorrelated}%
  \BibitemOpen
  \bibfield  {author} {\bibinfo {author} {\bibfnamefont {N.~H.}\ \bibnamefont
  {Nickerson}}\ and\ \bibinfo {author} {\bibfnamefont {B.~J.}\ \bibnamefont
  {Brown}},\ }\bibfield  {title} {\bibinfo {title} {Analysing correlated noise
  on the surface code using adaptive decoding algorithms},\ }\href
  {https://doi.org/10.22331/q-2019-04-08-131} {\bibfield  {journal} {\bibinfo
  {journal} {{Quantum}}\ }\textbf {\bibinfo {volume} {3}},\ \bibinfo {pages}
  {131} (\bibinfo {year} {2019})}\BibitemShut {NoStop}%
\bibitem [{\citenamefont {Tripathi}\ \emph {et~al.}(2022)\citenamefont
  {Tripathi}, \citenamefont {Chen}, \citenamefont {Khezri}, \citenamefont
  {Yip}, \citenamefont {Levenson-Falk},\ and\ \citenamefont
  {Lidar}}]{PhysRevApplied.18.024068}%
  \BibitemOpen
  \bibfield  {author} {\bibinfo {author} {\bibfnamefont {V.}~\bibnamefont
  {Tripathi}}, \bibinfo {author} {\bibfnamefont {H.}~\bibnamefont {Chen}},
  \bibinfo {author} {\bibfnamefont {M.}~\bibnamefont {Khezri}}, \bibinfo
  {author} {\bibfnamefont {K.-W.}\ \bibnamefont {Yip}}, \bibinfo {author}
  {\bibfnamefont {E.}~\bibnamefont {Levenson-Falk}},\ and\ \bibinfo {author}
  {\bibfnamefont {D.~A.}\ \bibnamefont {Lidar}},\ }\bibfield  {title} {\bibinfo
  {title} {Suppression of crosstalk in superconducting qubits using dynamical
  decoupling},\ }\href {https://doi.org/10.1103/PhysRevApplied.18.024068}
  {\bibfield  {journal} {\bibinfo  {journal} {Phys. Rev. Appl.}\ }\textbf
  {\bibinfo {volume} {18}},\ \bibinfo {pages} {024068} (\bibinfo {year}
  {2022})}\BibitemShut {NoStop}%
\bibitem [{\citenamefont {Ezzell}\ \emph {et~al.}(2023)\citenamefont {Ezzell},
  \citenamefont {Pokharel}, \citenamefont {Tewala}, \citenamefont {Quiroz},\
  and\ \citenamefont {Lidar}}]{ezzell2023dynamical}%
  \BibitemOpen
  \bibfield  {author} {\bibinfo {author} {\bibfnamefont {N.}~\bibnamefont
  {Ezzell}}, \bibinfo {author} {\bibfnamefont {B.}~\bibnamefont {Pokharel}},
  \bibinfo {author} {\bibfnamefont {L.}~\bibnamefont {Tewala}}, \bibinfo
  {author} {\bibfnamefont {G.}~\bibnamefont {Quiroz}},\ and\ \bibinfo {author}
  {\bibfnamefont {D.~A.}\ \bibnamefont {Lidar}},\ }\href@noop {} {\bibinfo
  {title} {Dynamical decoupling for superconducting qubits: a performance
  survey}} (\bibinfo {year} {2023}),\ \Eprint
  {https://arxiv.org/abs/2207.03670} {arXiv:2207.03670 [quant-ph]} \BibitemShut
  {NoStop}%
\bibitem [{\citenamefont {Chamberland}\ and\ \citenamefont
  {Campbell}(2022)}]{PRXQuantum.3.010331}%
  \BibitemOpen
  \bibfield  {author} {\bibinfo {author} {\bibfnamefont {C.}~\bibnamefont
  {Chamberland}}\ and\ \bibinfo {author} {\bibfnamefont {E.~T.}\ \bibnamefont
  {Campbell}},\ }\bibfield  {title} {\bibinfo {title} {Universal quantum
  computing with twist-free and temporally encoded lattice surgery},\ }\href
  {https://doi.org/10.1103/PRXQuantum.3.010331} {\bibfield  {journal} {\bibinfo
   {journal} {PRX Quantum}\ }\textbf {\bibinfo {volume} {3}},\ \bibinfo {pages}
  {010331} (\bibinfo {year} {2022})}\BibitemShut {NoStop}%
\bibitem [{\citenamefont {Gottesman}(1998)}]{gottesman1998heisenberg}%
  \BibitemOpen
  \bibfield  {author} {\bibinfo {author} {\bibfnamefont {D.}~\bibnamefont
  {Gottesman}},\ }\href@noop {} {\bibinfo {title} {The heisenberg
  representation of quantum computers}} (\bibinfo {year} {1998}),\ \Eprint
  {https://arxiv.org/abs/quant-ph/9807006} {arXiv:quant-ph/9807006 [quant-ph]}
  \BibitemShut {NoStop}%
\end{thebibliography}
%

\newpage
\clearpage
% \onecolumn
\appendix

\section{Background}\label{Quantum Computing Background}
Here, we will provide a brief introduction to the fundamentals of circuit-based quantum computing, sufficient to understand the majority of methods/results in this paper. For brevity, some important details will be omitted, but can be found in most larger quantum computing texts \cite{nielsen_chuang_2010}.

\subsection{Quantum Errors}\label{quantum_errors}
Quantum errors describe the imperfections in quantum states and quantum gates. Noisy quantum states are often represented with a density operator $\rho$. The density operator of a noise-free (pure) single qubit state $\ket{\psi}=\alpha\ket{0}+\beta \ket{1}$ is given by,
\begin{equation}
    \rho=\ket{\psi}\bra{\psi}=\begin{bmatrix} \alpha^*\alpha & \alpha \beta^*\\ \alpha^* \beta & \beta^*\beta \end{bmatrix}.
\end{equation}
Density operators corresponding to noisy quantum states can be constructed by taking linear combinations of the density operators of pure states. For example, if a state preparation procedure prepares a $\ket{0}$ state with probability 0.9 and a $\ket{1}$ state with probability 0.1, then the associated density operator would be,
\begin{equation}
    \rho = 0.9 \ket{0}\bra{0}+0.1 \ket{1}\bra{1}=\begin{bmatrix} 0.9 & 0\\ 0 & 0.1 \end{bmatrix},
\end{equation}
which does not take the general form of a pure state density operator. The vanishing of off-diagonal elements is a sign of coherence loss and that the density operator corresponds to a classical mixture of computational basis states.

The density operator representation can be used to define different noise channels. Some examples include the bit-flip channel,
\begin{equation}\label{bit-flip}
    \rho\rightarrow (1-p)\rho + p X \rho X,
\end{equation}
the phase-flip channel,
\begin{equation}\label{phase-flip}
    \rho\rightarrow (1-p)\rho + p Z \rho Z,
\end{equation}
and the depolarizing channel,
\begin{equation}\label{depolarizing}
    \rho\rightarrow (1-p)\rho + \frac{p}{3} X \rho X+ \frac{p}{3} Y \rho Y+ \frac{p}{3} Z \rho Z.
\end{equation}
Here, $p$ is the error rate of the channel and corresponds to the probability that a nontrivial operation is applied to the quantum state. Other channels exist which cannot be, in general, represented by the probabilistic application of Pauli errors. One example is the amplitude damping channel, which can be used to model energetic relaxation. The amplitude damping channel has an action on a single qubit density matrix given by,
\begin{equation}\label{amp-damp}
    \rho\rightarrow K_0\rho K_0^\dag + K_1\rho K_1^\dag,
\end{equation}
with,
\begin{equation}\quad K_0 = \begin{bmatrix}
1 & 0 \\
0 & \sqrt{1-\gamma} 
\end{bmatrix}, \quad K_1=\begin{bmatrix}
0 & \sqrt{\gamma} \\
0 & 0 
\end{bmatrix},
\end{equation}
where $\gamma$ is the decay probability.

When applied to a system of multiple qubits multi-qubit versions of these channels can be applied by applying the single qubit channels one qubit at a time. However, if the channel is used to describe the failure of a $n$-qubit gate then applying a $n$-qubit depolarizing channel directly may be preferred. The $n$-qubit depolarizing channel is defined as,
\begin{equation}
    \rho \rightarrow (1-p)\rho + \frac{p}{4^n-1}\sum_{P\in \{ I,X,Y,Z\}^{\otimes n}\setminus \{I^{\otimes n}\}}{P\rho P}.
\end{equation}
This operation can be applied after an $n$-qubit gate to represent it failing with probability $p$.

The depolarizing channel is sometimes described differently as,
\begin{equation}\label{depolarizing-lambda}
    \rho\rightarrow \left(1-p'\right) \rho+ \frac{p'}{4} I \rho I+ \frac{p'}{4} X \rho X+ \frac{p'}{4} Y \rho Y+ \frac{p'}{4} Z \rho Z,
\end{equation}
for the single qubit case and,
\begin{equation}
    \rho \rightarrow \left(1-p'\right)\rho + \frac{p'}{4^n}\sum_{P\in \{ I,X,Y,Z\}^{\otimes n}}{P\rho P}.
\end{equation}
for the $n$-qubit case.
Here, $p'$ is the depolarizing noise parameter and represents the probability that the state is depolarized. Complete depolarization ($p'=1$) just before measurement using this definition causes the measurement to have no correlation with the otherwise intended value. In contrast, if error events cause the application of $X$, $Y$ or $Z$ errors with equal probability $1/3$ then these would instead cause a measurement result to be anti-correlated with the otherwise intended value. Similar definitions can also be applied for the bit-flip channel and the phase-flip channel. The first type of definition can be argued to allow error events to be counted easier, while the second type of definition allows slightly simpler expressions to be written that describe the expectation values of noisy circuits. If error rate $p$ and depolarizing parameter $p'$ describe the same depolarizing channel then, for the single qubit case, we have the equation,
\begin{equation}\label{depol1}
    \frac{p}{3}=\frac{p'}{4}.
\end{equation}
Note however, the two-qubit case requires a different equality,
\begin{equation}\label{depol2}
    \frac{p}{15}=\frac{p'}{16}.
\end{equation}
Since Equations \ref{depol1} and \ref{depol2} only have simultaneous solutions for $p=p'=0$, in general the two families of circuit noise models they define are distinct. Again, the key benefit of error rate based models is that they make it easier to track the expected number of gate errors to occur, while the second definition more readily tracks how expectation values will decay towards zero.

More general noise models can also be defined, such as those with unequal rates of $X$, $Y$ and $Z$ errors,
\begin{equation}\label{pauli_noise1}
    \rho\rightarrow (1-p)\rho + p(r_X X \rho X+ r_Y Y \rho Y+ r_Z Z \rho Z),
\end{equation}
where to preserve the trace of $\rho$ we require,
\begin{equation}
    r_X+r_Y+r_Z=1.
\end{equation}
For multiple qubits, generalizing Equation \ref{pauli_noise1} leads to defining a system with arbitrary stochastic Pauli noise.

One special case is $Z$ biased noise \cite{Ataides_NComms_2021}, which when parameterized by $p$ and $\eta$, is given by,
\begin{multline}\label{biased_1}
    \rho\rightarrow (1-p)\rho + \frac{p}{2(\eta+1) }X \rho X + \frac{p}{2(\eta+1) }Y \rho Y \\+ \frac{\eta p}{\eta+1 }Z \rho Z .
\end{multline}
Here, $\eta$ is the relative probability of $Z$ errors compared to $X$ or $Y$ errors,
\begin{equation}\label{eta}
    \eta=\frac{r_Z}{r_X+r_Y},
\end{equation}
for which Equation \ref{biased_1} holds if $r_X=r_Y$ is true.
An $n$-qubit definition must also be given. One generalization of Equation \ref{biased_1} and Equation \ref{eta} is to treat the numerator of Equation \ref{eta} to correspond to the sum of probabilities of non-trivial errors containing only $Z$ gates. The denominator would then be the sum of probabilities of all other nontrivial errors. However, when inspecting the case for unbiased noise, the ratio of these sums does not maintain being equal to $1/2$ for $n>1$. We can accommodate this by including an additional factor which divides by the unbiased value of the ratio of the sums, as well as by 2. Thus, for the $n$-qubit case $\eta$ can be generalized as,
\begin{equation}\label{eta_n}
    \eta=2^{n-1}\frac{\sum_{P\in \{ I,Z\}^{\otimes n}\setminus \{ I\}^{\otimes n}} r_P}{\sum_{P'\in \{ I,X,Y,Z\}^{\otimes n}\setminus \{ I,Z\}^{\otimes n}} r_{P'}}.
\end{equation}
If all the terms of the numerator are each set to equal $r_H$ and all the terms of the denominator are each set to equal $r_L$ then the resulting expression would be,
\begin{equation}
    \eta = \frac{1}{2}\frac{r_H}{r_L}.
\end{equation}
Enforcing the trace of $\rho$ to be preserved yields the form of $n$-qubit $Z$ biased noise given by,

\begin{equation}
    r_H=\frac{\eta}{(2^n-1)(\eta+2^{n-1})},\quad r_L=\frac{1}{2}\frac{1}{(2^n-1)(\eta+2^{n-1})}.
\end{equation}
We chose to use the above result for two-qubit biased depolarizing noise in our simulations, but others are also frequently used \cite{PRXQuantum.3.010331}. There may exist more ways of generalizing by making other assumptions or requirements at the beginning of the derivation. These would ideally be chosen to match those empirically observed in the system of interest. One particular approach would be to consider instead generalizations akin to Equation \ref{depolarizing-lambda}, which may remove some $2^n-1$ factors.

The intensity of errors often varies among different qubits of the device and different points in time. A simple extension to uniform depolarizing noise is inhomogeneous depolarizing noise, where each quantum operation can be assigned an independent noise parameter. In the interest of keeping the set of parameters manageable, inhomogeneity can be introduced at the level of qubits. In this case, each qubit is given an independent parameter governing single qubit error channels (reset, identity and measurement), as well as assigning two-qubit error parameters for all two-qubit gates of each distinct pair of qubits. Care must be taken so the error model remains manageable, while capturing the most important details. Further examples of noise characteristics include non-Markovian effects, where the dynamics of noise may be sensitive to the events of the past and cross-talk, where the behaviour of noise on a given qubit is sensitive to the states or operations applied to other qubits. 

In general, an $n$-qubit density operator is written as a matrix with $2^n$ rows and $2^n$ columns. Thus, alternative representations are often required whenever simulating even moderately sized noisy quantum systems. One technique is to use Monte Carlo simulation, where the noise model is treated as a prescription to apply errors stochastically with given probabilities. When all errors are Clifford operations this can also allow efficient stabilizer based simulation.

\subsection{State Preparation and Measurement Errors}
State preparation, or reset, is the process of setting the state of a qubit to a particular state of interest. This is necessary at the start of all quantum algorithms and may also occur repeatedly if ``fresh ancillas'' are required. Theoretically, state preparation errors can be considered by applying a noise channel after perfect preparation of a particular state. For example, in the preparation of a $\ket{0}$ state, if a depolarizing noise channel with error rate $p$ were used, then the noisy prepared state would be,
\begin{equation}
\begin{split}
    \rho &= (1-p) \ket{0}\bra{0} + \frac{p}{3}\left( X\ket{0}\bra{0}X + Y\ket{0}\bra{0}Y \right.\\
    & \hspace{5cm} \left. + Z\ket{0}\bra{0}Z \right) \\
    &= \left(1-\frac{2p}{3}\right)\ket{0}\bra{0} + \frac{2p}{3}\ket{1}\bra{1}.
\end{split}
\end{equation}

This can be used as a prescription to model failure of state preparation resulting in preparation of $\ket{1}$ states with probability $\frac{2p}{3}$, which occurs because $Z$ errors only apply a global phase to their eigenstates. Measurement errors can be treated similarly and cause flips in measurement outcomes with probability $\frac{2p}{3}$. This can be used to retain the uniform depolarizing noise model's convenient dependence on the single parameter, $p$, the error rate. If the noise model is parameterised in terms of a depolarizing parameter, $p'$, then in terms of the depolarizing noise parameter the state-preparation and measurement error rate would be $\frac{p'}{2}$, as expected.

One can generalize measurement errors to account for the possibility that measurement error rates may depend on the state being measured. For a qubit measured in the $Z$ basis, this can be described by a measurement error matrix,
\begin{equation}
    \begin{bmatrix}
P(0|0) & P(1|0) \\
P(0|1) & P(1|1) 
\end{bmatrix},
\end{equation}
where $P(i|j)$ is the probability of recording a measurement to be associated with computational basis state $\ket{i}$ which should have, ideally, been recorded to correspond to computational basis state $\ket{j}$. This matrix can be parameterized as,
\begin{equation}
    \begin{bmatrix}
1-\frac{p}{2} & \frac{p}{2} \\
\frac{p}{2}+\Delta & 1-\frac{p}{2}-\Delta 
\end{bmatrix}.
\end{equation}
With this parametrization, when $\Delta=0$, $p$ corresponds to the probability that a measurement result is uncorrelated with what it would be under ideal situations. Positive $\Delta$ values can be used to model asymmetric readout error phenomena which have the effect to increase the rate at which eigenvalues of $\ket{1}$ states are incorrectly recorded.

Biased noise simulations also require an appropriate way to account for errors in state preparation and measurement. Using the same logic as earlier, the error rate when preparing a $\ket{0}$ state would depend on $p(r_X+r_Y)$, as the phase applied by $Z$ type errors does not change measurement probabilities. In the case of $Z$-biased noise, this would result in state preparation and measurement error rates of $\frac{p}{\eta+1}$. The biased noise literature has generally chosen to instead use a more conservative state preparation and measurement error rate of,
\begin{equation}
    p(r_X+r_Z)=\frac{p(1+2\eta)}{2(\eta+1)}.
\end{equation}
For a bias parameter of $\eta=\frac{1}{2}$, corresponding to isotropic depolarizing noise, this is equal to $\frac{2p}{3}$, as expected. However, it approaches $\frac{p}{2}$ instead of $0$ as $\eta \rightarrow \infty$. This may initially give unintuitive results when running biased noise simulations on circuits with states that remain computational basis states. However, it has the benefit of modelling a regime where noise is biased but state preparation and measurement errors remain without requiring the introduction of additional noise model parameters.

\subsection{Heavy-Hexagon Error Correction}
\begin{figure*}
     \centering
     \includegraphics{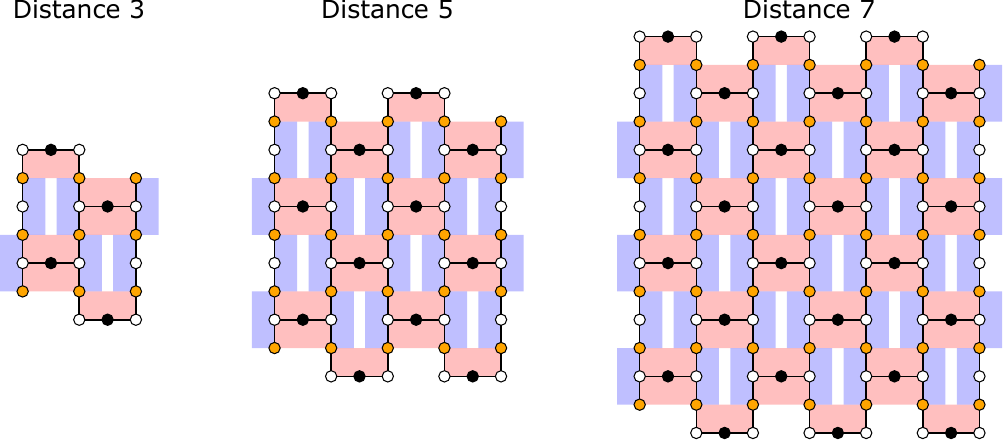}
     \caption{Heavy-hexagon codes of distances 3, 5 and 7. Data, measure and flag qubits are coloured orange, black and white respectively. $X$ and $Z$ gauge operators are represented by red blue tiles respectively. Two-qubit gate connectivity is represented by lines between qubit circles.}
     \label{fig:codes_3_5_7}
\end{figure*}

Here we will give a brief overview of the details of heavy-hexagon code error correction relevant to the experiments performed in the main text. Heavy-hexagon codes are gauge subsystem codes defined on a heavy-hexagon lattice. The adjective ``heavy'' denotes that qubits are placed on both the edges and the vertices of a hexagonal lattice. No qubits require connectivity to more than three neighbouring qubits. This minimizes the occurrence of frequency collisions during the fabrication of fixed-frequency transmon qubit quantum devices \cite{chamberland2020topological}. Examples of heavy-hexagon codes with distances 3, 5 and 7 are shown in Figure \ref{fig:codes_3_5_7}. A distance-$d$ code (for odd values of $d$) contains  $d^2$ data qubits, $\frac{d+1}{2}(d-1)$ measure qubits and $d(d+1)-2$ flag qubits. 
 
\subsubsection{Stabilizer Operators}
Stabilizers are operators which have $+1$ eigenvalues when applied to a state of interest. When described mathematically,
\begin{equation}
    S_i\ket{\psi}=+1\ket{\psi},
\end{equation}
it can be shown that the set of stabilizer operators of a state form an Abelian group. The set of all states which have $+1$ eigenvalues for each element of a stabilizer group of a QEC code form the code space. The stabilizer group of the heavy-hexagon code consists of Bacon-Shor style $X$ stabilizers and surface code style $Z$ stabilizers. Bacon-Shor style $X$ operators correspond to $X$ gates applied to all data qubits of two adjacent columns, while surface code style $Z$ operators correspond to $Z$ gates applied to four data qubits (two at the edges) in a checkerboard pattern. The majority of the stabilizer operators are however not measured directly, and instead are inferred by the result of the gauge operator measurements.

\subsubsection{Gauge Operators}
Gauge operators are the operators measured directly in heavy-hexagon code circuits. The group they form is called the gauge group and unlike the stabilizer group it consists of operators which do not necessarily commute with one another. The centre of the gauge group, the elements of the group which commute with all other elements of the gauge group, is the stabilizer group. States which differ by an element of the gauge group are considered logically equivalent. In other words, their exists additional degrees of freedom, referred to as gauge qubits, which are ignored.

\subsubsection{Logical Operators}
The logical Pauli operators of a heavy-hexagon code can be defined as the set of operators which commute with the stabilizer group, but are not elements of the stabilizer group. As these operators commute with all stabilizer operators they keep the system in the code space. Logical operators (when treated as unitary operations) can also be chosen to commute with the gauge operators, but this is not required, as no logical information is encoded in gauge operators. An example of a logical $Z$ operator is the product of single qubit $Z$ gates on each data qubit in the top row and an example of a logical $X$ operator is a product of single qubit $X$ operators in the leftmost column. Logical operators can be multiplied with stabilizer operators to result in additional definitions of logical operators.

\subsubsection{Error Correcting Properties}
When analysed analytically, it can be shown that heavy-hexagon codes of distance $d$ can correct any set of errors applied on $\mathrm{floor}(\frac{d-1}{2})$ data qubits. Monte Carlo calculations have been performed investigating the performance of heavy-hexagon codes when errors during syndrome measurement are modelled under a circuit level noise model. It was found that $X$ errors (detectable with the surface code style stabilizer operators) have a threshold behaviour at error rates lower than $0.0045$. No threshold behaviour was observed for $Z$ errors (detectable with the Bacon-Shor style stabilizer operators), which instead have a finite optimal code distance for each error rate \cite{chamberland2020topological}.

\section{Simulation}
Here we will describe the methods used to produce the simulated results. All circuit simulations were performed with the Qiskit Python package using the AerSimulator with the simulation method option set to ``statevector'' or ``stabilizer''.

\subsection{Statevector Simulators}
Statevector simulators store the state of n qubits in a vector of complex numbers of length $2^n$. Gates are stored as matrices of size $2^n \times 2^n$ and are applied by matrix multiplication with the current statevector of the system. Measurements operations can be applied with the use of projection operator matrices, or alternatively measurement probabilities can be calculated using the statevector amplitudes at the end of the circuit. This technique requires an amount of memory which scales exponentially with the number of qubits. 

\subsection{Stabilizer Simulators}
Rather than tracking the evolution of the statevector of length $2^n$, stabilizer simulators track the evolution of $2n$ $X$ and $Z$ operators through a quantum circuit. This simulation method supports Clifford circuits, which include many desirable circuits of QEC, and simulates them efficiently \cite{gottesman1998heisenberg}.

\subsection{Noisy Gate Simulations}
The effects of noise on quantum circuit performance can be investigated by simulating the behaviour of noisy gates, according to a given noise model. For stochastic noise, this can be done by applying undesired operations after each desired operation with a given probability. One example is to apply depolarizing noise of a given strength after each noisy operation. This can be a good way to capture the gate error rate dependence for the performance of a quantum circuit, especially when gate error is expected to be the dominant source of error. However, it neglects the additional occurrences of errors due to idle times often present in real quantum devices.

\subsection{Uniform Gate Time Simulations}
A simplifying approximation to the general structure of quantum circuits can be obtained by prescribing that each operation occurs during discrete time steps. Most quantum circuits are often drawn in a way similar to this convention, but may sometimes need to break it to avoid having multi-qubit operations with overlapping visuals. A time step where a qubit has no operations performed can be associated with an idle time step. Errors during idle time steps can be associated with an error operator associated with an identity gate. A benefit of this approach is that a single value can be varied to model the behaviour of a circuit at multiple different error rates, while not neglecting the consequences of non-zero gate times. This method does however neglect the variations in gate duration which may exist among different gates operating on different qubits.

\section{Further Detail For Main Text Methods}
\subsection{Implementation on IBM Devices}
\begin{figure*}
    \centering
    \includegraphics{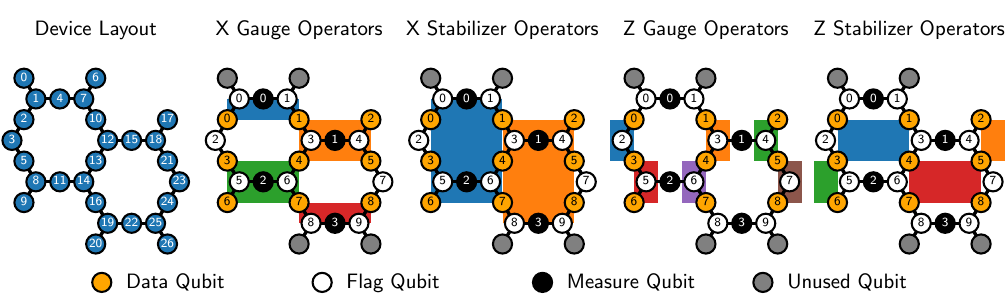}
    \caption{Device layout used for circuits, as well as definitions of $X$ and $Z$ gauge and stabilizer operators. Data, flag and measure qubits are orange, white and black respectively. Unused qubits are grey.}
    \label{fig:falcon_layout}
\end{figure*}

He we will describe how experimental results were retrieved from IBM superconducting qubit devices. Circuits implementing each of the operator measurements, all generalized from Figure \ref{fig:both_stabils}, were implimented using qiskit and transpiled with the ``alap'' (as late as possible) scheduling method. Some results were omitted which corresponded to days where device calibrations produced unusually high operator change rates. Results always used initial layouts corresponding to:
\begin{itemize}
    \item qubits 2, 10, 17, 5, 13, 21, 9, 16, 24 as data qubits,
    \item qubits 1, 7, 3, 12, 18, 8, 14, 23, 19, 25 as flag qubits,
    \item qubits 4, 15, 11, 22 as measure qubits,
\end{itemize}
as shown in Figure \ref{fig:falcon_layout}.

\subsubsection{Gauge Operator Measurements}\label{appendix:measurement_formulas}
The gauge operators of the heavy-hexagon code were evaluated using prepare and measure circuits. Each gauge operator measurement was evaluated in both $X$ operator form and $Z$ operator form. The data qubits were initialized in all possible product states of $\ket{0}$ and $\ket{1}$ for $Z$ operators and $\ket{+}$ and $\ket{-}$ for $X$ operators. With each weight $n$ operator requiring $2^{n+1}$ circuits,  a distance-$d$ code requires $16(d-1)+24(d-1)^2$ circuits to evaluate all its gauge operators with prepare and measure circuits. The number of shots for each circuit was set to $2048$.

The polynomials describing the gauge operator change rates, $R$, as a function of depolarization parameter, $p'$, are,

\begin{equation}
    R = \frac{1}{2}\left(1-(1-p')^{n_{\mathrm{conseq}}}\right),
\end{equation}
with $n_{\mathrm{conseq}}=6, 10, 15$ for $ZZ$ operators, $XX (\mathrm{flag})$ operators and $XXXX (\mathrm{flag})$ operators respectively. Here, $n_{\mathrm{conseq}}$ corresponds to the number of unique gates which can produce errors which can change the relevant operator measurement values as a consequence. Expanded forms of these functions were initially calculated using a symbolic simulation of the density matrix for circuits similar to those of Figure \ref{fig:fig_2}.

% \begin{equation}
%     p_{ZZ}=3p - \frac{15}{2}p^2+10p^3-\frac{15}{2}p^4+3p^5-\frac{1}{2}p^6,
% \end{equation}
% \begin{equation}
%     p_{XX (flag)}=5p-\frac{45}{2}p^2+60p^3-105p^4+126p^5-105p^6+60p^7-\frac{45}{2}p^8+5p^9-\frac{1}{2}p^{10},
% \end{equation}
% \begin{multline}
%     p_{XXXX (flag)}=\frac{15}{2}p-\frac{105}{2}p^2+\frac{455}{2}p^3-\frac{1365}{2}p^4+\frac{3003}{2}p^5-\frac{5005}{2}p^6+\frac{6435}{2}p^7-\frac{6435}{2}p^8+\frac{5005}{2}p^9\\-\frac{3003}{2}p^{10}+\frac{1365}{2}p^{11}-\frac{455}{2}p^{12}+\frac{105}{2}p^{13}-\frac{15}{2}p^{14}+\frac{1}{2}p^{15}.
% \end{multline}

\subsubsection{Stabilizer Operator Measurements}
The stabilizer operators of the heavy-hexagon code were evaluated using prepare and measure circuits. Each surface code stabilizer was evaluated similarly to the method described for gauge operators. With the increase in weight of Bacon-Shor stabilizers with distance further investigation may require operator change rates to be evaluated only for a random selection of inputs for code distances greater than 3. To fully characterize stabilizer operators across all product state inputs of Bacon-Shor stabilizers would require $2^{2d+1}(d-1)$ circuits which quickly becomes impractical for code distances beyond $d=3$. The number of circuits for surface code stabilizer evaluation instead grows much slower as $8(d-1)+16(d-1)^2$ and full characterization remains practical for larger code distances.

\subsubsection{Circuit Execution Details}
Experimental results used in Figure \ref{fig:fig_2} and Supplementary Figure \ref{fig:sup_gauge_ops} were obtained from jobs submitted to ibm\_montreal over a period from 14th September 2022 to 27th October 2022.

Experimental results used in Figure \ref{fig:fig_3} were obtained from jobs run on ibm\_montreal on 2nd February 2023 and 3rd February 2023.

Experimental results used in Figures \ref{fig:fig_4} involving circuits of $Z$ gauge operators, $X$ gauge operators and full heavy-hexagon code syndrome measurement were obtained from jobs run on ibm\_montreal on 2nd November 2022, 8th November 2022 and 25th August 2022 respectively.

Experimental results shown for additional correlation matrices of Supplementary Figure \ref{fig:sup_multi_correl_mat} were retrieved using circuits submitted to ibmq\_toronto on 27th October 2022, ibmq\_mumbai on 29th October 2022 and ibm\_geneva on 4th November 2022.

\subsection{Numerical Calculations}\label{appendix:numerical_calculations}
Here we will describe the equations used to calculate the numerical results present in the main text and Supplementary Information.

\subsubsection{Operator Change Rates}
These were calculated by dividing the number of shots of the incorrect value measured by the total number of shots run. Figure \ref{fig:fig_2} resolves this quantity for each unique input. For multi-cycle experiments, this must be calculated slightly differently for the first and final cycle. For the first cycle, a change is considered to have occurred if the operator value does not match what was expected based on initialization (and may be undefined if the preparation basis and measurement basis are not identical). For subsequent cycles, a change is considered to have occurred if the operator value measured does not match the value measured in the previous cycle. The final ``cycle'' corresponds to data qubit measurement. Here, a stabilizer change is considered to have occurred if the measurement result of products of data qubits is inconsistent with the previous stabilizer measurement (and again may be undefined if the measurement basis makes operator value determination impossible). The total change rate at any particular cycle is equal to the number of changes observed divided by the number of shots in consideration.

\subsubsection{Error Model Fitting}
When analytic forms are known governing the relationship between error model parameters and measurement statistics, such as in Figure \ref{fig:fig_2} and Figure \ref{fig:fig_3}, optimal parameters consistent with experiment can be found with root finding algorithms, such as Newton's method. When the number of parameters of a model is small, such as in uniform depolarizing noise or biased noise simulations of Figure \ref{fig:fig_4}, direct search can be used. When the number of parameters becomes large, such as in inhomogeneous noise modes, gradient-free optimization methods can be applied. An objective function can be defined which governs how good a set of parameters fits the data. This was initially applied for the inhomogeneous noise model of Figure \ref{fig:fig_4}, where the COBYLA optimization algorithm was used. Finding an adequate fit using Monte Carlo simulations was obstructed by the time required in simulation, as well as shot noise present when using 2048 shots. Significantly improved fits were found when analytic expressions were found for operator change rates as a function of error model parameters, described in the next section.

\subsubsection{Operator Change Rate Expressions}
The operator change rates in circuits similar to those we discussed were found to be well modelled by functions of the form,
\begin{equation}\label{change_rate_func}
    R = \frac{1}{2}\left(1-\prod_i(1-p'_i)\right),
\end{equation}
where the product is applied over all gates able to change the measurement value and $p'_i$ is a depolarization parameter or measurement/reset $p$ parameter. The equality appears to be exact based on our numerical testing and is consistent with similar expressions used in randomized benchmarking \cite{PhysRevA.77.012307}. The expression in outer brackets is the probability that no errors occur, and the full operator change rate is half this quantity. This can be understood to follow from the fact that a single depolarization event is sufficient to make the measurement result maximally mixed. A special case can occur when an error may cancel out with itself as it propagates to later parts of the circuit. In that case, it is excluded from the product. To generalize the formula for biased noise, one much replace $p'$ with the equivalent depolarization probability for the measurement of interest, which equals double the probability that an error inverting the measurement outcome occurs.

Operator change rates and system depolarization parameters can be related with a linear relationship using logarithms. Additionally, collecting $p'_k$ values which appear $n_k$ times yields the relationship,
\begin{equation}
    \log (1-2R) = \sum_k n_k\log(1-p'_k).
\end{equation}
For convenience we will change variables using,
\begin{equation}
    F=\log(1-2R), \quad f_k = \log(1-p'_k),
\end{equation}
which allows the relationships between operator change rates and noise model error rates to be written in a linear system,
\begin{equation}
    \Vec{F} = A \Vec{f},
\end{equation}
where $\Vec{F}$ is an $n_R$ element vector containing change rate data, expressed as measurement fidelities, $\Vec{f}$ is an $n_{p'}$ element vector containing error model data, expressed as gate fidelities. $A$ is a $n_R \times n_p$ matrix where each row contains the number of times a particular error has a chance to increase a particular change rate. The systems investigated in this work were found to be under-determined (with respect to inhomogeneous noise models defined by single-qubit and two-qubit errors) if only individual operator change rates are considered. When also including change rates of products of individual operators, systems were found to be fully determined, allowing single-qubit and two-qubit error parameters to be determined by direct inversion of the linear relationship. However, care must be taken that the final set of probabilities are physical, which, for experimental data, was achieved by instead fitting numerically using COBYLA. If average error rates are instead desired, Equation \ref{change_rate_func} can be directly inverted algebraically to find average error rates of gates of interest.

\subsubsection{Correlation Matrices}
Given two stochastic events $x_i$ and $x_j$, a measure of correlation can be calculated using the formula,
\begin{equation}\label{correlation_matrix_elements}
    p_{ij}=\frac{\braket{x_ix_j}-\braket{x_i}\braket{x_j}}{(1-2\braket{x_i})(1-2\braket{x_j})}.
\end{equation}
In the context of syndrome measurements $x_i$ and $x_j$ correspond to two particular syndrome bit detection events returning values of $0$ or $1$. See \cite{chen2021nature} for more details. Applying this formula to a repeated syndrome measurement experiment reveals signs of spatial and temporal correlation in error events. The data can be represented in a matrix. The row and column indices can be defined as $i=nS+c$, where  $n$ is the number of syndrome measurement cycles, $c$ is the cycle of interest and S denotes the stabilizer of interest.

Individual matrix elements are not expected to be of the form given by Equation \ref{change_rate_func}. Instead, quantities in the numerator and denominator of Equation \ref{correlation_matrix_elements} can be calculated one at a time and then combined to calculate the full expression. Quantities of the form $\braket{x_i}$ can be calculated using Equation \ref{change_rate_func}, as they correspond directly to an operator change rate. Quantities of the form $\braket{x_ix_j}$ correspond the rate of changes simultaneously observed in detection event $i$ and $j$. This was found to depend linearly on the change rates corresponding to $\braket{x_i}$, $\braket{x_j}$ and $\braket{x_i \oplus x_j}$, where $\oplus$ is addition modulo 2. The relationship is,
\begin{equation}
    \braket{x_ix_j}=\frac{\braket{x_i}+\braket{x_j}-\braket{x_i \oplus x_j}}{2}.
\end{equation}
Other properties related to correlations, such as conditional probabilities, can also be calculated with similar expressions.

\section{Additional Figures}

\begin{figure*}[h!]
    \centering
    \includegraphics[scale=0.3]{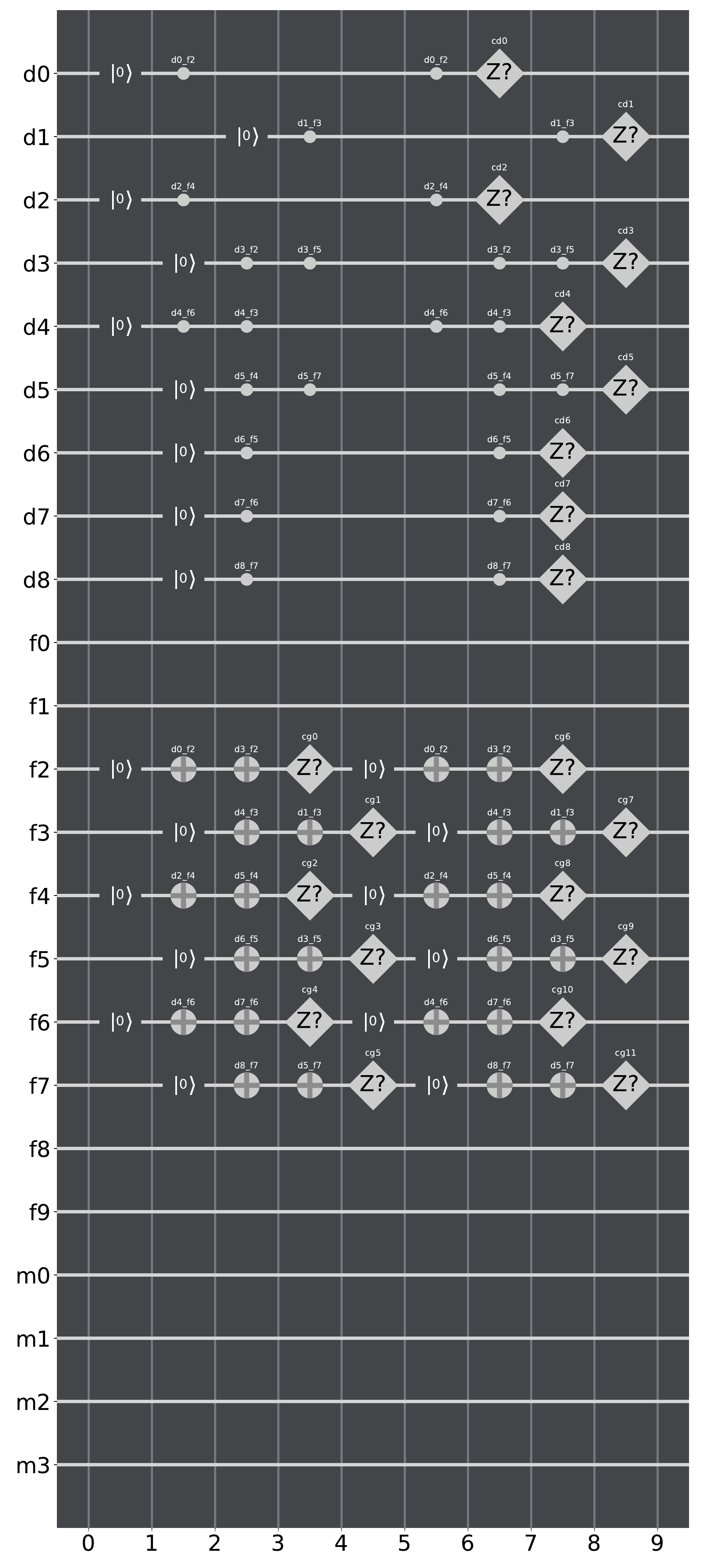}
    \caption{Circuit corresponding to two syndrome measurement cycles of a distance-3 heavy-hexagon code where only $Z$ gauge operators are measured. Lines connecting CNOT controls and targets were omitted to allow multiple simultaneous CNOT gates to be drawn on the same time step. The style of  circuit elements was based on the \href{https://qui.science.unimelb.edu.au/}{QUI simulator}.}
    \label{fig:z_stabils_only}
\end{figure*}

\begin{figure*}
    \centering
    \includegraphics[scale=0.3]{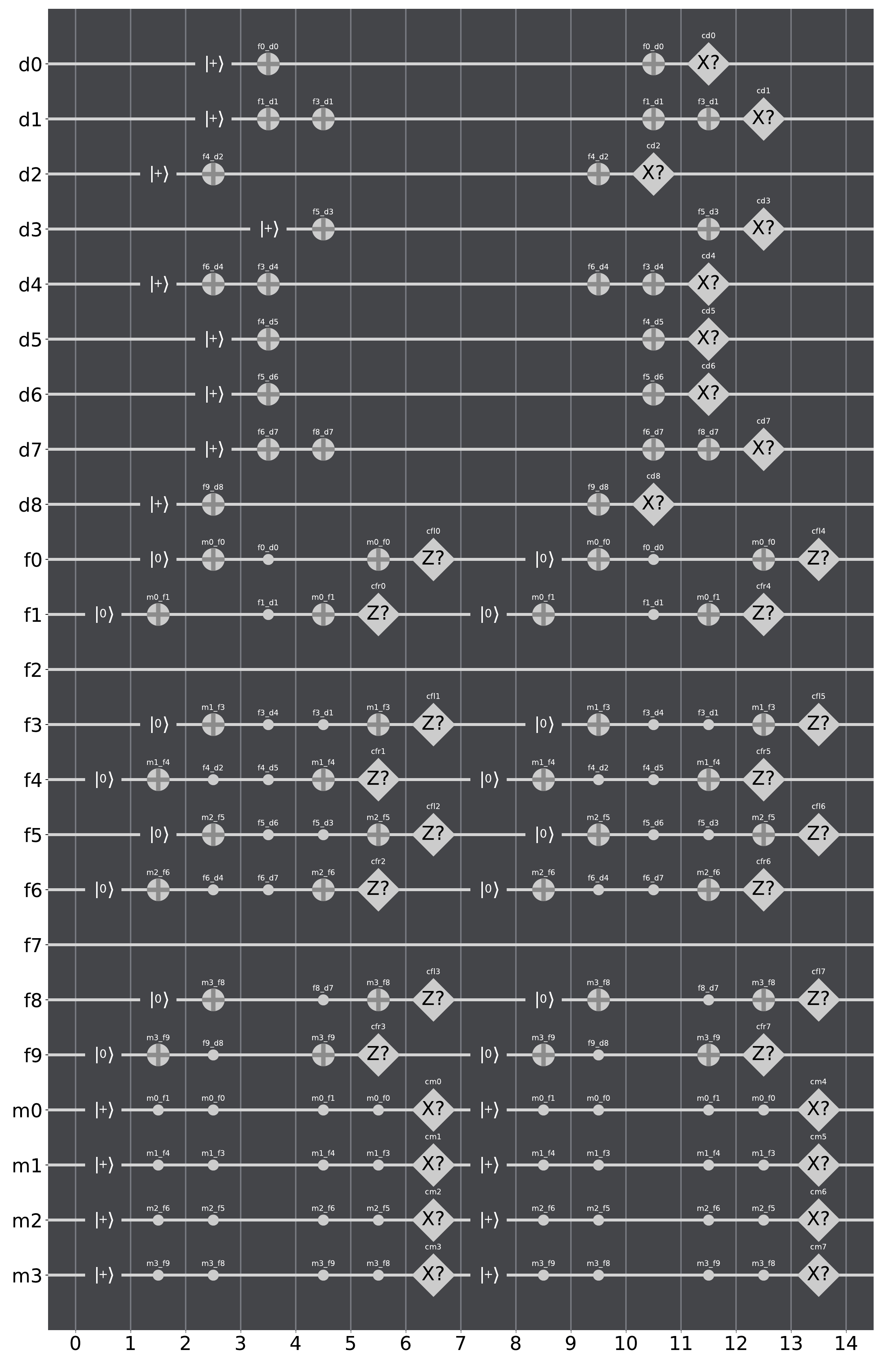}
    \caption{Circuit corresponding to two syndrome measurement cycles of a distance-3 heavy-hexagon code where only $X$ gauge operators are measured. Lines connecting CNOT controls and targets were omitted to allow multiple simultaneous CNOT gates to be drawn on the same time step. The style of  circuit elements was based on the \href{https://qui.science.unimelb.edu.au/}{QUI simulator}.}
    \label{fig:x_stabils_only}
\end{figure*}

\begin{figure*}
    \centering
    \includegraphics[scale=0.3]{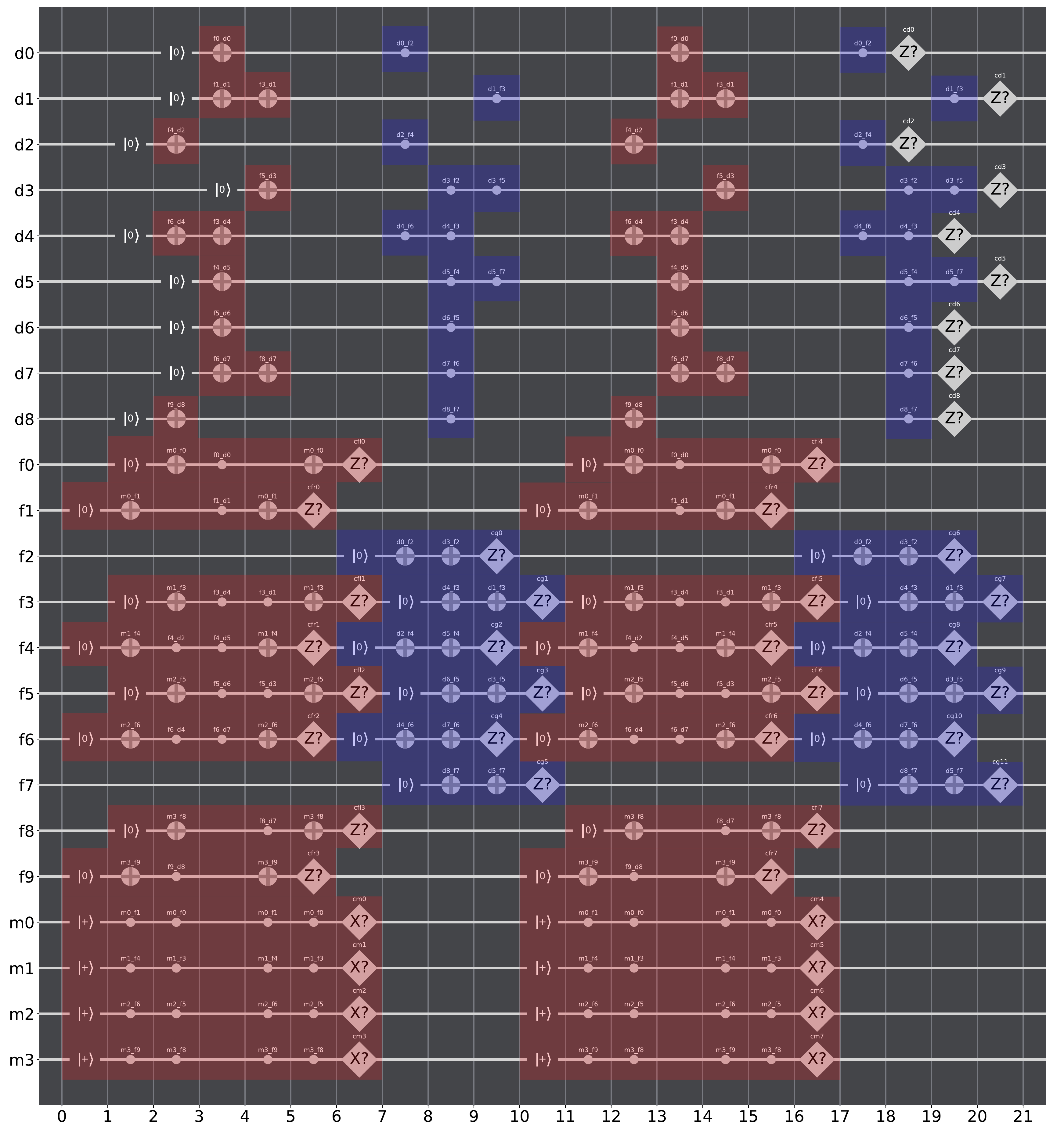}
    \caption{Circuit corresponding to two syndrome measurement cycles of a distance-3 heavy-hexagon code where both $X$ and $Z$ gauge operators are measured. Lines connecting CNOT controls and targets were omitted to allow multiple simultaneous CNOT gates to be drawn on the same time step. The style of  circuit elements was based on the \href{https://qui.science.unimelb.edu.au/}{QUI simulator}.}
    \label{fig:both_stabils}
\end{figure*}

\begin{figure*}[htp]
    \centering
    \includegraphics[width=1.0\textwidth, keepaspectratio]{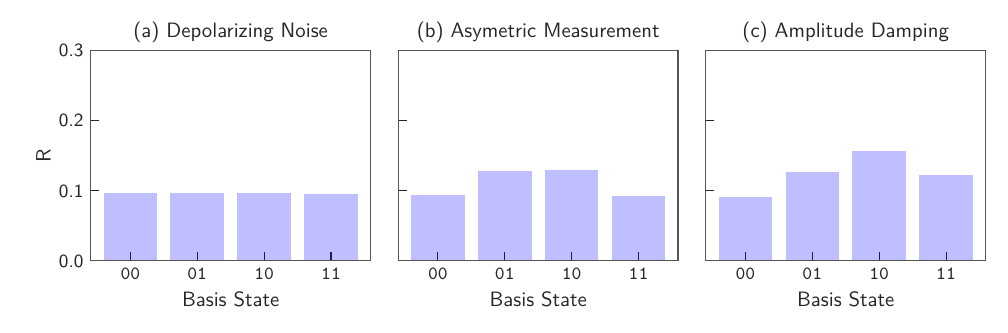}
    \caption{$ZZ$ gauge operator prepare and measure circuits simulated with additional noise models to demonstrate qualitative differences in change rates. (a) Shows the operator change rates when simulated with a uniform depolarizing noise model with depolarizing parameter $p'=0.03$. The noise model in (b) additionally includes asymmetric measurement noise. The noise model in (c) additionally includes amplitude damping noise.}
    \label{fig:sup_amp_damp}
\end{figure*}

\begin{figure*}[htp]
    \centering
    \includegraphics[width=1.0\textwidth, keepaspectratio]{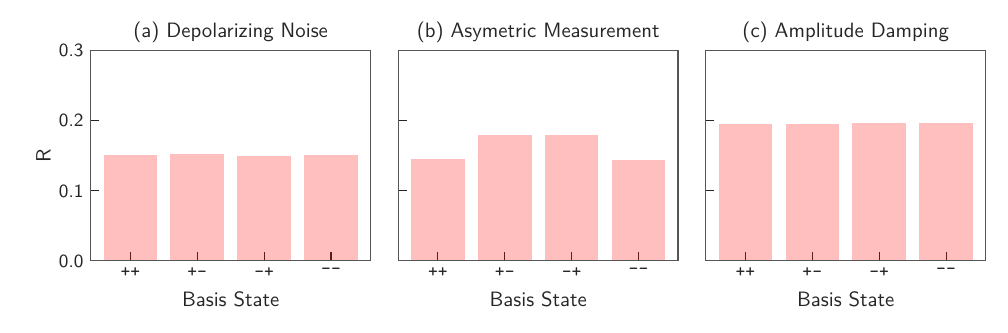}
    \caption{$XX$ flagged gauge operator prepare and measure circuits simulated with additional noise models to demonstrate qualitative differences in change rates. (a) Shows the operator change rates when simulated with a uniform depolarizing noise model with depolarizing parameter $p'=0.03$. The noise model in (b) additionally includes asymmetric measurement noise. The noise model in (c) additionally includes amplitude damping noise.}
    \label{fig:sup_amp_damp_XX}
\end{figure*}

\begin{figure*}[htp]
    \centering
    \includegraphics[width=1.0\textwidth, keepaspectratio]{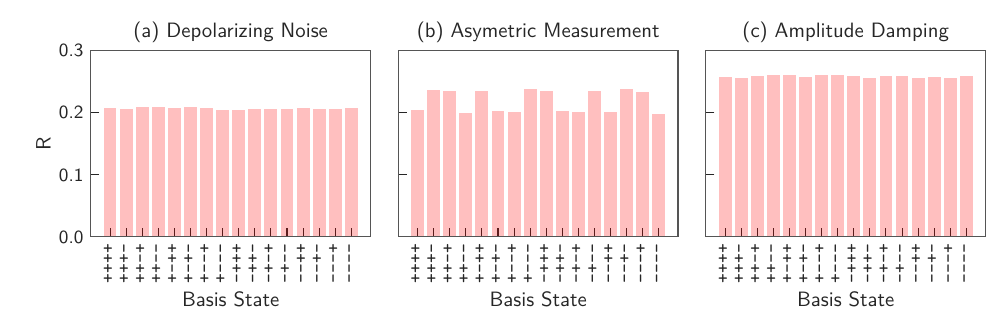}
    \caption{$XXXX$ flagged gauge operator prepare and measure circuits simulated with additional noise models to demonstrate qualitative differences in change rates. (a) Shows the operator change rates when simulated with a uniform depolarizing noise model with depolarizing parameter $p'=0.03$. The noise model in (b) additionally includes asymmetric measurement noise. The noise model in (c) additionally includes amplitude damping noise.}
    \label{fig:sup_amp_damp_XXXX}
\end{figure*}

\begin{figure*}[htp]
    \centering
    \includegraphics{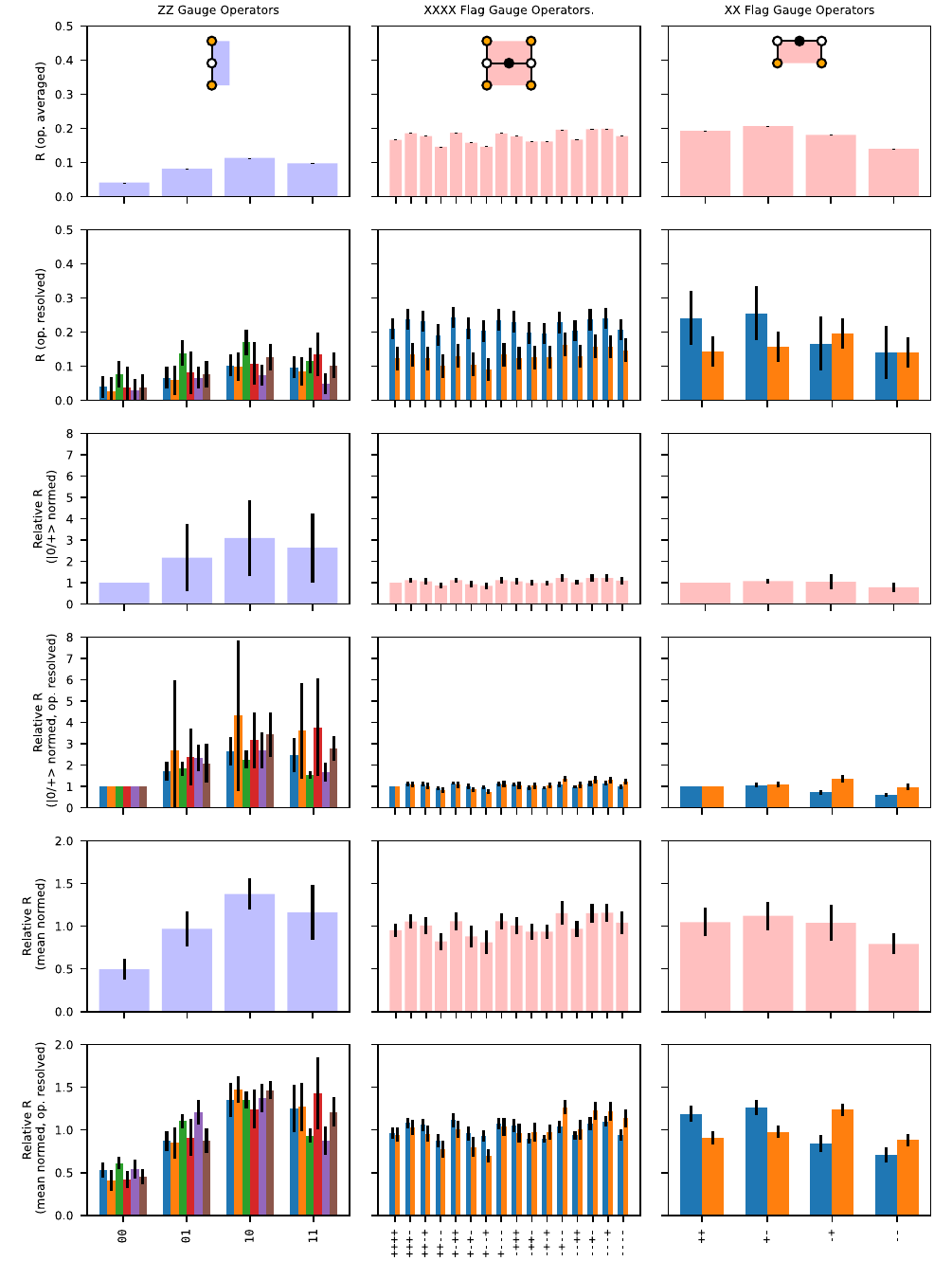}
    \caption{\footnotesize Gauge operator change rates, R, plot differently compared to Figure \ref{fig:fig_2} in the main text. The top rows show the change rates with error bars corresponding to uncertainty in estimating the mean across all calibration cycles. The second row instead resolves change rates for operators at different locations. The last four rows show different normalizations. ``$\ket{0/+}$ normed'' normalizes the change rate to that observed for the $\ket{0}^{\otimes n}$ or $\ket{+}^{\otimes n}$ input states. ``Mean normed'' normalizes such that the relevant dataset of change rates has a zero mean. All error bars are one standard deviation.}
    \label{fig:fig_2_comparison_2}
\end{figure*}

\begin{figure*}[htp]
    \centering
    \includegraphics[width=\textwidth, keepaspectratio]{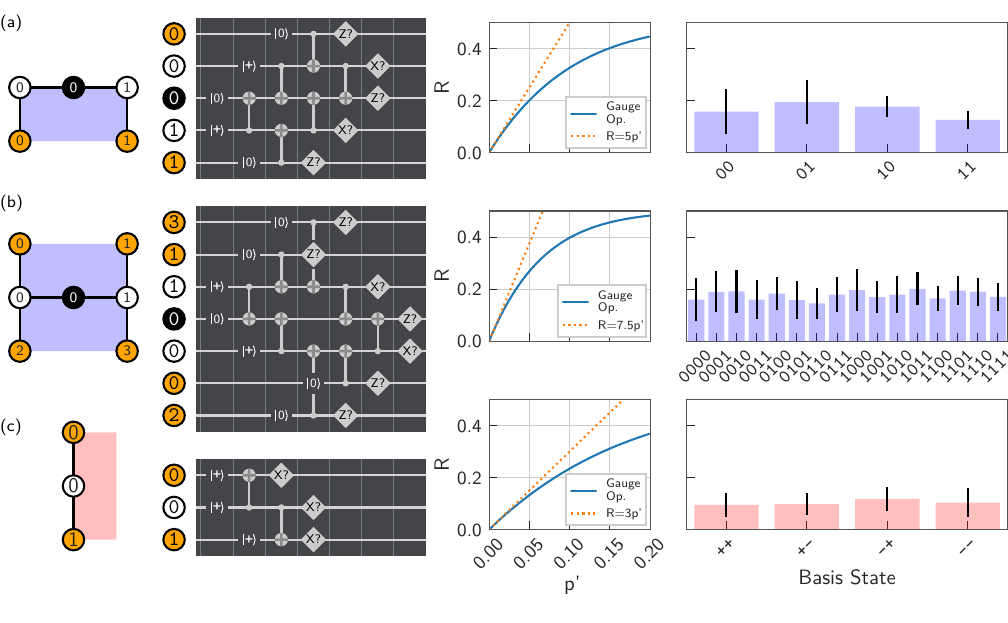}
    \caption{Gauge operator measurement circuit evaluation for the ibmq\char`_montreal device. The three rows (a), (b) and (c) correspond to the benchmarking of $ZZ$ flagged gauge operators, $ZZZZ$ flagged gauge operators and $XX$ gauge operators. The first column shows the diagram tile representation of each operator. The second column shows circuits, shown with the first data qubit input state, which were used for theoretical simulation. The third column shows theoretical operator change rates as a function of depolarizing error parameter for each operator measurement circuit. The fourth column shows the experimentally measured operator change rates for each separable eigenstate of the operator under investigation. Values are calculated by averaging over 34 calibrations and all error bars represent one standard deviation.}
    \label{fig:sup_gauge_ops}
\end{figure*}

\begin{figure*}[htp]
    \centering
    \includegraphics[width=1.0\textwidth, keepaspectratio]{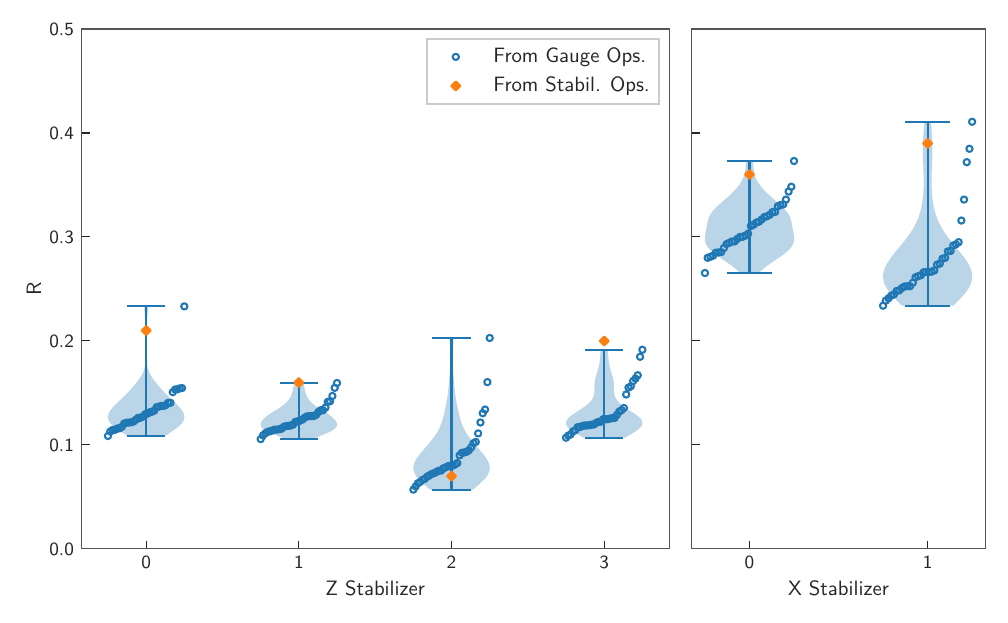}
    \caption{Comparing stabilizer operator change rates expected from gauge operator measurements (blue outlined circles) and measured  from stabilizer measurement circuits comprised of simultaneous gauge operator measurement (orange diamonds).}
    \label{fig:enter-label}
\end{figure*}

\begin{figure*}
    \centering
    \includegraphics{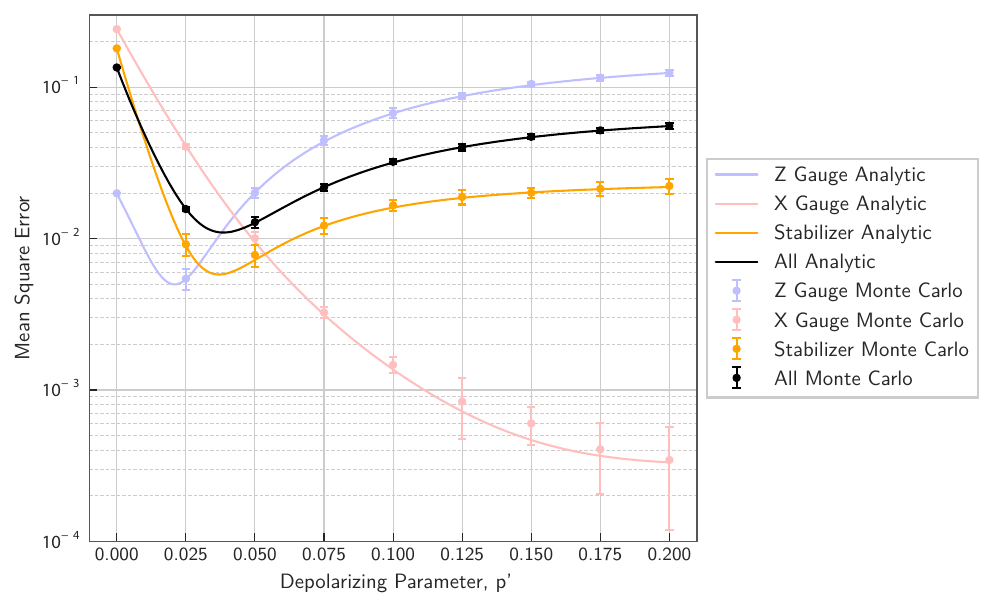}
    \caption{Uniform depolarizing noise model mean square error of fit as a function of depolarizing noise parameter, $p'$. Data is shown when fitting to $Z$ gauge operator change rates, $X$ gauge operator change rates, all stabilizer operator change rates, and all data simultaneously. The minimum mean squared error for the simultaneous fit occurs for $p'=0.039$. Monte Carlo data was sampled across 10 trials, each with 2048 shots. Error bars correspond to 1.96 standard deviations of the data of each set of 10 trials.}
    \label{fig:uniform_depol_fit}
\end{figure*}

\begin{figure*}
    \centering
    \includegraphics{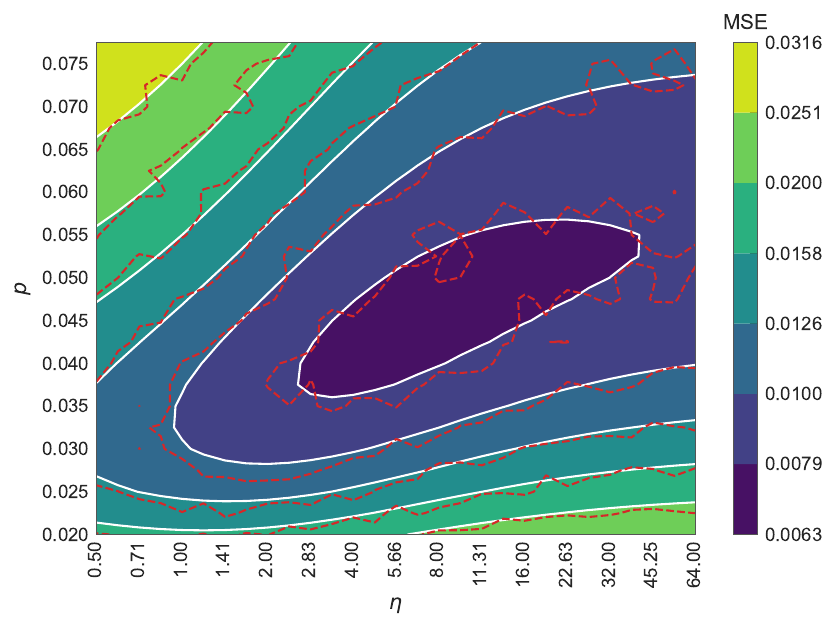}
    \caption{Uniform biased noise model mean square error of fit as a function of error rate, $p$, and bias parameter $\eta$. Data is shown when fitting to all data (individual gauge operators and stabilizer operators) simultaneously. The solid white lines show level curves from calculations with analytic functions. The minimum mean squared error occurs at approximately $(p, \eta)=(0.045, 6.5)$. The red dashed lines show the level curves generated from Monte Carlo data obtained using simulations with 2048 shots. Monte Carlo level curves are expected to oscillate around analytic level curves and have some regions of high curvature due to shot noise due to finite samples.}
    \label{fig:uniform_biased_fit}
\end{figure*}

\begin{figure*}[htp]
    \centering
    \includegraphics[width=\textwidth, keepaspectratio]{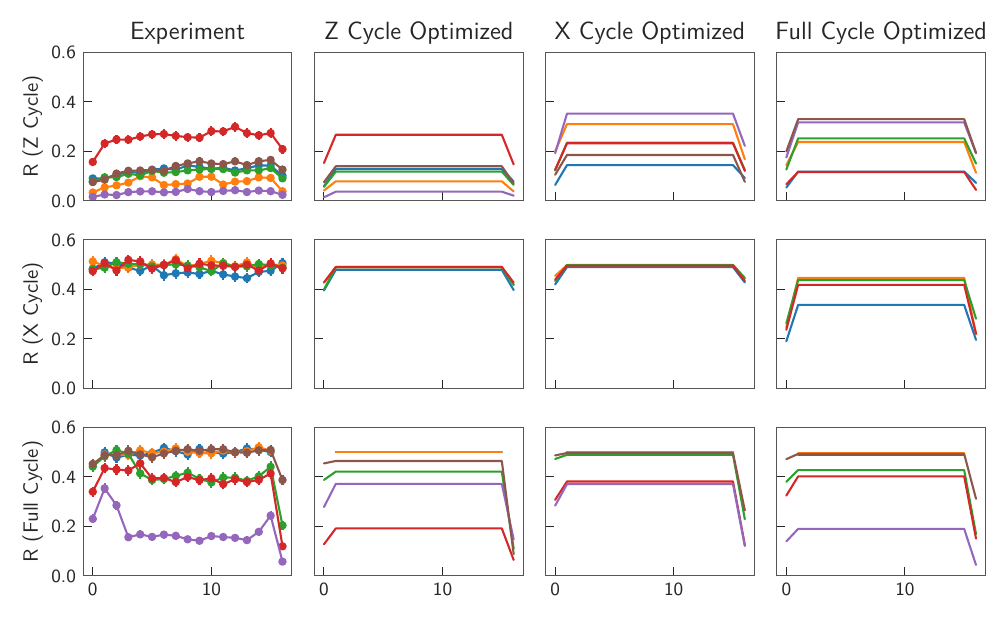}
    \caption{Experimental operator change rates fit to inhomogeneous depolarizing noise. Each column corresponds to fitting a noise model to only one experiment, with the simulated change rates also shown for the other experiments which were not primarily considered in optimization. Qubits which have no influence on results (for $Z$ cycle and $X$ cycle results) were optimized secondarily.}
    \label{fig:inhomogeneous_appendix}
\end{figure*}

\begin{figure*}[htp]
    \centering
    \includegraphics[width=\textwidth, keepaspectratio]{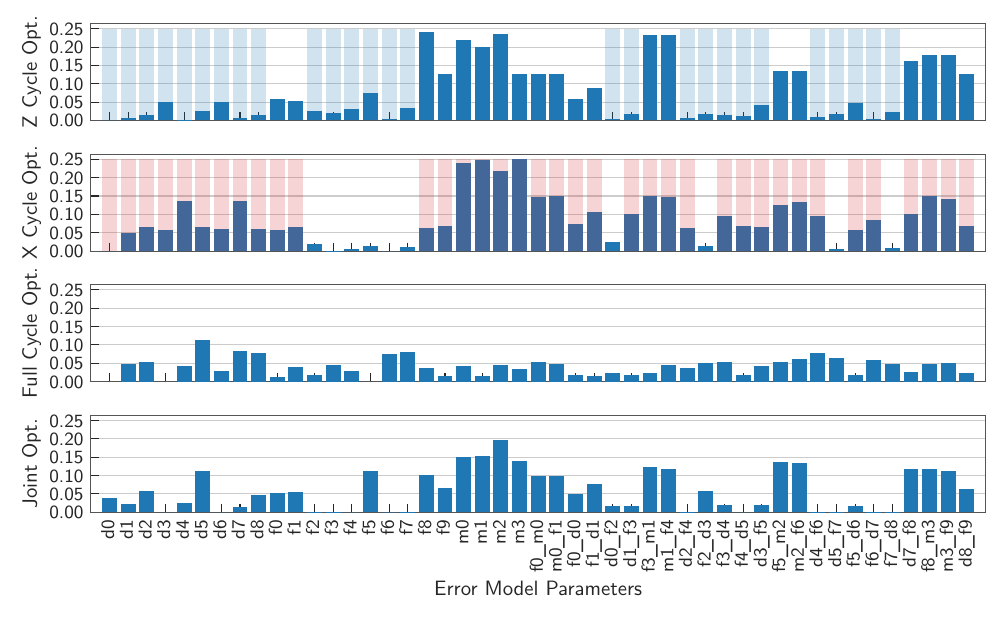}
    \caption{Depolarization parameters used for single-qubit and two-qubit operations to produce inhomogeneous error model fits of Figure \ref{fig:fig_4} and Supplementary Figure \ref{fig:inhomogeneous_appendix}. For $Z$ and $X$ cycle optimization, optimization was performed in two steps. Parameters effecting the primary circuit of interest were found first found (highlighted), and then frozen while joint optimization was performed on the remaining parameters. L2 norm regularization was used for the error parameters with a coefficient of $0.05$.}
    \label{fig:error_rates_appendix}
\end{figure*}

\begin{figure*}[htp]
    \centering
    \includegraphics[width=\textwidth, keepaspectratio]{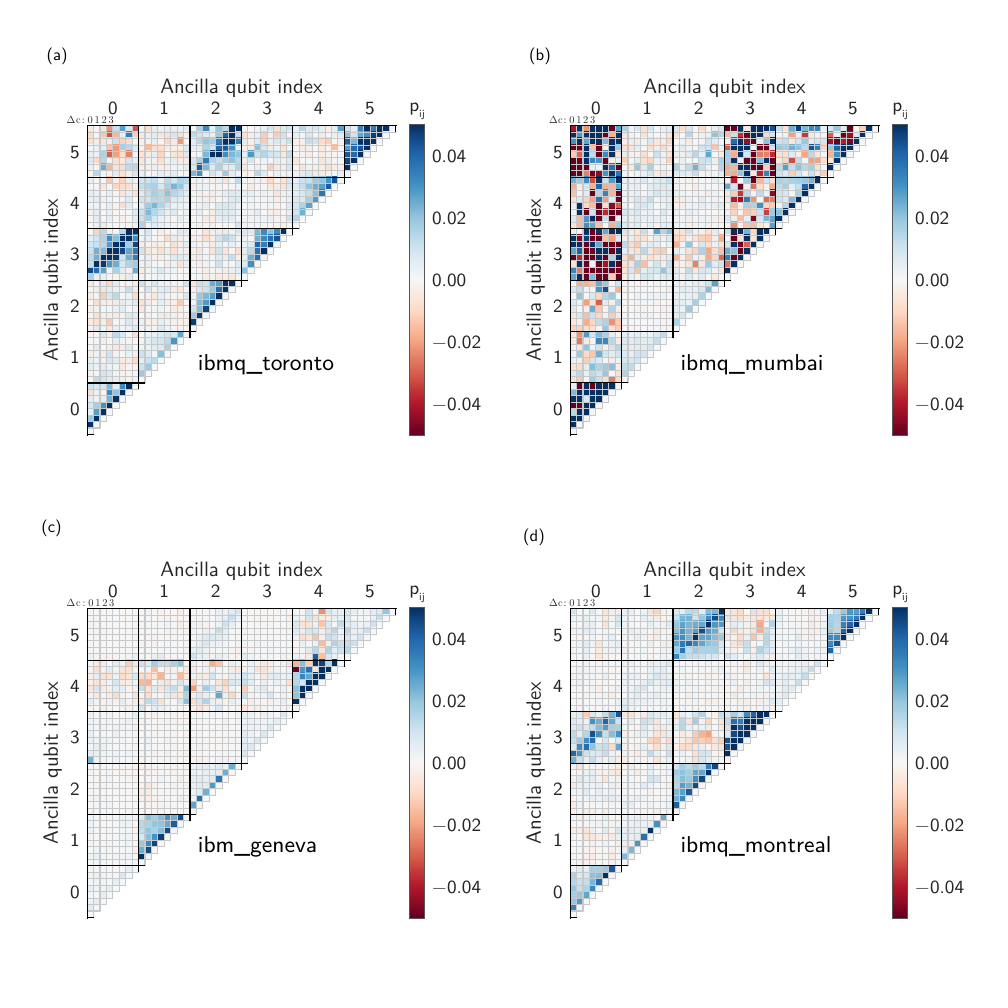}
    \caption{Correlation matrices for eight repeated measurements of $Z$ gauge operators of distance-3 heavy-hexagon codes on four different devices. The initial state of all data qubits was $\ket{0}$. Black and grey lines separate different $Z$ operator ancilla qubits and different measurement cycles respectively. Matrix values with magnitude above 0.05 have been clipped.}
    \label{fig:sup_multi_correl_mat}
\end{figure*}

\begin{figure*}[htp]
    \centering
    \includegraphics[width=\textwidth, keepaspectratio]{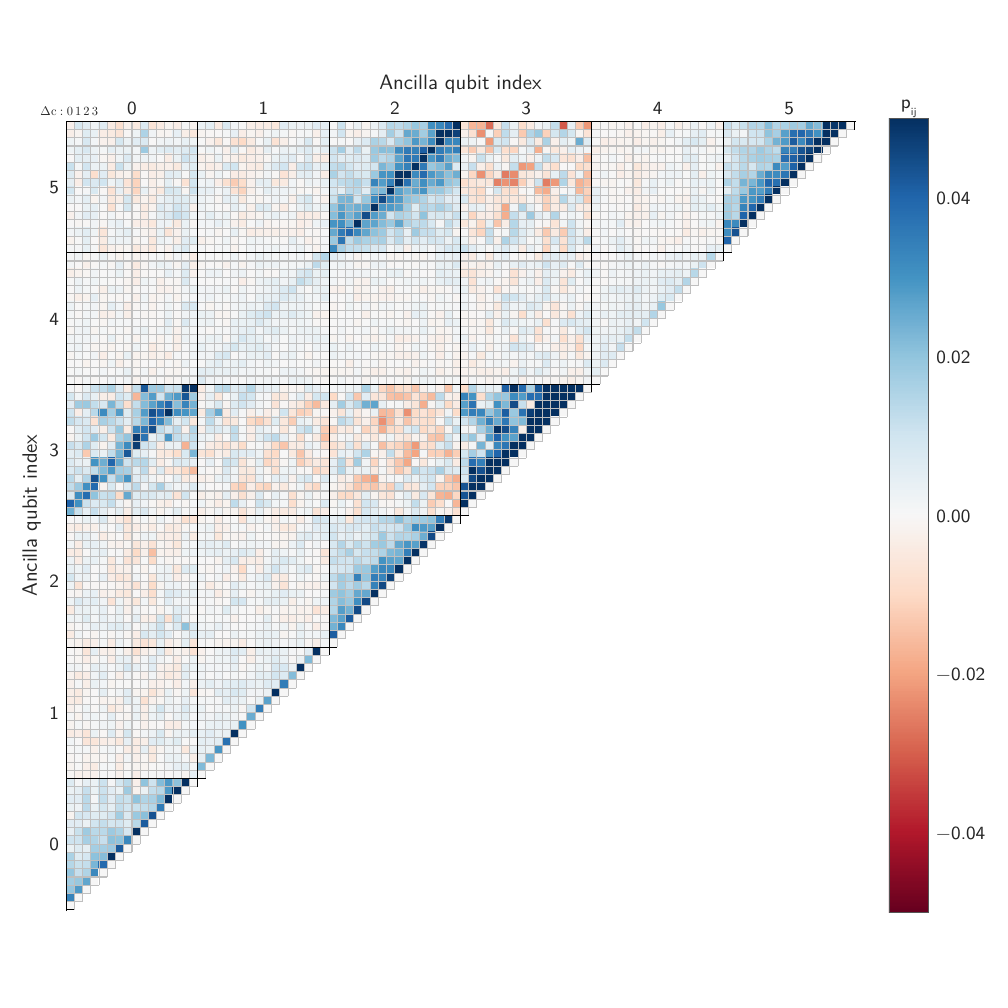}
    \caption{Full correlation matrix for 16 repeated measurements of $Z$ gauge operators of a distance-3 heavy-hexagon code on ibmq\char`_montreal. The initial state of all data qubits was $\ket{0}$. Black and grey lines separate different $Z$ operator ancilla qubits and different measurement cycles ($\Delta c$) respectively. Matrix values with magnitude above 0.05 have been clipped.}
    \label{fig:sup_correl_mat_16_montreal}
\end{figure*}

\begin{figure*}[htp]
    \centering
    \includegraphics[width=\textwidth, keepaspectratio]{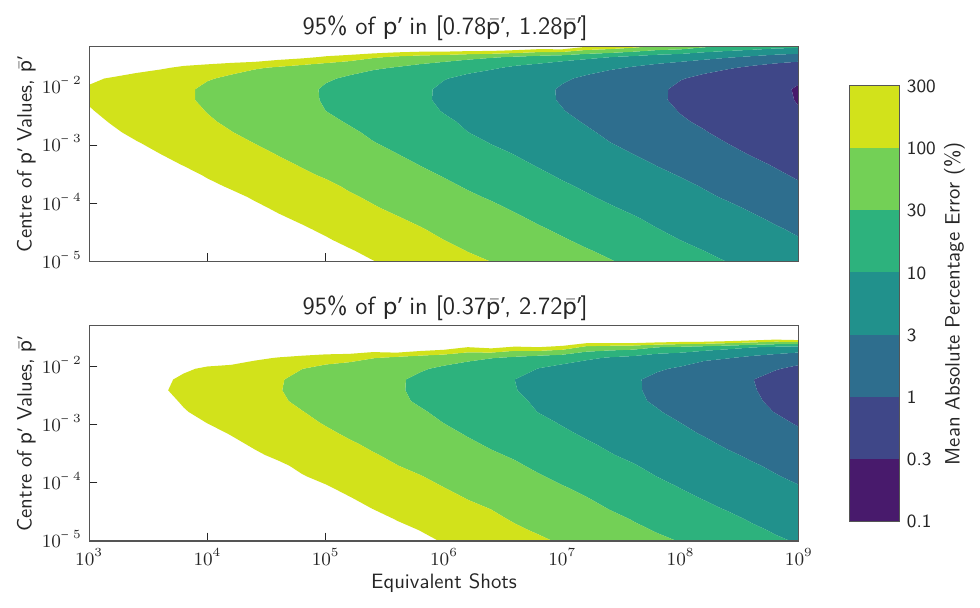}
    \caption{Contour plot of mean absolute percentage error as a function of centre of depolarizing parameters $\Bar{p'}$ and equivalent shots for heavy-hexagon code stabilizer measurement circuits under inhomogeneous depolarizing noise. Random inhomogeneous noise was sampled from a log-normal distribution with centre at $\log(\Bar{p'})$. The spread is represented by the 95\% interval around $p'$ (however depolarization parameters were clipped to fall between $e^{-2}\Bar{p'}$ and $e^{2}\Bar{p'}$). Each circuit was modelled to have effectively 2 cycles, giving one measurement of each of the initial, steady-state and final stabilizer change rates. Operator parity change rates were first calculated exactly, with shot noise approximated by sampling from the associated binomial distribution. Parities of all pairs of detection events were used.}
    \label{fig:error_rates_map_appendix}
\end{figure*}

\end{document}